\documentclass[aps,prd,preprintnumbers,showpacs,showkeys,nofootinbib,
superscriptaddress,fleqn,floatfix,tightenlines,10pt]{revtex4-1}
\usepackage{amsmath,amsfonts,amssymb,amscd,amsxtra,amsthm}
\usepackage{graphicx}  % Include figure files
\usepackage{epstopdf}
\usepackage{dcolumn}  % Align table columns on decimal point
\usepackage{bm}          % bold math
\usepackage{slashed}
\usepackage{cancel}
\usepackage{float}
\usepackage{mathtools}
\usepackage{amsbsy}
\usepackage{amstext}

\usepackage[utf8]{inputenc}
\usepackage{booktabs}
\usepackage[normalem]{ulem} % \sout{old text} for strikeout
\usepackage[dvipsnames]{xcolor} % For blue in-text comments and
\usepackage{tabularx}
\usepackage{enumitem}
\usepackage{array}
\usepackage{slashed}
\usepackage{tikz}
\usepackage{float}
\usepackage{multirow}
\renewcommand\sout{\bgroup \color{red} \ULdepth=-.5ex \ULset}

\makeatletter

\begin{document}
\preprint{INHA-NTG-04/2020}
\title{Strong force fields and stabilities of the nucleon and singly
  heavy baryon $\Sigma_c$}
%--------------------------------------------------
%--------------------------------------------------
\author{June-Young Kim}
\email[E-mail: ]{Jun-Young.Kim@ruhr-uni-bochum.de}
\affiliation{Institut f\"ur Theoretische Physik II, Ruhr-Universit\"at
  Bochum, D-44780 Bochum, Germany}
\author{Hyun-Chul Kim}
\email[E-mail: ]{hchkim@inha.ac.kr}
\affiliation{Department of Physics, Inha University, Incheon 22212,
Republic of Korea}
\affiliation{School of Physics, Korea Institute for Advanced Study
  (KIAS), Seoul 02455, Republic of Korea}
\author{Maxim V. Polyakov}
\email[E-mail: ]{maxim.polyakov@rub.de}
\affiliation{Petersburg Nuclear Physics Institute, Gatchina, 188300,
  St. Petersburg, Russia}
\affiliation{Institut f\"ur Theoretische Physik II, Ruhr-Universit\"at
  Bochum, D-44780 Bochum, Germany}
\author{Hyeon-Dong Son}
\email[E-mail: ]{hyeon-dong.son@ruhr-uni-bochum.de}
\affiliation{Institut f\"ur Theoretische Physik II, Ruhr-Universit\"at
  Bochum, D-44780 Bochum, Germany}
%--------------------------------------------------
\date{\today}
\begin{abstract}
We investigate the strong force fields and stabilities of the nucleon
and the singly heavy baryon $\Sigma_c$ within the framework of the
chiral quark-soliton model. Having constructed the pion mean fields in
the presence of the $N_c-1$ level quarks self-consistently, we are
able to examine the gravitational form factors of $\Sigma_c$. We
mainly focus in the present work on the stability conditions for both
the nucleon and $\Sigma_c$ and discuss the strong force fields and
their physical implications. We also present the results for the
gravitational form factors and relevant observables, emphasising the
difference between the nucleon and $\Sigma_c$. 
\end{abstract}
\maketitle
%--------------------------------------------------
\section{Introduction}
While the hadron form factors of the energy-momentum
tensor (EMT) have been considered as
academic quantities~\cite{Kobzarev:1962wt, Pagels:1966zza} for
decades, the modern concept of the form factors based on the
generalized parton distributions (GPDs)~\cite{Mueller:1998fv, Ji:1996ek,
  Radyushkin:1996nd, Goeke:2001tz, Diehl:2003ny, Belitsky:2005qn} shed
light on the novel understanding of the EMT form
factors~\cite{Teryaev:1999su,Polyakov:2002yz}.   
The EMT makes it possible to couple the gravity
to matter, which is the reason why the EMT form factors are also called the
gravitational form factors (GFFs). The hadronic matrix
elements of the EMT show how the mass and spin of the hadron are
distributed, and moreover provide critical information on how the
corresponding hadron acquires the mechanical stability (see 
recent review~\cite{Polyakov:2018zvc}).
The conservation of the EMT yields a stability condition for 
a hadron. This stability condition, also
known as the von Laue condition, is a nontrivial one, since it plays a
role of the touchstone to disclose the validity of any theories or
models of hadrons. The global stability condition constrains the
3D integral of the pressure inside a hadron should vanish. In fact,
the stability conditions of hadrons are deeply rooted in chiral
symmetry and its spontaneous breakdown. For example, the pressure
inside the pion may hold a clue on understanding the chiral symmetry
and its spontaneous breakdown~\cite{Polyakov:1999gs, Son:2014sna}. 
Yet another form factors, which are called the $D$-term form factors
$D(t)$ and are related to the spatial components of the EMT, are
given in terms of the pressure and shear force,  indicating that the
$D$-term form factors characterize the mechanical properties of the
hadrons~\cite{Polyakov:2002yz}. The first measurement of the nucleon GFF
$D(t)$  in deeply virtual Compton scattering was reported
in Ref.~\cite{Burkert:2018bqq}. See also Ref.~\cite{Kumericki:2019ddg} for
discussion of subtleties in extraction of GFFs from DVCS data.

A singly heavy baryon consists of the light-quark pair and a singly
heavy quark. In the limit of the infinite heavy-quark mass ($m_Q\to
\infty$), the heavy quark spin is preserved. This leads also to the
conservation of the spin of the light-quark degrees of freedom. This
is often called heavy-quark spin symmetry. In this heavy-quark mass
limit, the physics becomes independent of the flavor of a heavy
quark, which is called heavy-quark flavor symmetry. In the nonstrange
sector, we have two different representations: the isospin singlet and
triplets. Since the spin state of the light-quark pair is either a spin
singlet or a spin triplet, the isospin singlet that consists of a
single member $\Lambda_c$ becomes naturally a spin singlet. On the
other hand, the isospin triplets have degeneracy between the
spin-$1/2$ and -$3/2$ states, which correspond to $\Sigma_c$ with spin
1/2 and $\Sigma_c^*$ with spin 3/2, respectively. The chromomagnetic
hyperfine interaction is responsible for the removal of this
degeneracy. In the present work, we will consider the GFFs of
$\Sigma_c$ with spin 1/2. Since the light-quark pair inside
$\Lambda_c$ is in spin-zero state and $\Sigma_c^*$ has spin 3/2, it is
more instructive to consider the GFFs of $\Sigma_c$ in order to
compare them with those of the nucleon.

The GFFs of the nucleon have been extensively
investigated within various
approaches~\cite{Ji:1997gm,Polyakov:2002wz,Schweitzer:2002nm, 
  Goeke:2007fp, Goeke:2007fq, Wakamatsu:2007uc, Cebulla:2007ei,
  Jung:2013bya, Hagler:2003jd, Gockeler:2003jfa, Hagler:2007xi,
  Pasquini:2007xz, Hwang:2007tb, Abidin:2008hn, Brodsky:2008pf,
  Pasquini:2014vua,  Chakrabarti:2015lba,Teryaev:2016edw,
  Lorce:2018egm,Shanahan:2018nnv,Shanahan:2018pib, Neubelt:2019sou,
  Anikin:2019kwi,Anikin:2019ufr,Alharazin:2020yjv,Varma:2020crx}. 
  The nucleon
GFFs have been also examined within nuclear matter~\cite{Kim:2012ts,
  Jung:2014jja}. In the present work, we want for first time to study the
GFFs of the singly heavy baryon $\Sigma_c$ that contains a charm quark
and two light quarks, emphasizing the comparison with those of the
nucleon~\cite{Goeke:2007fp}, based on the chiral quark-soliton model
($\chi$QSM). The model was successfully extended to the description of
the singly heavy baryons~\cite{Yang:2016qdz}, being motivated by
Ref.~\cite{Diakonov:2010tf}. The model reproduced well the masses of
the singly heavy baryons~\cite{Yang:2016qdz, Kim:2018xlc, Kim:2019rcx}.
A singly heavy baryon within the $\chi$QSM is viewed as a bound
state of $N_c-1$ level quarks. The presence of $N_c-1$ level 
quarks creates the vacuum polarization or the pion mean fields that
affects self-consistently the $N_c-1$ level quarks. On the other
hand, a heavy quark can be regarded as a mere static color source in
the limit of the infinitely heavy quark mass ($m_Q\to\infty$). In a
recent work~\cite{Kim:2019rcx}, it was found that the presence of
$N_c-1$ valence quarks produces weaker pion mean fields in comparison
with the case of a light baryon that consists of the $N_c$ valence
quarks. While Refs.~\cite{Yang:2016qdz, Kim:2018xlc} have assumed that
this modification of the pion mean fields can be neglected and a
simple strategy was taken, in which the color factor $N_c$ is replaced
merely by $N_c-1$ for the valence contributions to the mass spectra of
the singly heavy baryons. However, as we will describe later, the
modification of the pion mean fields is crucial in computing the GFFs
of a singly heavy baryon, since the stability condition is
otherwise not satisfied. Therefore, we will adopt in the present work
the modified pion mean fields to study the GFFs of the heavy baryon.
The chiral soliton that consists of $N_c-1$ valence quark has either
spin 0 or spin 1. The soliton with spin 0 will construct $\Lambda_c$
by combining itself with a heavy quark, whereas that with spin 1 will
produce $\Sigma_c$ and $\Sigma_c^*$ together with a heavy quark. Since
we are interested in comparing the GFFs of the singly heavy baryon
with those of the nucleon, we will concentrate in the present work on
the GFFs of $\Sigma_c$.

We sketch how the present work is organized: In Section~\ref{sec:2}
and ~\ref{sec:3}, we recapitulate the general formalism for the GFFs
and the stability condition for spin 
$1/2$ baryons briefly and explain how the GFFs of the singly heavy baryons
 can be computed within the framework of the $\chi$QSM.
 In Section~\ref{sec:4}, we present the numerical
results for the GFFs of the heavy baryon $\Sigma_c$, emphasizing the
comparison of them with those for the nucleon. We first discuss the
energy densities of the $\Sigma_c$. Then we proceed to describe the
densities of the angular momentum. Since the pressure and shear force
are the essential quantities to reveal the mechanical stability of
the $\Sigma_c$, we discuss them in detail. We also discuss the
mechanical equation of states, i.e. a functional relation between the
energy density and the pressure. Finally, we present the main results
for the GFFs of the charmed baryon $\Sigma_c$.
In the final Section we make summary of the present work, draw
conclusions, and give outlook.

\section{Gravitational form factors and stability
  conditions\label{sec:2}}
\subsection{Gravitational form factors of a baryon with spin 1/2}
The unpolarized GPDs $H^{q}(x,\xi,t)$ and $E^{q}(x,\xi,t)$ parametrize
the matrix element of nonlocal two-quark operators on the
light-cone. In the leading twist order, it can be decomposed in terms
of the unpolarized GPDs as follows
\begin{align}
&\int \frac{d\lambda}{2\pi} e^{i\lambda x} \langle p',\sigma' |
                \overline{\psi}_{q}\left(-\frac{\lambda n}{2}\right)
                \slashed{n}  \psi_{q}\left(\frac{\lambda n}{2}\right)
                | p,\sigma  \rangle \cr
  &  = H^{q}(x,\xi,t) \overline{u}(p',\sigma')\slashed{n} u(p,\sigma )
    + E^{q}(x,\xi,t) \overline{u}(p',\sigma' ) \frac{i\sigma^{\mu\nu}
    n_{\mu}    \Delta_{\nu}}{2M_{B}} u(p,\sigma),
\label{eq:GPDs}
\end{align}
where ${\psi}_{q}$ is the quark field with flavor $q$ on the light
cone. $\sigma_{\mu\nu}$ is defined by $\sigma_{\mu\nu}=i
[\gamma_\mu,\,\gamma_\nu]/2$.  $M_B$ is the corresponding mass of a
baryon. $\sigma$($\sigma'$) denotes the helicity of the 
  initial (final) baryon state. The baryon states and Dirac spinors are
  normalized by $\langle p',\sigma' |  p,\sigma \rangle =
  2p^{0}\delta_{\sigma'\sigma}(2\pi)^{3}\delta(\bm{p}'-\bm{p})$ and
  $\overline{u}(p)u(p) =2M_{B} \delta_{\sigma'\sigma}$, respectively.
The average of the baryon momenta and the momentum transfer are
respectively defined by $P=(p+p')/2$ and $\Delta=(p'-p)$. $t$
designates the square of the momentum transfer,
$t=\Delta^{2}$. We leave out the gauge connection between the quark
operators because it becomes unity in the light-cone gauge. Note that
we have suppressed the renormalization scale dependence for
simplicity. The light-like vector $n$ satisfies $n^2=0$ and $n\cdot
(p'+p)= 2$. $x$ denotes the longitudinal momentum fraction of a baryon
carried by a parton whereas the $\xi$ stands for the skewedness,
defined as $n \cdot \Delta=-2\xi$.

Form factors of a baryon can be defined  by the Mellin
moments of the GPDs with respect to $x$. The first Mellin moments of
the unpolarized GPDs given in Eq.~\eqref{eq:GPDs} yield the
electromagnetic form factors of a baryon with spin 1/2 as follows:
\begin{align}
&\int^{1}_{-1} dx \sum_{q} H^{q}(x,\xi,t) = F_{1}(t),  \ \ \ \ \
 \int^{1}_{-1} dx \sum_{q} E^{q}(x,\xi,t) = F_{2}(t),
\end{align}
where $F_1(t)$ and $F_2(t)$ are the well-known Dirac and Pauli form
factors of a baryon respectively. These two form factors are
traditionally defined by the baryonic matrix element of the
electromagnetic current
\begin{align}
\langle p', \sigma' | \overline{\psi}_{q}(0) \gamma^{\mu} \psi_{q}(0)
  | p, \sigma \rangle = \overline{u}(p',\sigma') \left[ F^{q}_{1} (t)
  \gamma^{\mu} + F^{q}_{2}(t)\frac{i
  \sigma^{\mu\nu}\Delta_{\nu}}{2M_{B}} \right] u(p,\sigma).
\end{align}
The first Mellin moments of the unpolarized GPDs for the singly heavy
baryons, i.e., the electromagnetic form factors of the singly heavy
baryons, were already investigated in Ref.~\cite{Kim:2018nqf}.
The second Mellin moments of the GPDs are as equally important as
the first Mellin moments, because they provide the GFFs of a baryon,
which reveal mechanical properties of a baryon.
Thus, the GFFs of a baryon can be defined as the second Mellin Moments
as follows:
\begin{align}
  \int^{1}_{-1} dx \ x \ \sum_{q}  H^{q}(x,\xi,t) &=
 A^{Q}(t)+D^{Q}(t) \xi^{2}, \cr
\int^{1}_{-1} dx \ x \ \sum_{q} E^{q}(x,\xi,t) &=
2J^{Q}(t)-A^{Q}(t) -D^{Q}(t) \xi^{2},
\end{align}
where $J^{Q}(t)$, $A^{Q}(t)$ and $D^{Q}(t)$ are respectively
defined by $J^{Q}=\sum_{q} J^{q}$, $A^{Q}=\sum_{q} A^{q}$ and
$D^{Q}=\sum_{q} D^{q}$. The superscript $Q$ emphasizes the
quark part of the GFFs. Note that a GPD should satisfy a
polynomiality, which means that the $n$th Mellin moment of a GPD can
be expressed in terms of a polynomial given by
only even power of the skewedness $\xi$. The GFFs of a baryon with
spin $1/2$ can be also defined by the matrix element of the EMT
current
\begin{align}
\langle p',\sigma' | \hat{T}^{a}_{\mu\nu}(0) | p ,\sigma \rangle =
  \overline{u}(p',\sigma')
  &\bigg{[} A^{a}(t) \frac{P_{\mu}P_{\nu}}{M_{B}} +
    J^{a}(t) \frac{i(P_{\mu}\sigma_{\nu\rho} +
    P_{\nu}\sigma_{\mu\rho})\Delta^{\rho}}{2M_{B}} \cr
  &+ D^{a}(t) \frac{\Delta_{\mu}\Delta_{\nu}-g_{\mu\nu}\Delta^{2}}{4M_{B}}
 + \overline{c}^{a}(t) M_{B}g_{\mu\nu}\bigg{]} u(p,\sigma),
\end{align}
where $\hat{T}^{a=Q(G)}_{\mu\nu}$ stands for the quark (gluon) part of the
QCD EMT current. $A^{a}$, $J^{a}$, and $D^{a}$ represent
respectively the mass form factor, the angular momentum form factor, and
the $D$-term form factor. Once these quark and gluon GFFs are separate
from each other, the corresponding currents are not anymore conserved,
since 
\begin{align}
\partial^{\mu}\hat{T}_{\mu\nu} = 0, \ \ \ \ \ \hat{T}_{\mu\nu} =
  \hat{T}^{Q}_{\mu \nu} + \hat{T}^{g}_{\mu \nu} =
  \sum_{q}\hat{T}^{q}_{\mu \nu} + \hat{T}^{g}_{\mu \nu}.
\label{eq:EMTcon}
\end{align}
Thus, the quark or gluon part of the GFFs should only depend on a
specific scale $\mu$. However, we will suppress the scale dependence
of the form factors for brevity.  Actually, Eq.~\eqref{eq:EMTcon}
constrains the non-conservation term $\overline{c}^{a}(t)$ to be
  $\overline{c}^{a}(t)=\sum_{q=Q,G}\overline{c}^{a}(t,\mu) = 0$, and
  the total GFFs turn out to be renormalization scale invariant, i.e,
  $D(t)=\sum_{a}D^{a}(t,\mu)$. In fact, since the gluon degrees
of freedom have been already integrated out in the present effective
approach, for example, through instantons, we do not
have any contributions from the gluon EMT current.

In the Breit frame, both the quark and gluon parts of the GFFs are defined
by
 \begin{align}
 T^{a}_{\mu\nu}(\bm{r},\sigma',\sigma)=\int
   \frac{d^{3}\Delta}{(2\pi)^{3}2E} e^{-i\bm{\Delta}\cdot \bm{r}}
   \langle p',\sigma' | \hat{T}^{a}_{\mu\nu}(0) | p, \sigma \rangle.
\end{align}
The temporal component $T^{a}_{00}$ is related to the energy
density of partons inside a baryon
\begin{align}
\frac{1}{M_{B}} T^{a}_{00}(\bm{r},\sigma',\sigma) =  \int
  \frac{d^{3}\Delta}{(2\pi)^{3}} e^{-i\bm{\Delta} \cdot \bm{r}} \left[
  A^{a}(t) + \overline{c}^{a}(t) - \frac{t}{4M_{B}^{2}}\left(
  A^{a}(t) - 2J^{a}(t) + D^{a}(t) \right) \right]
  \delta_{\sigma' \sigma}.
\end{align}

By Integrating $T^{a}_{00}$ over space, one gets the mass of a
spin-1/2 baryon in the rest frame
\begin{align}
 \int d^{3}r \sum_{a=Q,G}  T^{a}_{00}(\bm{r},\sigma',\sigma) = M_{B}
  A(0)=M_{B},
\end{align}
with the normalized mass form factor $A(0)=1$, where the
contribution of $\overline{c}^{a}$ to $T_{00}$ is found to be zero by
Eq.~\eqref{eq:EMTcon}. 

The mixed components $T^{a}_{0i}$ are relevant to the linear
momentum and total angular momentum (spin + orbital angular momentum)
densities, carried by the partons inside a baryon. According to the
angular momentum operator in QCD, we define the total angular momentum
distributions inside a baryon as
  \begin{align}
J^{a,i}_{\sigma'\sigma}(\bm{r})=\epsilon^{ijk}r_{j}T^{a}_{0k}(\bm{r},\sigma',\sigma)
  = 2S^{j}_{\sigma'\sigma}\int \frac{d^{3}\Delta}{(2\pi)^{3}}
  e^{-i\bm{\Delta} \cdot \bm{r}} \left[ \left(J^{a}(t)+ \frac{2}{3}t
  \frac{J^{a}(t)}{dt}\right)\delta^{ij} + \left(\Delta^{i}\Delta^{j}-
  \frac{1}{3}\bm{\Delta}^{2} \delta^{ij}\right) \frac{J^{a}(t)}{dt}
  \right].
\end{align}
Integrating $J^{a,i}_{\sigma'\sigma}(\bm{r}) $ over space gives the
spin of the baryon as follows
\begin{align}
\int dr^{3} \sum_{a=Q,G} J^{a,i}_{\sigma'\sigma}(\bm{r}) =
  2\hat{S}^{i}_{\sigma'\sigma} J(0) = \hat{S}^{i}_{\sigma'\sigma},
\end{align}
which is just the spin operator of a baryon.

The spatial components $T^{a}_{ij}$ charaterize mechanical properties
of a baryon such as the pressure $p(r)$ and shear force $s(r)$
distributions inside a baryon. $T^{a}_{ij}$ is decomposed in terms of
the irreducible tensors, so that the pressure and shear force
distributions are expressed as
\begin{align}
T^{a}_{ij}(\bm{r},\sigma',\sigma) =  p^{a}(r) \delta^{ij} \delta
_{\sigma'\sigma}+ s^{a}(r)\bigg{(} \frac{r^{i}r^{j}}{r^{2}} -
  \frac{1}{3} \delta^{ij}\bigg{)} \delta_{\sigma'\sigma}.
\end{align}
where the pressure and shear force distributions are defined as
\begin{align}
  p^{a}(r)=\frac{1}{6M_{B}}\frac{1}{r^{2}}\frac{1}{dr}r^{2}\frac{d}{dr}
  \tilde{D}^{a}(r)-M_{B}\int
  \frac{d^{3}\Delta}{(2\pi)^{3}}e^{-i\bm{\Delta} \cdot
  \bm{r}}\overline{c}^{a}(t), \ \ \
  s^{a}(r)=-\frac{1}{4M_{B}}r\frac{d}{dr}\frac{1}{r}\frac{d}{dr}\tilde{D}^{a}(r),
\end{align}
with
\begin{align}
\tilde{D}^{a}(r) = \int \frac{d^{3}\Delta}{(2\pi)^{3}}
  e^{-i\bm{\Delta}\cdot \bm{r}} D^{a}(t).
\end{align}
Equivalently, the form factor $D(t)$ can be obtained by the
Fourier transform
\begin{align}
D(t) = 4M_{B} \int d^{3}r \frac{j_{2}(r\sqrt{-t})}{t} s(r) =
  12M_{B} \int d^{3}r \frac{j_{0}(r\sqrt{-t})}{2t} p(r).
\label{eq:d-term2}
\end{align}
Note that the shear force distributions of the gluon and
quark separately are independent of $\overline{c}^{a}(t)$ whereas
one should know it to determine the pressure distributions.
In addition, we introduce a new form factor $F(t)$ that is defined by the
matrix element of the trace of the total EMT operator
\begin{align}
\langle p',\sigma' | \hat{T}^{\mu}_{\mu}(0) | p, \sigma \rangle =
  M_{B}F(t) \overline{u}(p',\sigma'){u}(p,\sigma),
\end{align}
where $F(t)$ contains all the GFFs, given by
\begin{align}
F(t) =A(t) +\frac{t}{4M^{2}_{B}} \bigg{(}2J(t) - A(t)\bigg{)}
  - \frac{3t}{4M^{2}_{B}}D(t).
\end{align}
Then, the mean square radius $\langle r^{2}_{F} \rangle$ is obtained
to be
\begin{align}
\langle r^{2}_{F} \rangle = 6\frac{d F(t)}{dt}\bigg{|}_{t=0}
  =6\left(\frac{d A(t)}{dt} - \frac{3D(t)}{4M^{2}_{B}}\right)
  \bigg{|}_{t=0}.
\end{align}

\subsection{Stability conditions for a baryon with spin 1/2}
In the static
case, the spatial part of the EMT current satisfies the following conservation
law
\begin{align} \partial^{i}T_{ij} = \frac{r_{j}}{r}\left[ \frac{2}{3}\frac{
\partial s(r)}{ \partial r}+ \frac{2s(r)}{r} + \frac{ \partial p(r)}{ \partial
  r}\right] = 0.
  \label{eq:diffEq}
\end{align}
Thus, the shear force and
pressure distributions are related each other by the differential equation
given in Eq.~\eqref{eq:diffEq}. By integrating Eq.~\eqref{eq:diffEq} over
space, we have one of the most important stability conditions, that is, the
so-called von Laue stability condition,
\begin{align}
  \int^{\infty}_{0} dr \   r^{2} p(r)=0,
  \label{eq:stability}
\end{align}
which implies that the pressure
distribution has at least one node. In addition to the condition of the
global stability given by Eq.~\eqref{eq:stability}, one can consider the
conditions of the local stability~\cite{Perevalova:2016dln, Polyakov:2018rew,
Lorce:2018egm} by introducing a concept of strong force field. Given the EMT
densities, strong force fields can be defined as
\begin{align}
\label{eq:strong_force} dF^{i}_{(r,\theta,\phi)} = T^{ij}
  dS_{(r,\theta,\phi)}\bm{e}^{j}_{(r,\theta,\phi)},
\end{align}
where the normal
and tangential pressure densities corresponding to\footnote{The
  normal and tangential force fields
$F_{r}$ and $F_{\phi}$ are respectively defined as $4\pi
r^{2}p_{r}(r)$ and $4\pi r^{2}p_{\phi}(r)$ acting on 
the spherical shell of the radius $r$ in a hadron.} $F^i$ are defined
respectively by
\begin{align}
  \label{eq:force_comp}
p_r(r) :=  \frac{dF_{r}}{dS_{r}} =
  \frac{2}{3}s(r) + p(r), \;\;\;
p_\theta(r) :=  \frac{dF_{\theta}}{dS_{\theta}} =
-\frac{1}{3}s(r) + p(r),\;\;\;
p_\phi (r):=\frac{dF_{\phi}}{dS_{\phi}} = -\frac{1}{3}s(r) + p(r) ,
\end{align}
where $p_\theta=p_\phi$. Using Eq.~\eqref{eq:strong_force}, Perevalova
et al.~\cite{Perevalova:2016dln} examined a local criterion for the
stability and found that at any distance
$r$ the normal force should be directed outwards. This is often called the
mechanical stability of a hadron. This leads to the explicit local criterion
for the mechanical stability formulated as
\begin{align}
p_r(r) > 0.
  \label{eq:mecstab1}
\end{align}
This condition allows one to introduce the mechanical radius of a
hadron
\begin{align}
\langle r^{2} \rangle_{\mathrm{mech}} = \frac{\int d^{3} r~r^{2}
p_r(r) }{\int d^{3} r~ p_r(r)} =
  \frac{6D(0)}{\int^{0}_{-\infty}  D(t) dt} .
  \label{eq:mecradius}
\end{align}
Meanwhile, one can establish an additional stability condition by interpreting
the tangential force as a two-dimensional (2D) subsystem of the whole
three-dimensional (3D) system~\cite{Polyakov:2018zvc}. Then the 2D von Laue
stability condition can be derived as
\begin{align}
  \int^{\infty}_{0} dr \; r \; p_\phi =0.
  \label{eq:stability2D}
\end{align}

\section{Gravitational form factors of $\Sigma_c$ within the SU(2)
  chiral quark-soliton model\label{sec:3}}
We will now briefly show how to compute the GFFs of the singly heavy
baryon $\Sigma_c$ within the framework of the SU(2) $\chi$QSM.
We begin from the low-energy effective QCD partition function in
Euclidean space
\begin{align}
&Z_{\mathrm{eff}} = \int \mathcal{D}\psi
                \mathcal{D}\psi^{\dagger}\mathcal{D}U
                \mathrm{exp}\left[-\int d^{4}x \psi^{\dagger}
                D(U)\psi\right]= \int \mathcal{D}U
                \mathrm{exp}[-S_{\mathrm{eff}}],
\end{align}
where $S_{\mathrm{eff}}$ is the effective chiral action
\begin{align}
&S_{\mathrm{eff}}(U) = -N_{c}\mathrm{Tr\;ln}\;D(U).
\label{eq:effecXac}
\end{align}
$N_{c}$ denotes the number of colors. The Dirac operator $D(U)$ in
Eq.~\eqref{eq:effecXac} is defined by
\begin{align}
D(U)= i \slashed{\partial} + i \hat{m} + i  MU^{\gamma_{5}},
\end{align}
where $U^{\gamma_5}$ is called the chiral field, defined as
\begin{align}
U^{\gamma_5} = U\frac{1+\gamma_5}{2} + U^\dagger \frac{1-\gamma_5}{2},
\end{align}
with
\begin{align}
U = \exp\left[i{\pi^a \tau^a}\right] = \exp\left[i{P(r)\bm{n}\cdot
  \bm{\tau}}\right]. 
\end{align}
$\pi^a$ represents the pseudo-Nambu-Goldstone field and $\hat{m}$ is
the flavor matrix of the current quark masses, written as
$\hat{m}=\mathrm{diag}(m_{\mathrm{u}},\,m_{\mathrm{d}})$. We assume in
the present work isospin symmetry,
i.e. $m=m_{\mathrm{u}}=m_{\mathrm{d}}$.
Note that we introduce the hedgehog ansatz $\pi^a=P(r) n^a$ with the
profile function of the chiral soliton $P(r)$, which will be
determined by solving the classical equation of motion. $\bm{n}$ is
the normal unit vector along the radial direction. 
The Dirac Hamiltonian $h(U)$
is defined as
\begin{align}
h(U)= i \gamma_{4}\gamma_{i}\partial_{i} -
  \gamma_{4}MU^{\gamma_{5}}-\gamma_{4}m.
\end{align}
The corresponding eigenenergies and eigenfunctions are obtained by
diagonalizing the Dirac Hamiltonian as follows
\begin{align}
h(U) \Phi_{n}(\bm{r}) = E_{n} \Phi_{n}(\bm{r}),
\label{eq:one_enregy}
\end{align}
where $E_n$ denote the eigenenergies of the Hamiltonian $h(U)$ and
$\Phi_{n}(\bm{r})$ stand for the quark eigenfunctions.
Similarly, Likewise, the free Dirac Hamiltonian ${h}_{0}$ can be
defined by replacing the chiral field by the unity. The eigenvalues of
$h_0$ are expressed as $E_{n^0}$.
The classical soliton for singly heavy baryons consist of $N_c-1$
discrete level quarks bound by the pion mean field. Thus, the classical
equation of motion can be derived by minimizing the energy of the
classical $N_{c}-1$ soliton
\begin{align}
\left.\frac{\delta}{\delta U(\bm{r})}[ (N_c-1)  E_{\mathrm{lev}} +
  E_{\mathrm{con}}]\right|_{U_c} = 0.
\label{eq:saddle}
\end{align}
where $E_{\mathrm{lev}}$ stands for the energy of the discrete bound
level, and $E_{\mathrm{con}}$ is the sum of the lower Dirac continuum
energies. Here, $U_c$ is the solution of the classical equation of
motion, so that it is identified as the pion mean field. Thus, the
$N_{c}-1$ soliton mass is finally derived as
\begin{align}
M_{\mathrm{sol}} = (N_c-1) \theta(E_{\mathrm{lev}})
  E_{\mathrm{lev}}(U_c) + E_{\mathrm{con}}(U_c).
\label{eq:solnc}
\end{align}
The detailed calculations are presented in
Ref.~\cite{Kim:2019rcx}. Note that as discussed in
Ref.~\cite{Goeke:2007fp}, the stability condition is secured by the
solution of the classical equation of motion.

The EMT current can be written as
\begin{align}
\hat{T}_{\mu\nu}^{\mathrm{(eff)}} =& -\frac{i}{4}\psi^\dagger (x)\left(
                                   i\gamma^{\mu}
                                   \overrightarrow{\partial}^{\nu}
                                   +i\gamma^{\nu}
                                   \overrightarrow{\partial}^{\mu} -
                                   i\gamma^{\mu}
                                   \overleftarrow{\partial}^{\nu} -
                                   i\gamma^{\nu}
                                   \overleftarrow{\partial}^{\mu}
                                   \right)
                                     {\psi}(x) \cr
&
                                     -\frac{i}{4}\Psi^\dagger
                                     (x) \left(
                                   i\gamma^{\mu}
                                   \overrightarrow{\partial}^{\nu}
                                   +i\gamma^{\nu}
                                   \overrightarrow{\partial}^{\mu} -
                                   i\gamma^{\mu}
                                   \overleftarrow{\partial}^{\nu} -
                                   i\gamma^{\nu}
                                   \overleftarrow{\partial}^{\mu}
                                   \right) {\Psi}(x),
\label{eq:EMT_current}
\end{align}
where $\Psi$ denotes the heavy-quark field.
Since we are interested in the GFFs of the singly heavy baryons, we
need to explain how the heavy quark inside in them is treated. In
the limit of the infinitely heavy-quark mass, i.s., $m_Q\to\infty$,
the heavy quark can be regarded as a mere static color source. Thus,
it does not play any important role in computing the heavy baryon
GFFs for which the light quarks govern dynamics of quarks inside a
singly heavy baryon.

The matrix element of the EMT current given in
Eq.~\eqref{eq:EMT_current} can be computed by considering the
following baryonic correlation function
\begin{align}
\langle B, p' | \hat{T}_{\mathrm{eff}}^{\mu\nu}(0) | B, p \rangle
  =\frac{1}{Z_{\mathrm{eff}}} \lim_{T\to \infty}
  \mathrm{exp}\left(ip_{4}\frac{T}{2}-ip'_{4}\frac{T}{2}\right) \int
  d^{3}\bm{x} d^{3}\bm{y} \ \mathrm{exp}\left(-i\bm{p}'\cdot
  \bm{y}+i\bm{p}\cdot \bm{x}\right)  \cr
\times \int \mathcal{D} \psi \mathcal{D} \psi^{\dagger} \mathcal{D}U
  J_{B}(\bm{y},T/2) T_{\mu\nu}^{\mathrm{(eff)}}(0) J^{\dagger}_{B}
  (\bm{x},-T/2) \mathrm{exp}\left[-\int d^{4}
  \psi^{\dagger}D(U)\psi\right],
\label{eq:correlftn}
\end{align}
where $J_{B}$ denotes the Ioffe-type baryonic current for the $N_c-1$
discrete level quarks, which is expressed as
\begin{align}
J_B(x) = \frac1{(N_c-1)!} \epsilon_{i_1\cdots i_{N_c-1}} \Gamma_{JJ_3
  TT_3}^{\alpha_1\cdots \alpha_{N_c-1}} \psi_{\alpha_1 i_1} (x)
  \cdots \psi_{\alpha_{N_c-1} i_{N_c-1}}(x).
\end{align}
Here, $\alpha_1\cdots \alpha_{N_c-1}$ represent spin-flavor indices whereas
$i_1\cdots i_{N_c-1}$ color indices. The matrices $\Gamma_{JJ_3
  TT_3}^{\alpha_1\cdots \alpha_{N_c-1}}$ are projection operators that
will pick up a light-quark component of the singly heavy baryon with proper
quantum numbers $JJ_3TT_3$. The creation operator $J_B^\dagger$ can be
constructed in a similar way.
The baryon states $| B,p \rangle$and $\langle B,p
|$ are, respectively, defined by
\begin{align}
| B,p \rangle &= \lim_{x_{4}\to-\infty}
                \mathrm{exp}(-ip_{4}x_{4})\frac{1}{\sqrt{Z_{\mathrm{eff}}}}
                \int d^{3}x\;\mathrm{exp}(i \bm{p}\cdot \bm{x})
                J_{B}^{\dagger}(\bm{x},x_{4}) | 0 \rangle, \cr
\langle B,p | &= \lim_{y_{4}\to \infty}
                \mathrm{exp}(-ip'_{4}x_{4})\frac{1}{\sqrt{Z_{\mathrm{eff}}}}
                \int d^{3}y\;\mathrm{exp}(-i \bm{p}'\cdot \bm{y})
                \langle 0 | J_{B}(\bm{y},y_{4}).
\end{align}
As for a detailed formalism of the zero-mode quantization and the
techniques of computing the baryonic correlation function given in
Eq.~\eqref{eq:correlftn}, we refer to Ref.~\cite{Christov:1995vm}.
The final form of the collective wave functions should be constructed
by combining the $N_c-1$ light-quark component of the singly heavy
baryon with a singly heavy quark according to a standard algebra of
the angular momentum addition.

The collective baryon wave functions are finally obtained by coupling
the collective $N_c-1$ light-quark wave functions $\psi_{J \lambda
  TT_{3}}(A)$ with the heavy quark spinor $\chi_{mm_{3}}$, as
\begin{align}
\Psi_{\frac{1}{2}\sigma TT_{3}}(A) = \sum_{J \lambda mm_3} C^{\frac{1}{2}
  \sigma}_{J \lambda m m_{3}} \psi_{J \lambda TT_{3}}(A) \chi_{m
  m_{3}},
  \label{eq:collWf}
\end{align}
where $\psi_{J \lambda  TT_{3}}(A)$ are expressed as
\begin{align}
\psi_{J \lambda TT_{3}}(A) = (-1)^{T+T_{3}}\sqrt{2T+1}D^{J=T}_{-T_{3},
  \lambda }(A).
\end{align}
$\sigma$ in Eq.~\eqref{eq:collWf} denotes the third component of the
singly heavy baryon spin. The $T$ and
$T_{3}$ stand for the isospin and its third component,
respectively. The $J$ and $\lambda$ represent respectively the total
spin of the $N_c-1$ light-quark and its third component. The $m$ and
$m_{3}$ designate the heavy-quark spin and its third component,
respectively.

The matrix elements of the temporal, spatial, and mixed components of
the EMT current are expressed in the large $N_c$ limit in terms of the GFFs
\begin{align}
\langle p', \sigma' | \hat{T}^{00}_{\mathrm{eff}} | p, \sigma
  \rangle &= 2M^{2}_{B}  \bigg{(} A(t)- \frac{t}{4M^{2}_{B}
            }D(t)\bigg{)}\delta_{\sigma'\sigma}, \cr
\langle p', \sigma' | \hat{T}^{ik}_{\mathrm{eff}} | p, \sigma
            \rangle &= \frac{1}{2} \bigg{(} \Delta^{i}\Delta^{k} -
                      \delta^{ik} \bm{\Delta}^{2} \bigg{)}
                      D(t) \delta_{\sigma'\sigma}, \cr
\langle p', \sigma' | \hat{T}^{0k}_{\mathrm{eff}} | p, \sigma
                      \rangle &= -2 i M_{B} \varepsilon^{klm}
                                \hat{S}^{m}_{\sigma'\sigma} \Delta^{l} J(t).
\end{align}
The matrix elements of the light-quark part of the current are
obtained in the $\chi$QSM as
\begin{align}
\langle p', \lambda' | \hat{T}^{00}_{\mathrm{eff}} | p, \lambda
  \rangle&=2M_{B}  \delta_{\lambda' \lambda} \left( (N_{c}-1)
           E_{\mathrm{lev}} \langle \mathrm{lev} | e^{i \bm{\Delta}
           \cdot \bm{x}}  | \mathrm{lev} \rangle + N_{c}\sum_{n}
           R_{1}(E_{n},\Lambda) \langle n | e^{i \bm{\Delta} \cdot \bm{r}}  |
           n \rangle \right), \cr
\langle p', \lambda' | \hat{T}^{ik}_{\mathrm{eff}} | p, \lambda
           \rangle&= \frac{M_{B}}{2} \delta_{\lambda' \lambda}\left(
                    (N_{c}-1)\langle \mathrm{lev} |
                    \{e^{i\bm{\Delta}\cdot\bm{r}},\gamma^{0}\gamma^{i}{p}^{k}\}
                    | \mathrm{lev} \rangle +
                    N_{c}\sum_{n}R_{2}(E_{n},\Lambda)\langle n |
                    \{e^{i\bm{\Delta}\cdot\bm{r}},\gamma^{0}\gamma^{i}{p}^{k}\}
                    | n \rangle\right. \cr
                    & \hspace{3cm} + (i \leftrightarrow k) \bigg), \cr
\langle p', \lambda' | \hat{T}^{0k}_{\mathrm{eff}} | p, \lambda
                    \rangle&=  \frac{M_{B}}{4I} \hat{S}^{l}_{\lambda'
                             \lambda}\bigg{(}(N_{c}-1)\sum_{ j \neq
                             \mathrm{lev}}\frac{\langle \mathrm{lev} |
                             \tau^{l} | j
                             \rangle}{E_{\mathrm{lev}}-E_{j}} \langle
                             j | \left\{ e^{i\bm{\Delta}\cdot
                             \bm{r}},p^{k} \right\}+
                             (E_{\mathrm{lev}}+E_{j})
                             \gamma^{0}\gamma^{k} e^{i \bm{\Delta}
                             \cdot \bm{r}}  | \mathrm{lev} \rangle
                             \cr
&+N_{c}\sum_{m\neq j} R_{3}(E_m,E_j,\Lambda)\langle m | \tau^{l} | j \rangle
                                    \langle j | \left\{
                                    e^{i\bm{\Delta}\cdot \bm{r}},p^{k}
                                    \right\}+ (E_{m}+E_{j})
                                    \gamma^{0}\gamma^{k} e^{i
                                    \bm{\Delta} \cdot \bm{r}}  | m
                                    \rangle  \bigg{)},
\label{eq:EMT_light}
\end{align}
where $R_1(E_n,\Lambda)$, $R_2(E_n,\Lambda)$ and
$R_3(E_m,E_j,\Lambda)$ denote the regularization functions with a
cutoff mass $\Lambda$ for the Dirac-continuum contributions.
The corresponding expressions can be found in Appendix~\ref{app:a}.
The expression for the moment of inertia $I$ can be found in
Ref.~\cite{Christov:1995vm}.
Here, we restrict ourselves to take $\sigma'=\sigma=1/2$ without loss
of any generality. Then, the expressions for the GFFs of the light-quark
part are simplified to be
\begin{align}
&A(t) - \frac{t}{4M^{2}_{B}}D(t) =
                 \int\frac{d\Omega_{\Delta}}{4\pi}\frac{1}{M_{B}}
                 \left( (N_{c}-1) E_{\mathrm{lev}} \langle
                 \mathrm{lev} | e^{i \bm{\Delta} \cdot \bm{r}}  |
                 \mathrm{lev} \rangle + N_{c}\sum_{n} R_{1}(E_{n},\Lambda)
                 \langle n | e^{i \bm{\Delta} \cdot \bm{r}}  | n
                 \rangle \right), \cr
&D(t)=\int\frac{d\Omega_{\Delta}}{4\pi} \frac{M_{B}}{t}  \left(
                                         (N_{c}-1)\langle \mathrm{lev}
                                         |
                                         \{e^{i\bm{\Delta}\cdot\bm{r}},\gamma^{0}\bm{\gamma}
                                         \cdot \bm{p}\}  |
                                         \mathrm{lev} \rangle +
                                         N_{c}\sum_{n}R_{2}(E_{n},\Lambda)\langle
                                         n |
                                         \{e^{i\bm{\Delta}\cdot\bm{r}},\gamma^{0}\bm{\gamma}
                                         \cdot \bm{p}\}  | n \rangle
                                         \right), \cr
&J(t)=   \int\frac{d\Omega_{\Delta}}{4\pi}
                                                         \frac{i\varepsilon^{k3m}\Delta^{k}}{2It}
                                                         \bigg{(}
                                                         (N_{c}-1)\sum_{
                                                         j \neq
                                                         \mathrm{lev}}\frac{\langle
                                                         \mathrm{lev}
                                                         | \tau^{3} |
                                                         j
                                                         \rangle}{E_{j}-E_{\mathrm{lev}}}
                                                         \langle j |
                                                         \left\{
                                                         e^{i\bm{\Delta}\cdot
                                                         \bm{r}},p^{m}
                                                         \right\}+
                                                         (E_{\mathrm{lev}}+E_{j})
                                                         e^{i
                                                         \bm{\Delta}
                                                         \cdot \bm{r}}
                                                         \gamma^{0}\gamma^{m}
                                                         |
                                                         \mathrm{lev}
                                                         \rangle  \cr
&\hspace{3.5cm}+N_{c}\sum_{m\neq j} R_{3}(E_m,E_j,\Lambda)\langle m | \tau^{3}
 | j \rangle \langle j | \left\{ e^{i\bm{\Delta}\cdot \bm{r}},p^{m}
\right\} + (E_{m}+E_{j})  e^{i \bm{\Delta} \cdot \bm{x}}
  \gamma^{0}\gamma^{m} | m \rangle  \bigg{)}.
\label{eq:EMT_total}
\end{align}

When it comes to the heavy-quark part in Eq.~\eqref{eq:EMT_current}
in the limit of $m_{Q}\to \infty$, the heavy quark can be regarded as
a mere static color source as discussed in
Ref.~\cite{Hudson:2017oul}. Therefore, we will adopt a
naive quark model to evaluate the heavy-quark part. In this regard,
one can obtain the following contributions of the singly heavy quark
to the GFFs
\begin{align}
A(t)=1, \ \ J(t)=\frac{1}{2}, \ \ D(t)=0.
\label{eq:EMT_free}
\end{align}
We want to emphasize that this heavy-quark contribution to each form
factor satisfies the constraints on the form factors. That is,
$A(0)$ and $J(0)$ should yield 1 and 1/2 as they should be, and the
contribution to the $D$-term should vanish since the heavy quark is
regarded as a free quark in the limit of $m_Q\to \infty$. Thus, we are
able to introduce the heavy-quark mass, angular momentum and pressure
distributions as the Dirac delta-function types
\begin{align}
\varepsilon^{Q}(\bm{r})  =    m_{Q}\delta(\bm{r}), \;\;\;
\rho^{\mathrm{Q}}_{J}(\bm{r})  =  -\frac{1}{6}  \delta(\bm{r}), \;\;\;
p^{\mathrm{Q}}(\bm{r})  =   0,
\end{align}
which are coupled with the light-quark pair\footnote{The angular
  momentum of the singly heavy baryon
  $J_{\Sigma_{c}}$ is decomposed into the soliton $J_{\mathrm{sol}}$
  and heavy-quark $J_{Q}$ contributions, and the corresponding angular
  momenta for $J_{3,\Sigma_{c}}=1/2$ are found to be $2/3$ and $-1/6$,
  respectively. }.
Based on these assumptions, the expressions for the GFFs are
obtained as follows:
\begin{align}
A(t) - \frac{t}{4M^{2}_{B}}D(t) &= \frac{1}{M_{B}} \int
                                          d^{3} r \ \varepsilon(r)
                                          j_{0}(r\sqrt{-t}) , \cr
D(t) &=  6M_{B} \int d^{3} r \ p(r)
           \frac{j_{0}(r\sqrt{-t})}{t}, \cr
J(t) &= 3\int d^{3} r \ \rho_{J}(r)
       \frac{j_{1}(r\sqrt{-t})}{r\sqrt{-t}},
\label{eq:EMTFF}
\end{align}
where
\begin{align}
\varepsilon(r) &= (N_{c}-1) E_{\mathrm{lev}}
              \phi^{*}_{\mathrm{lev}}(\bm{r})\phi_{\mathrm{lev}}(\bm{r})
              +N_{c}  \sum_{n}
              R_{1}(E_{n},\Lambda)\phi^{*}_{n}(\bm{r})\phi_{n}(\bm{r}) +
              m_{Q} \delta(\bm{r}), \cr
p(r) &= (N_{c}-1)\frac{1}{3}  \phi^{*}_{\mathrm{lev}}(\bm{r})
       (\gamma^{0} \bm{\gamma}\hat{\bm{p}})
       \phi_{\mathrm{lev}}(\bm{r}) + N_{c}\frac{1}{3}  \sum_{n}
       R_{2}(E_{n},\Lambda) \phi^{*}_{n}(\bm{r}) (\gamma^{0}
       \bm{\gamma}\hat{\bm{p}}) \phi_{n}(\bm{r}) , \cr
\rho_{J}(r) &= -(N_{c}-1)\frac{1}{6 I} \sum_{j\neq\mathrm{lev}}
              \epsilon^{ab3} r^{a} \phi^{*}_{j}(\bm{r}) \bigg{(} 2
              \hat{p}^{b} + (E_{\mathrm{lev}} + E_{j}
              \gamma^{0}\gamma^{b}) \bigg{)}
              \phi_{\mathrm{lev}}(\bm{r}) \frac{\langle \mathrm{lev} |
              \tau^{3} | j \rangle}{ E_{j} - E_{\mathrm{lev}}} \cr
&-\frac{N_{c}}{6 I} \sum_{\substack{n\neq j}} R_{3}(E_{n},E_{j},\Lambda)
\epsilon^{ab3}   r^{a}    \phi^{*}_{j}(\bm{r})        \bigg{(}
 2   \hat{p}^{b}  +  (E_{n} +   E_{j}  \gamma^{0}\gamma^{b})
 \bigg{)}   \phi_{n}(\bm{r}) \langle   n | \tau^{3}  |  j \rangle
              -\frac16 \delta(\bm{r}).
\label{eq:EMT_dens}
\end{align}

\section{Results and discussion\label{sec:4}}
Before we present the numerical results for the GFFs of the singly
heavy baryons, we first explain how the parameters are fixed.
The dynamical quark mass $M$ is the only free parameter in
the $\chi$QSM. Its value was already determined by computing
various properties of the proton, so we use the same value $M =
420$~MeV.  The current quark mass $m$ and cut-off mass $\Lambda$ are
fixed by reproducing the experimental data on the pion mass
$m_{\pi}=140$~MeV and the pion decay constant $f_{\pi}=93$~MeV. 
The detailed procedure for fixing these parameters can be found in
Ref.~\cite{Goeke:2005fs}.

\subsection{Energy density}
We start with examining the energy density for the mass form factor.
The energy density $\varepsilon(r)$ arises from the temporal component of
the EMT $T^{00}$. The integration of $\varepsilon(r)$ over space will
give the mass of a singly heavy baryon as follows
\begin{align}
\int d^{3} r\,  \varepsilon(r) = M_{\mathrm{sol}} + m_{Q} = M_B,
                \label{eq:enden}
\end{align}
where $M_{B}$ stands for the mass of a singly heavy baryon.
The mass form factor $A$(t) is usually normalized by Eq.~\eqref{eq:enden} 
\begin{align}
A(0) &= \frac{1}{M_{B}}\int d^{3} r \  \varepsilon(r) = 1,
\end{align}
This normalization condition coincides with the constraint on
the nucleon mass form factor from Ref.~\cite{Goeke:2007fp}. Obviously,
the entire momentum of the singly heavy baryon is carried by quarks
and antiquarks, since there are no gluons within the $\chi$QSM.

\begin{figure}
\includegraphics[scale=0.253]{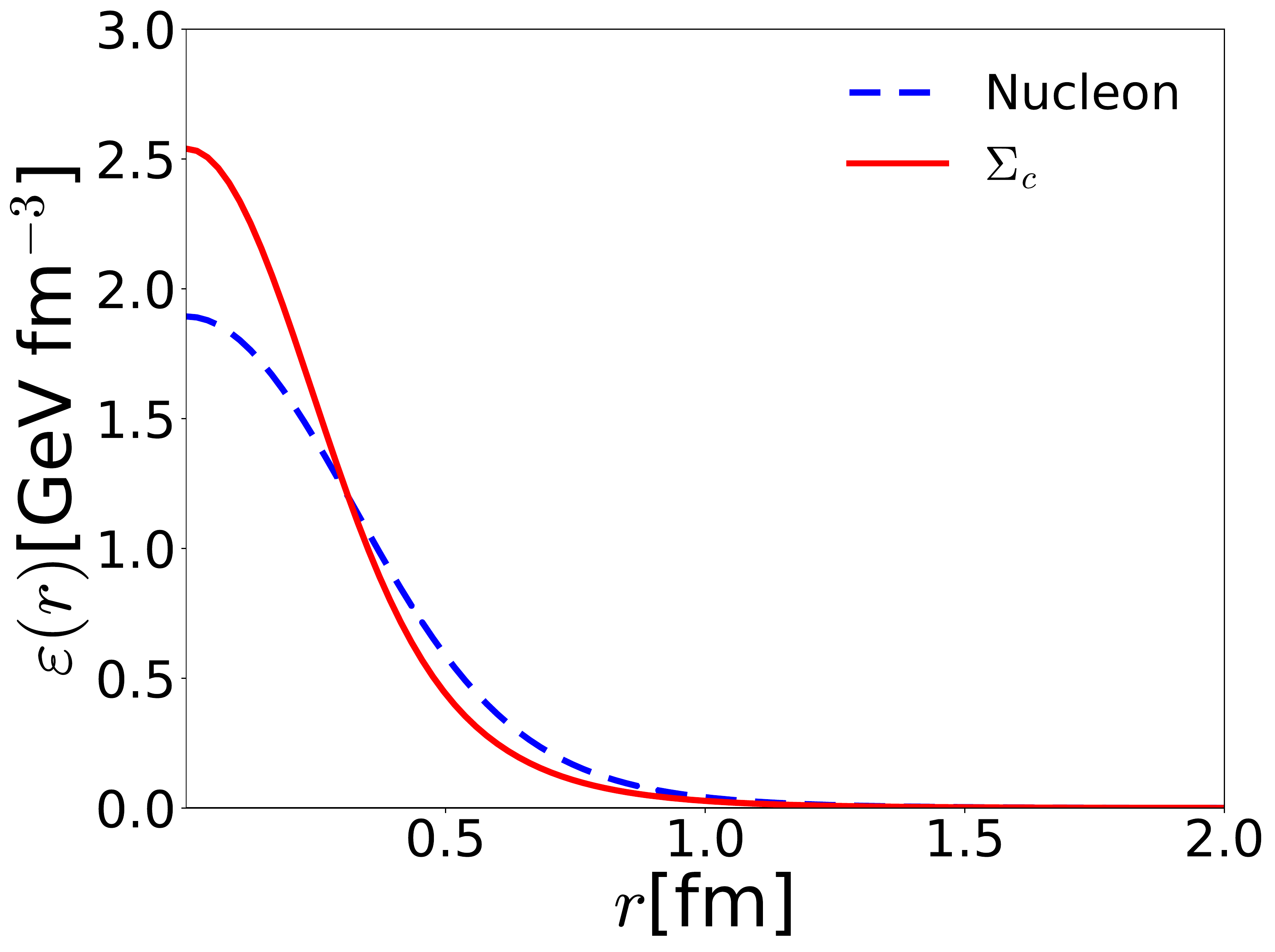}
\includegraphics[scale=0.253]{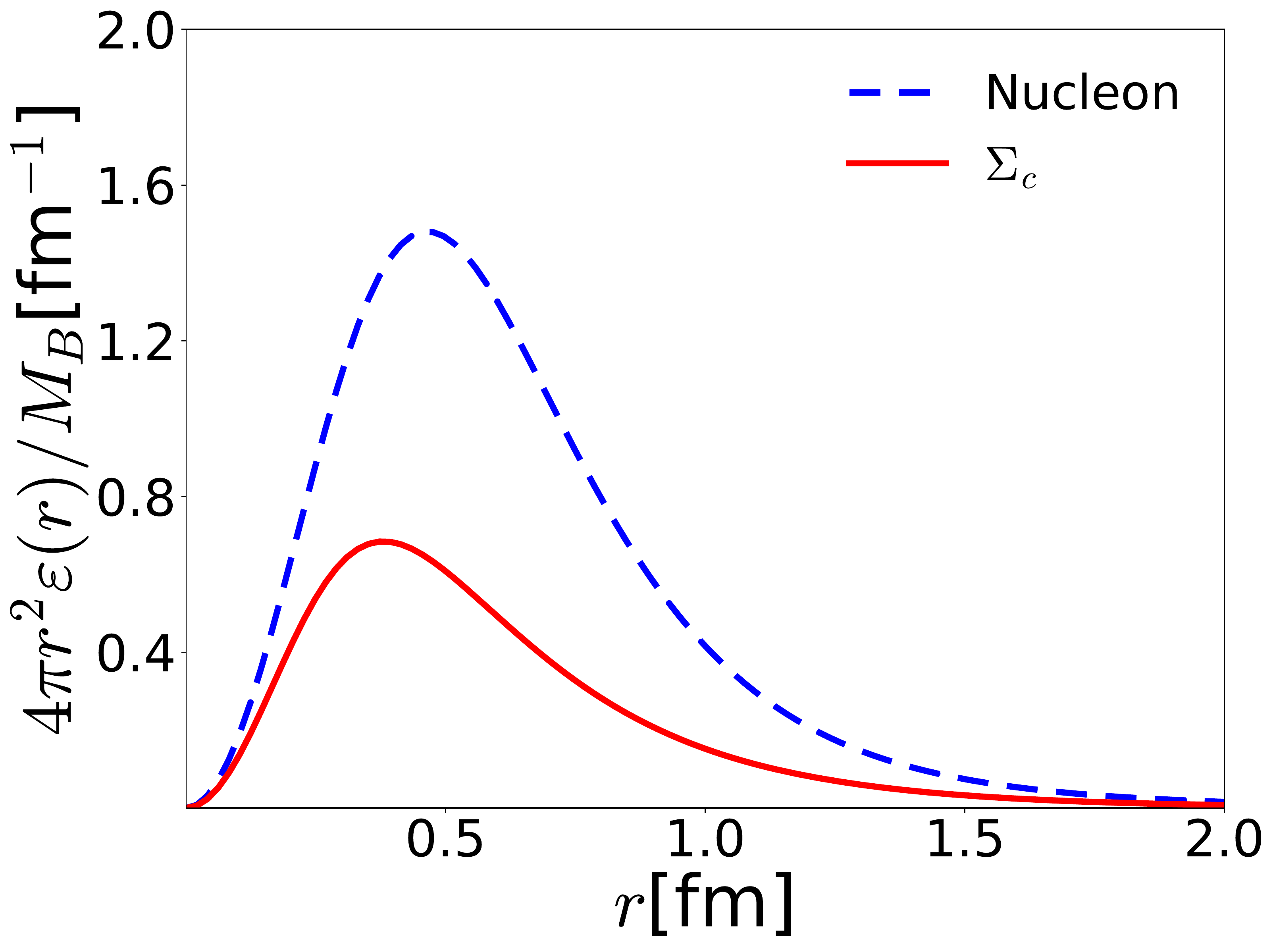}
\caption{The comparison of the $N_c-1$ light-quark energy density for
  the singly heavy baryon $\Sigma_c$ with the full energy density for the
  nucleon. The left panel depicts the energy densities as functions
  of radius $r$ whereas the right panel shows the energy densities
  multiplied by $4\pi r^2$ as functions of $r$. The solid and dashed
  curves draw respectively $\varepsilon(r)$ for $\Sigma_c$ and for the nucleon.
}
\label{fig:1}
\end{figure}
In the left panel of Fig.~\ref{fig:1}, we compare the results for the
light-quark energy density of the singly heavy baryon $\Sigma_c$
with that for the full energy density of the nucleon. As shown in
Fig.~\ref{fig:1}, the result for $\Sigma_c$ is narrower than that for
the nucleon. This indicates that $\Sigma_c$ is more compact than the
nucleon. A similar tendency was already found in the case of the
electromagnetic form factors of the singly heavy
baryons in a previous work~\cite{Kim:2018nqf} where the singly heavy
baryons turn out electromagnetically compact objects.
Integrating the density over space yields the mass of the $N_{c}-1$
light-quark contribution to the $\Sigma_c$ mass, which is around
$1$~GeV. Note that we find the value of the density in the center of
$\Sigma_c$ to be $\varepsilon(0) = 2.54\,\mathrm{GeV\cdot fm}^{-3}$, whereas
its value of the nucleon is found to be $\varepsilon(0) =
1.89~\mathrm{GeV\cdot fm}^{-3}$, which is smaller than that for
$\Sigma_c$ by about 30~$\%$. On the other hand, the $\Sigma_c$ energy
density falls off faster than that of the nucleon as $r$ increases,
which results in the narrower shape of the $\Sigma_c$ energy density.

In the right panel of Fig.~\ref{fig:1}, we compare the weighted
$\Sigma_c$ energy density by the usual factor $4\pi r^{2}$ with that
of the nucleon. The energy density of $\Sigma_c$
tends to be more centered in comparison with that of the nucleon. This
implies again that $\Sigma_c$ is more compact than the nucleon. The
calculation of the radius squared will explicitly show that the
$\Sigma_c$ is more compact than the nucleon. The
radius squared of the $\Sigma_c$ for the mass distribution is obtained
by
\begin{align}
\langle r^{2}_{E} \rangle  = \frac{\int d^{3}{r} \ r^{2}
                \varepsilon(r) }{ M_{\mathrm{sol}} + m_{Q} }=
                \frac{1}{M_B}\int d^{3}{r} \ r^{2} \varepsilon(r),
\end{align}
which is identical to the derivative of the $A(t)$ form factor with
respect to the momentum squared
\begin{align}
  \label{eq:3}
\langle r^{2}_{E} \rangle  =  \left. \frac{6}{A(0)} \frac{\partial A
  (t)}{\partial t}\right|_{t=0}.
\end{align}
The results for the $A(t)$ form factor will be presented soon.  The
corresponding result for the mass radius squared is evaluated as
follows $\langle r_E^2\rangle_{\Sigma_c} =0.21\,\mathrm{fm}^2$. On the
other hand, that for the nucleon is $\langle r_E^2\rangle_{\Sigma_c}
=0.54\,\mathrm{fm}^2$. Thus, we find that $\Sigma_c$ is indeed more
compact than the nucleon.

\subsection{Angular momentum density}

The density $\rho_{J}(r)$ refers to the total angular momemtum density
which arises from the mixed component of the EMT current,
$T^{0i}(\bm{r},\sigma',\sigma)$, and is normalized as
\begin{align}
J(0) = J_{\mathrm{sol}}(0)+J_{Q}(0) &= \int d^{3} r \rho_{J}(r) = \frac{1}{2}.
\end{align}
The total angular momentum for the constituents of a baryon comes from
the spin and orbital angular momenta of quarks and antiquarks, so that
it should be the same as the spin of a baryon $2J(0)=1$.
Concerning the heavy quark inside a singly heavy baryon, it is assumed
to be static. Thus, its total angular momentum is identified as its
spin. The distribution for the total angular momentum is again
governed by the light-quark pair inside a singly heavy baryon.

\begin{figure}
\includegraphics[scale=0.253]{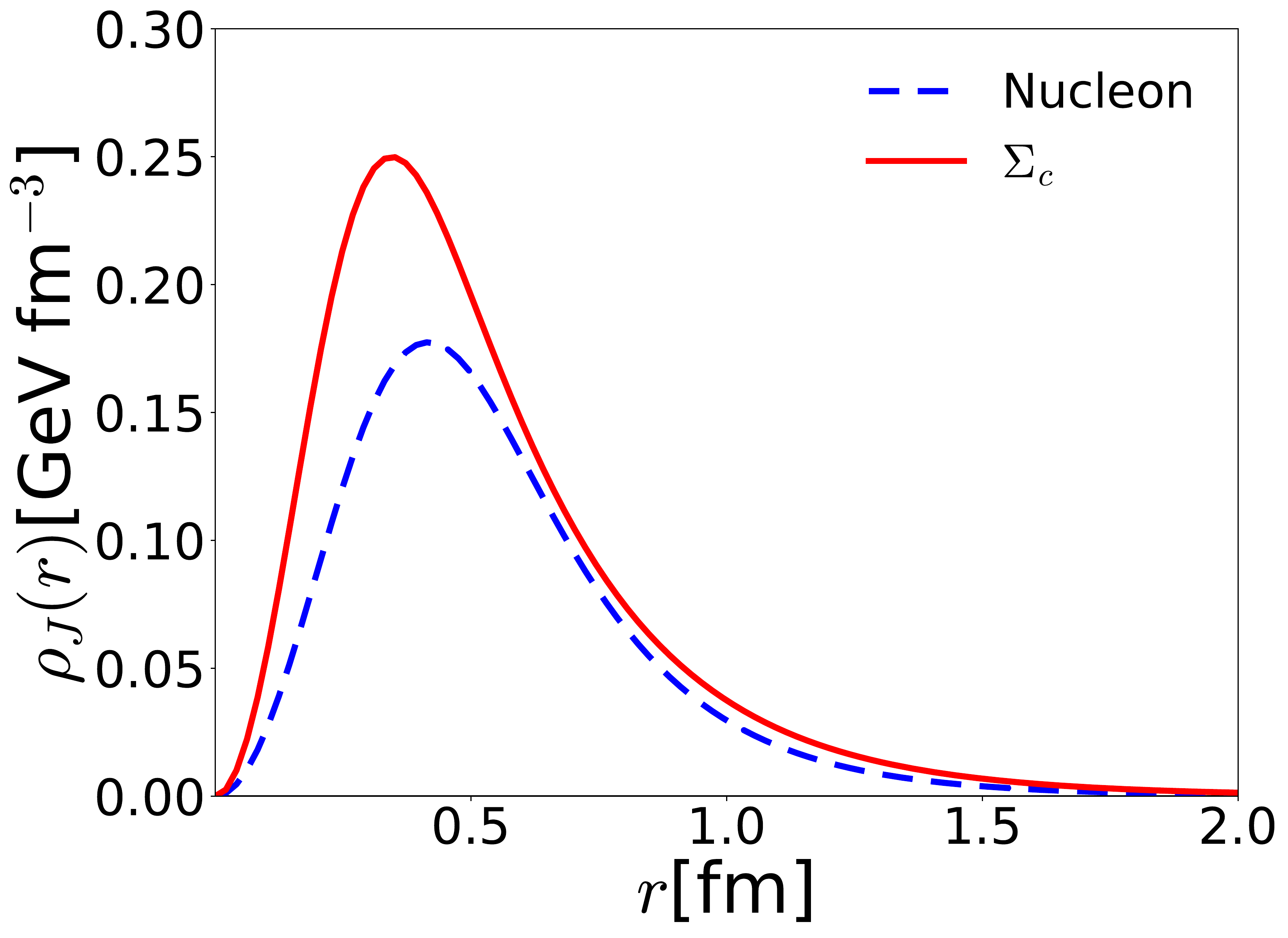}
\includegraphics[scale=0.253]{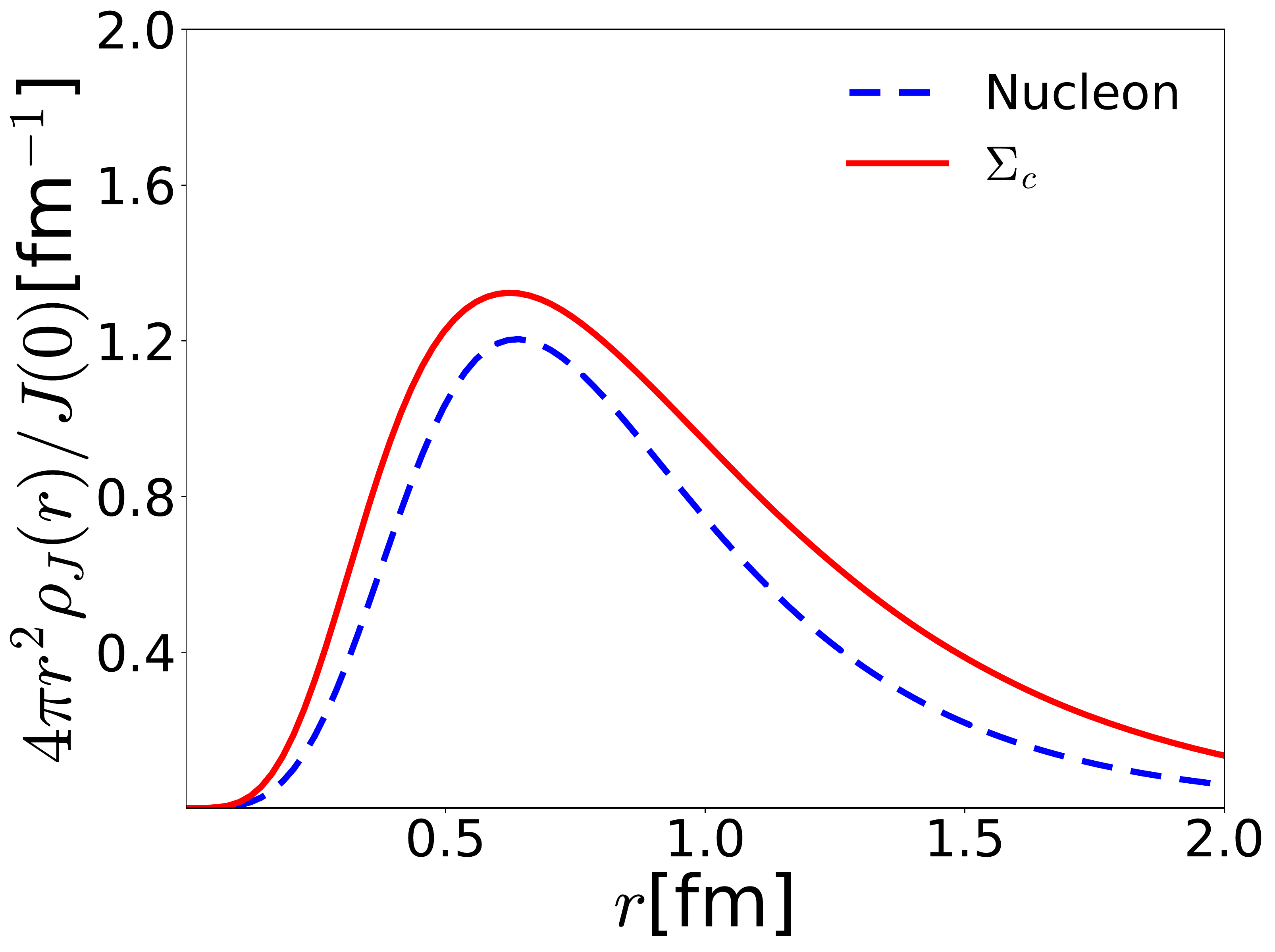}
\caption{The comparison of the $N_c-1$ light-quark density for the
  total angular momentum of the singly heavy baryon $\Sigma_c$ with
  that for the nucleon. The left panel depicts the total
  angular-momentum densities as functions of radius $r$ whereas the
  right panel shows the total angular-momentum  densities   multiplied
  by $4\pi r^2$ as functions of $r$. The solid and dashed
  curves draw respectively $\rho_J(r)$ for $\Sigma_c$ and for the
  nucleon. }
\label{fig:2}
\end{figure}
In the left panel of Fig.~\ref{fig:2}, we depict the numerical result
for the spin distribution of $\Sigma_c$ in comparison with that of the
nucleon. At first sight, the result seems peculiar, since the
result for the spin distribution of $\Sigma_c$ turns out larger than
that of the nucleon. However, as noted previously, the $N_c-1$
light-quark density corresponds to the spin 1. The spin of the
$\Sigma_c$ baryon, which is 1/2, will be obtained by coupling that of
the $N_c-1$ light-quark or the soliton with the singly heavy quark
spin 1/2. Thus, it is natural for the spin distribution of the
$N_c-1$ light quarks to be larger than that of the nucleon. The right
panel of Fig.~\ref{fig:2} draws the spin distribution wighted by $4\pi
r^2$. One can make a quantitative comparison by considering the mean
radius squared for the spin distribution, which is defined by
\begin{align}
&\langle r^{2}_{J} \rangle  = \frac{\int d^{3}{r} \ r^{2} \rho_{J}(r)
                }{ \int d^{3}{r} \ \rho_{J}(r)}=\frac{1}{J(0)}\int
                d^{3}{r} \ r^{2} \rho_{J}(r).
\label{eq:Eradius2}
\end{align}
The results are obtained to be $\langle r^{2}_{J}
\rangle_{\Sigma_{c}}=1.56\,\mathrm{fm}^2$ for $\Sigma_c$ and $\langle r^{2}_{J}
\rangle_{N}=1.02\,\mathrm{fm}^2$ for the nucleon, respectively.
This indicates that the spin distribution of $\Sigma_c$ is spreaded
more widely than that of the nucleon. Note that in the chiral limit
the angular momentum density is proportional to $r^{-4}$ at large $r$,
so the radius diverges, which is very similar to the
isovector radius of the nucleon~\cite{Beg:1973sc, Adkins:1983ya}.

\subsection{Strong force fields and stability conditions}
Using the conservation of the EMT current, we can derive the global stability
condition for $\Sigma_c$ in Eq.~\eqref{eq:stability} within the framework of the
$\chi$QSM
\begin{align}\label{eq:vL_xqsm}
\int dr \  r^{2} \ p(r) &=\frac{N_{c}-1}{12\pi}\int d^{3}r \;
                          \phi^{*}_{\mathrm{val}}(\bm{r}) (\gamma^{0}
                          \bm{\gamma}\hat{\bm{p}})
                          \phi_{\mathrm{val}}(\bm{r}) +
                          \frac{N_{c}}{12\pi}\int d^{3}r \;
                          \sum_{n} R_{2}(E_{n}) \phi^{*}_{n}(\bm{r})
                          (\gamma^{0} \bm{\gamma}\hat{\bm{p}})
                          \phi_{n}(\bm{r}) = 0.
\end{align}
In Ref.~\cite{Kim:2018xlc}, the
pion mean field was not newly computed but the $N_c$ factor was simply
replaced by $N_c-1$ for the level parts. This brings about violation
of the stability condition. To make this condition 
satisfied, we have to modify the pion mean field in the presence of
$N_c-1$ light quarks, which was performed in
Ref.~\cite{Kim:2019rcx}. In the present work, thus, we employ the 
improved pion mean field derived in Ref.~\cite{Kim:2019rcx} to compute
the GFFs and the pressure density that complies with the von Laue
condition. The shear force is obtained by solving the differential
equation given in Eq.~\eqref{eq:diffEq} with the boundary conditions
$s(r)=0$ at $r\to0$ and $r\to\infty$ imposed.

\begin{figure}[htp]
\includegraphics[scale=0.253]{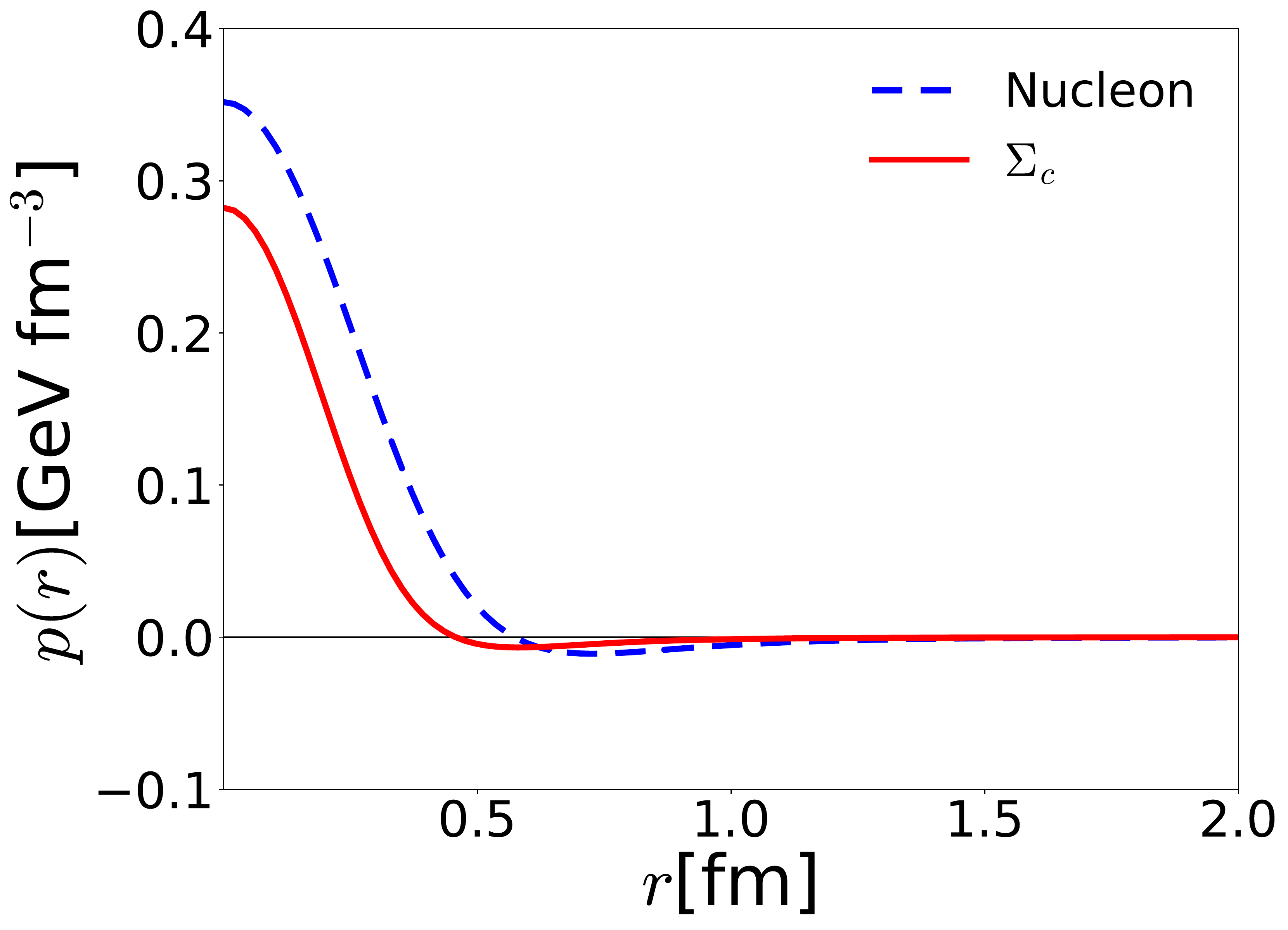}
\caption{The comparison of the pressure distributions for
  $\Sigma_c$ with those for the nucleon. The numerical results for the
  pressure densities of the nucleon and $\Sigma_c$ are drawn as
  functions of radius $r$. The solid and dashed curves depict
  respectively those for $\Sigma_c$ and for the nucleon.}
\label{fig:3}
\end{figure}
In Fig~\ref{fig:3}, we evaluate the pressure density for $\Sigma_c$ in
comparison with that for the nucleon. As expected, the magnitude of the
$\Sigma_c$ pressure density becomes smaller than that of the nucleon
one, because the $N_c-1$ pion mean field is weaker than the $N_c$ pion
mean field. To satisfy the global stability condition, the densities
must have at least one node as shown in Fig~\ref{fig:3}.
The pressures at the center of the $\Sigma_c$ and nucleon were
estimated as $p(0)\big{|}_{\Sigma_{c}}=0.282$~GeV fm$^{-3}$ and
$p(0)\big{|}_{N}=0.352$~GeV fm$^{-3}$, respectively.

We examine numerically the pressure densities of the $\Sigma_c$ and 
nucleon satisfy the following global stability conditions
\begin{align}
  \int dr \ r^{2} p(r) &= \frac{N_{c}-1}{12\pi}  \langle \widetilde{\mathrm{val}}|
  \gamma^{0} \bm{\gamma}\cdot \bm{p} | \widetilde{\mathrm{val}} \rangle
  +\frac{N_{c}}{12\pi} \sum_{n} R_{2}(E_{n},\Lambda) \langle \tilde{n} | \gamma^{0}
  \bm{\gamma}\cdot \bm{p} | \tilde{n} \rangle = 0\;\;\mbox{ for $\Sigma_c$},\cr
\int dr \ r^{2} p(r) &= \frac{N_{c}}{12\pi}  \langle \mathrm{val}|
  \gamma^{0} \bm{\gamma}\cdot \bm{p} | \mathrm{val} \rangle
  +\frac{N_{c}}{12\pi} \sum_{n} R_{2}(E_{n},\Lambda) \langle n | \gamma^{0}
  \bm{\gamma}\cdot \bm{p} | n \rangle = 0 \;\; \mbox{ for the nucleon},
\label{eq:stabcon}
\end{align}
where $|\widetilde{\mathrm{val}}\rangle$ and $|\tilde{n}\rangle$
emphasize the level and Dirac continuum eigenstates under the influence
of the $N_c-1$ pion mean field. 

\begin{figure}[ht]
\includegraphics[scale=0.253]{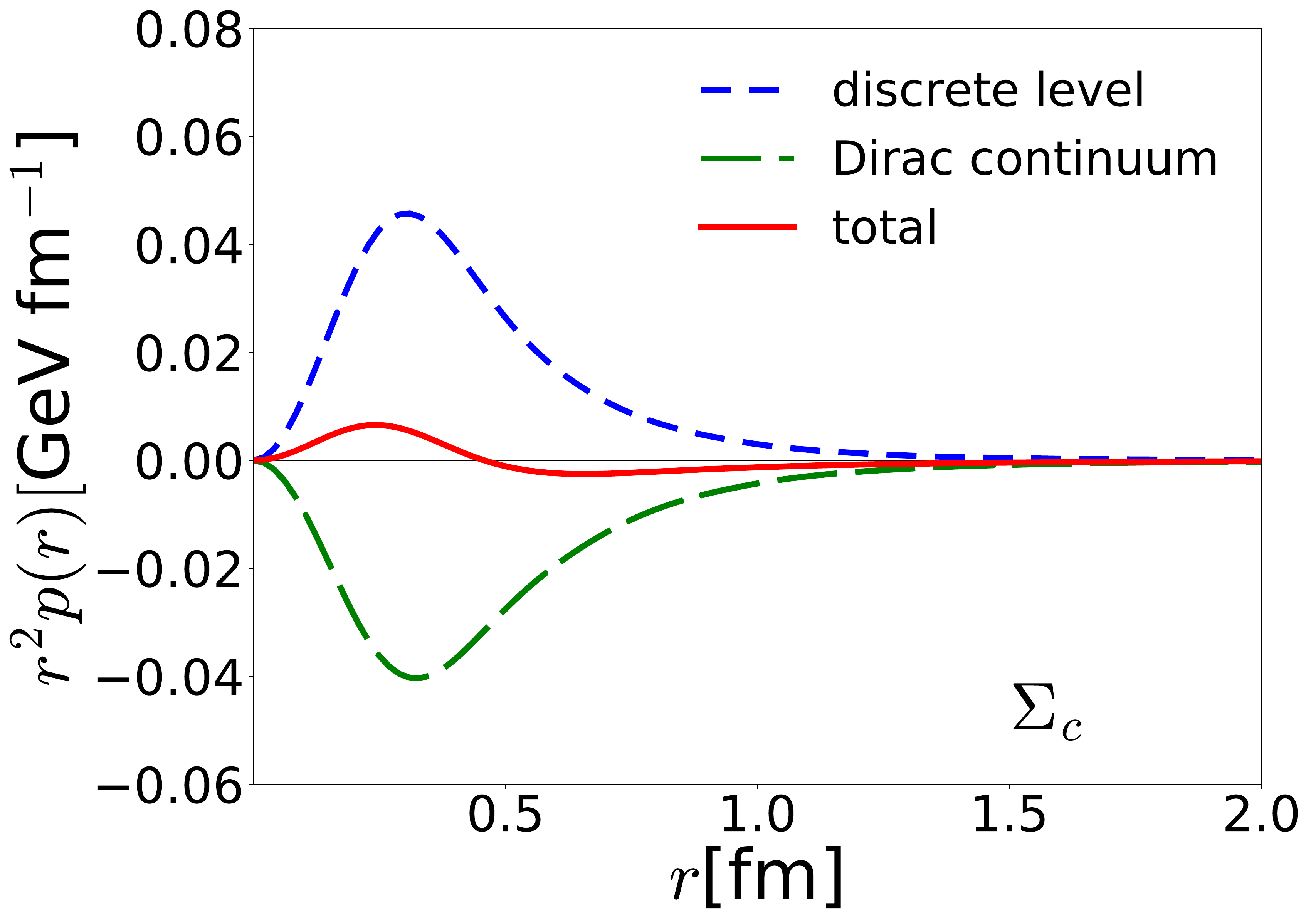}
\includegraphics[scale=0.253]{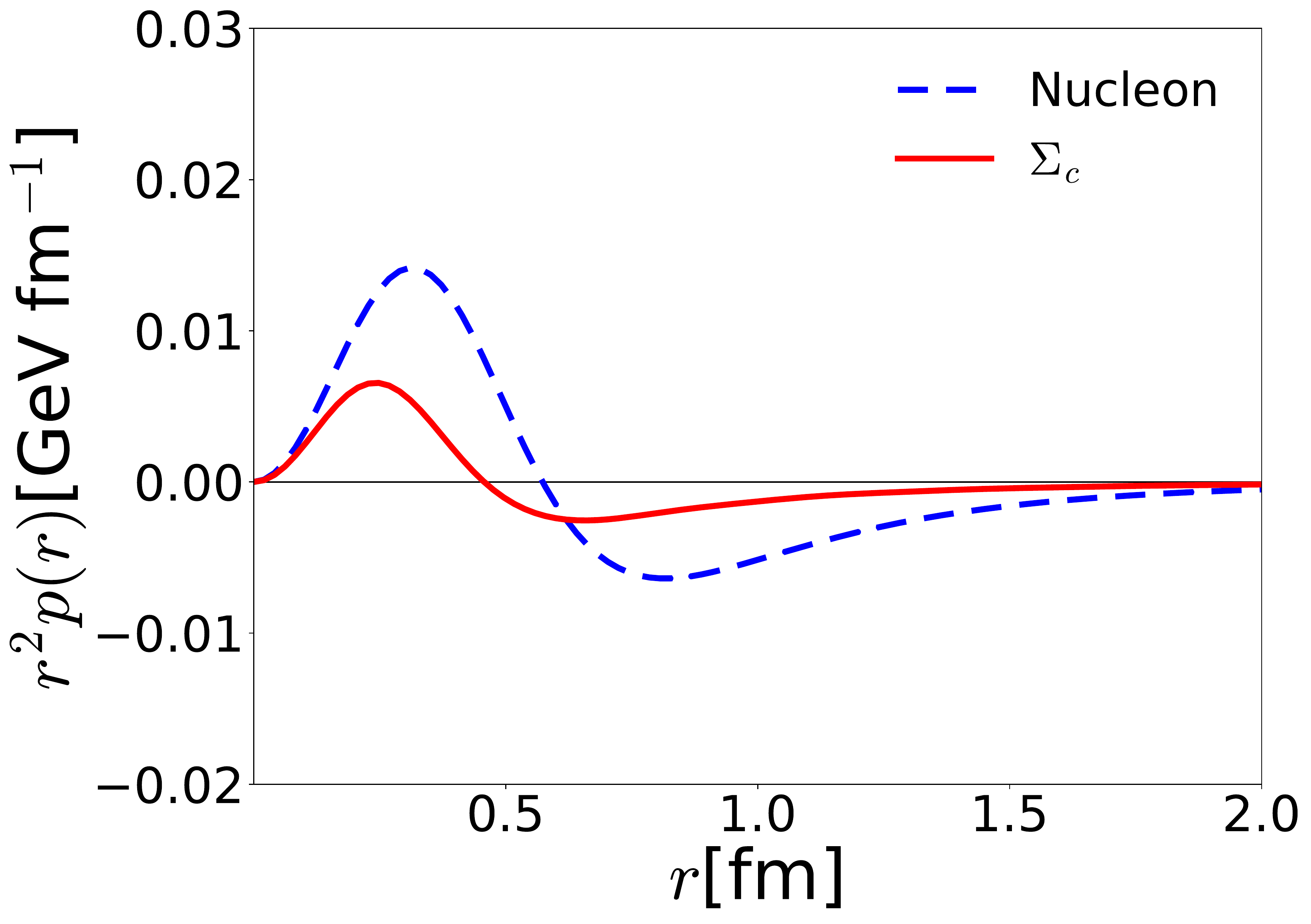}
\caption{The left panel presents decomposition of the level and
  Dirac continuum contributions to the pressure. The right panel shows
  $r^{2}$-weighted pressures for both the light and singly heavy
  baryons.}
\label{fig:4}
\end{figure}
In the left panel of Fig~\ref{fig:4}, we show the level and Dirac
continuum contributions to the pressure density of the $\Sigma_c$, 
weighted by $r^2$. As in the case of the nucleon, the level quarks
contribute dominantly the core part of the pressure density and are
positive over the whole region of $r$, while the Dirac continuum
becomes dominant in the outer part and negative. This implies that
while the level quarks tend to be driven away from the center, the
Dirac continuum keeps them bound in the core part. This picture is the
very same as in the case of the nucleon. This gives a possible
conjecture that the level quarks inside a hadron may be confined by
strong vacuum fluctuations.

In the right panel of Fig~\ref{fig:4}, we compare the present results
for the pressure density of $\Sigma_c$ with that of the nucleon. This
shows that the pressure density for $\Sigma_c$ is overall weaker than
that for the nucleon. Moreover, the comparison tells us that the size
of $\Sigma_c$ is more compact than that of the nucleon. It can be
shown clearly by introducing $r_{0}$ where the pressure density
vanishes. We find $r_{0}=0.46$~fm for $\Sigma_c$ and
$r_{0}=0.57$~fm for the nucleon. Indeed, $\Sigma_c$ is a more compact
object than the nucleon.

\begin{figure}[htp]
  \includegraphics[scale=0.24]{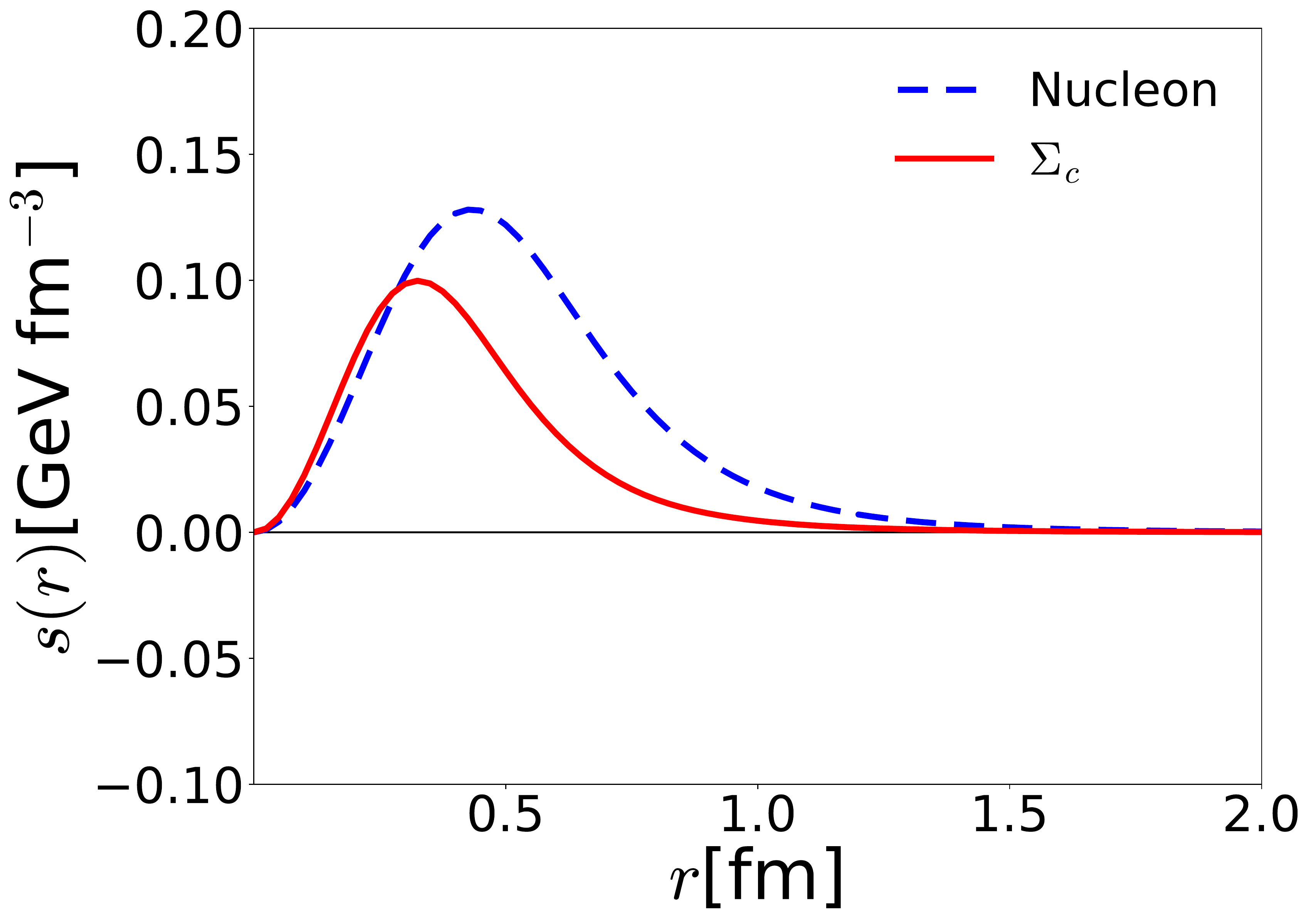} \hspace{0.2cm}
  \includegraphics[scale=0.24]{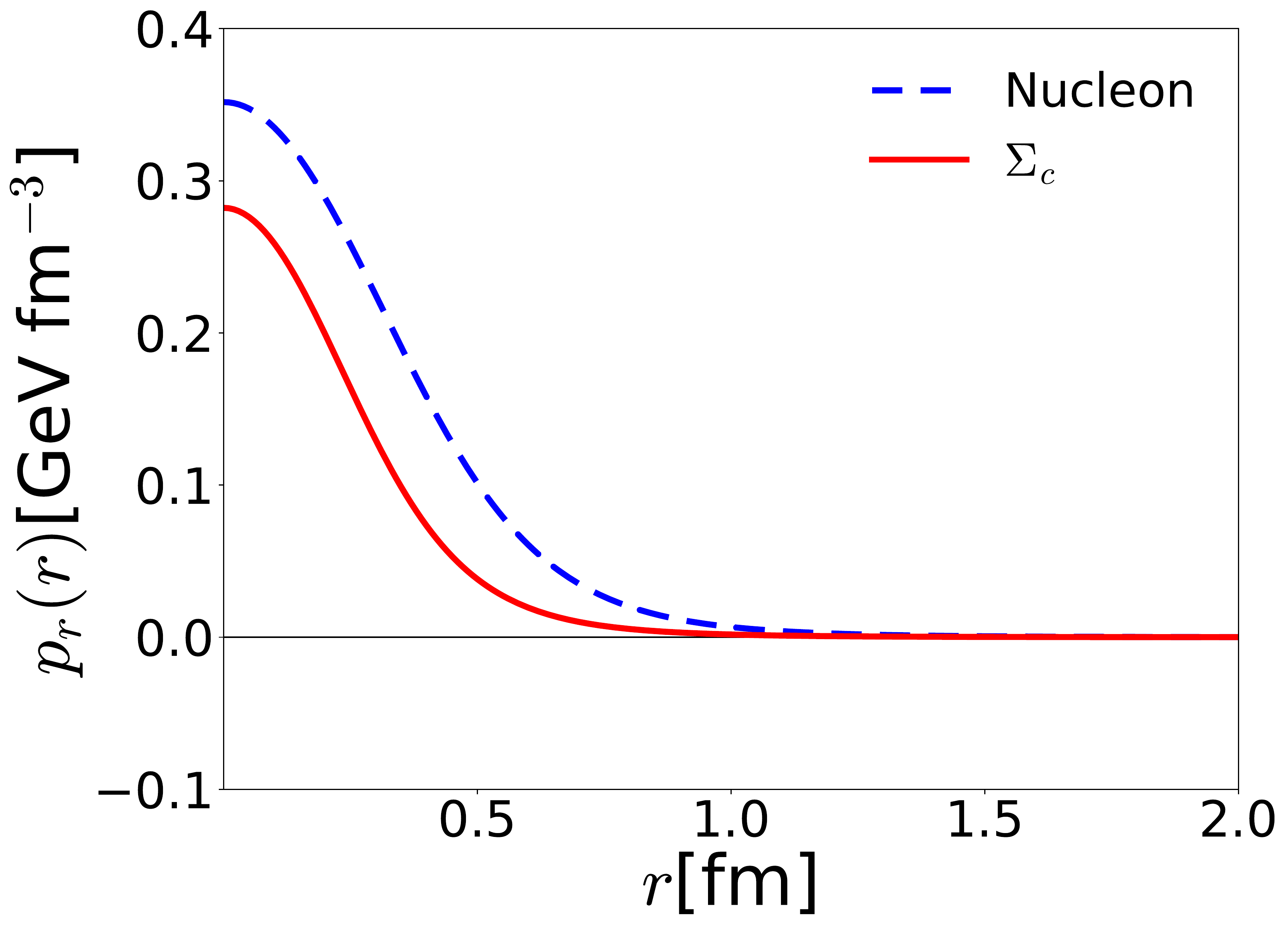}
\caption{Comparison of the results for the shear-force density $s(r)$
  and $p_r(r)$ of $\Sigma_c$ with those for the nucleon. The solid
  curves depict those of $\Sigma_c$ whereas the dashed ones draw those
  of the nucleon. }
\label{fig:5}
\end{figure}
The left panel of Fig.~\ref{fig:5} illustrates the result for the
shear-force density of $\Sigma_c$ in comparison with that for the
nucleon. The result for $s(r)$ of $\Sigma_c$ is closer to its center
than that for the nucleon. The magnitude of $s(r)$ of the $\Sigma_c$
is smaller than that of the nucleon. The $D$-term form factor, of
which the expression is given in Eq.~\eqref{eq:d-term2}, gives a clue
on the signature of the shear-force density. Since $D(0)$ should be
negative to comply with the stability condition, the shear-force
density should be positive for all values of $r$. If we take the limit
of $t\to 0$ in Eq. ~\eqref{eq:d-term2}, $D(0)$ has the following
expression
\begin{align}
  \label{eq:d-term3}
D(0) = -\frac{4}{15} M_B \int d^3 r \,r^2 \,s(r)  .
\end{align}
Indeed, the results for $s(r)$ of both the nucleon and $\Sigma_c$
satisfy the condition $s(r)>0$.
In the right panel of Fig.~\ref{fig:5}, we draw the results for
$p_r(r)$ of the nucleon and $\Sigma_c$. Being similar to the case of
$s(r)$, $p_r(r)$ is also positive for all the values of
$r$. $p_r(r)>0$ is just a local mechanical stability
condition given in Eq.~\eqref{eq:mecstab1}. As shown
in Fig.~\ref{fig:5}, $p_r(r)$ is positive definite. Note that
the result for the stability density of $\Sigma_c$ is again weaker
than that of the nucleon. This can be explained by the weaker pion
mean field for singly heavy baryons. The mechanical radius is defined
in terms of the stability density, which is given in
Eq.~\eqref{eq:mecradius}. The numerical
results for the mechanical radii of $\Sigma_c$ and nucleon are
obtained as $\langle r^{2} \rangle^{\Sigma_{c}}_{\mathrm{mech}}=
0.45$~fm$^{2}$ and $\langle r^{2} \rangle^{N}_{\mathrm{mech}}=
0.55$~fm$^{2}$.  This indicates that $\Sigma_c$ is also mechanically a
more compact object than the nucleon. The size of $\Sigma_c$ is
reduced by approximately $25~\%$ in comparison with that of the
nucleon.

\begin{figure}[htp]
\includegraphics[scale=0.28]{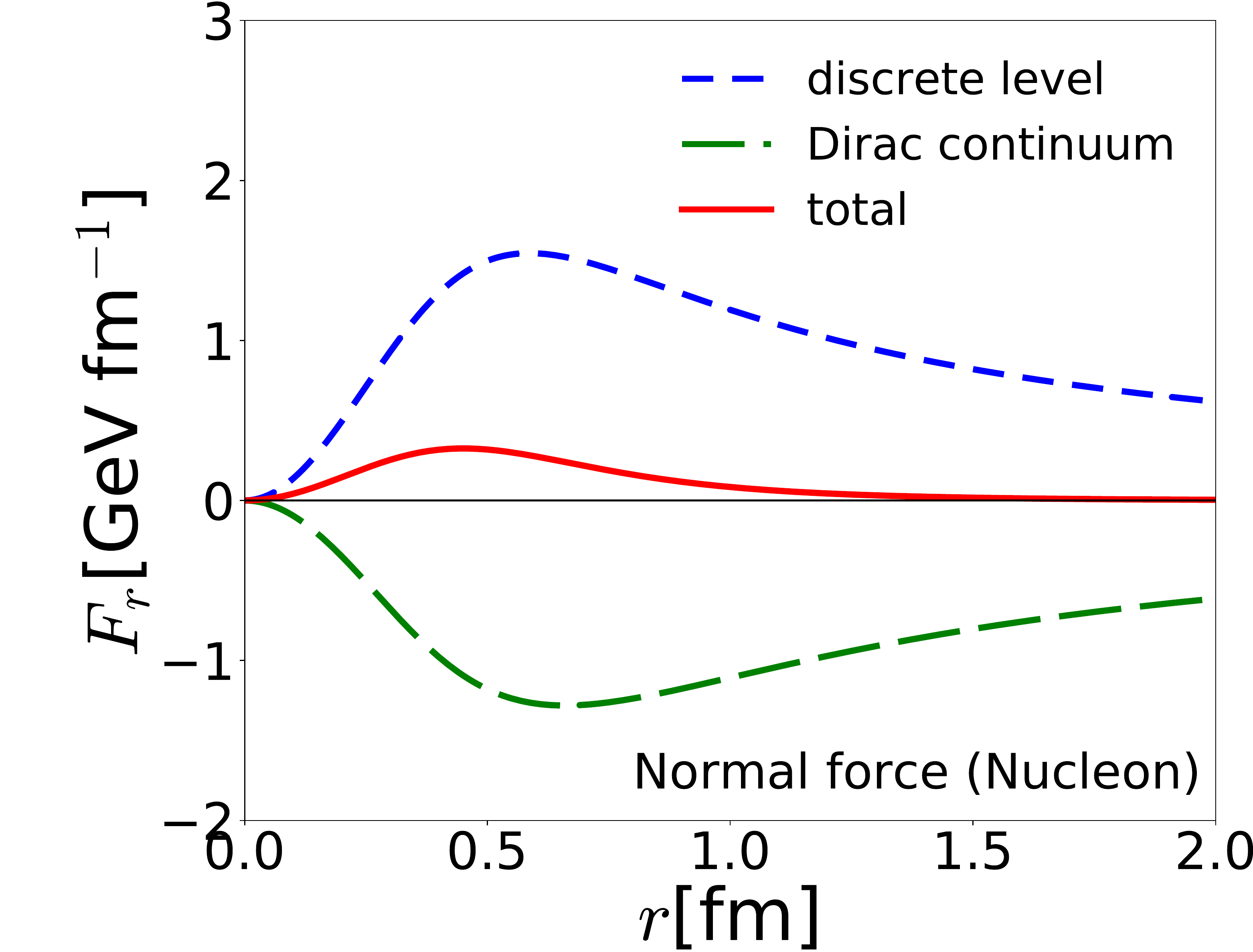}
\includegraphics[scale=0.28]{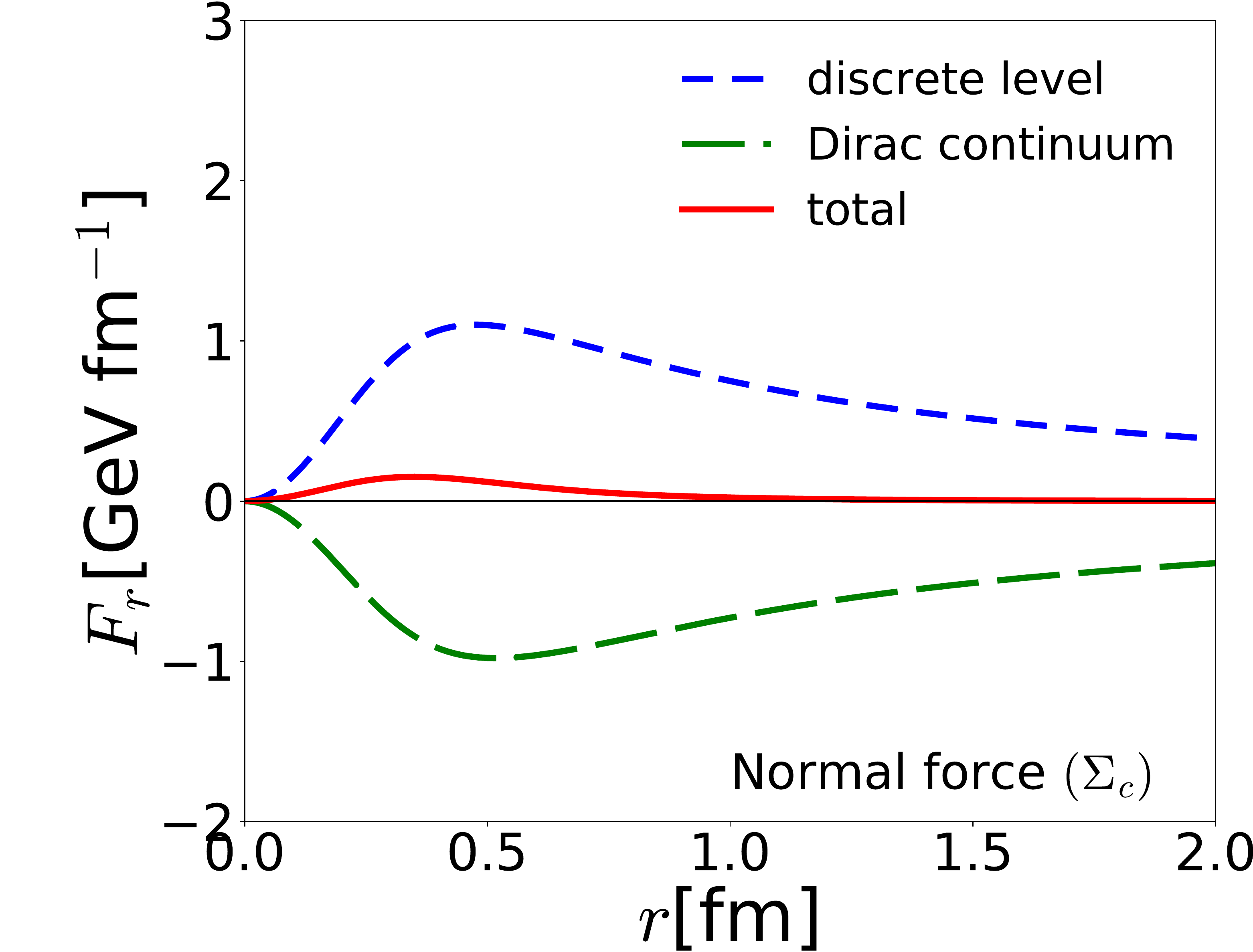}
\includegraphics[scale=0.28]{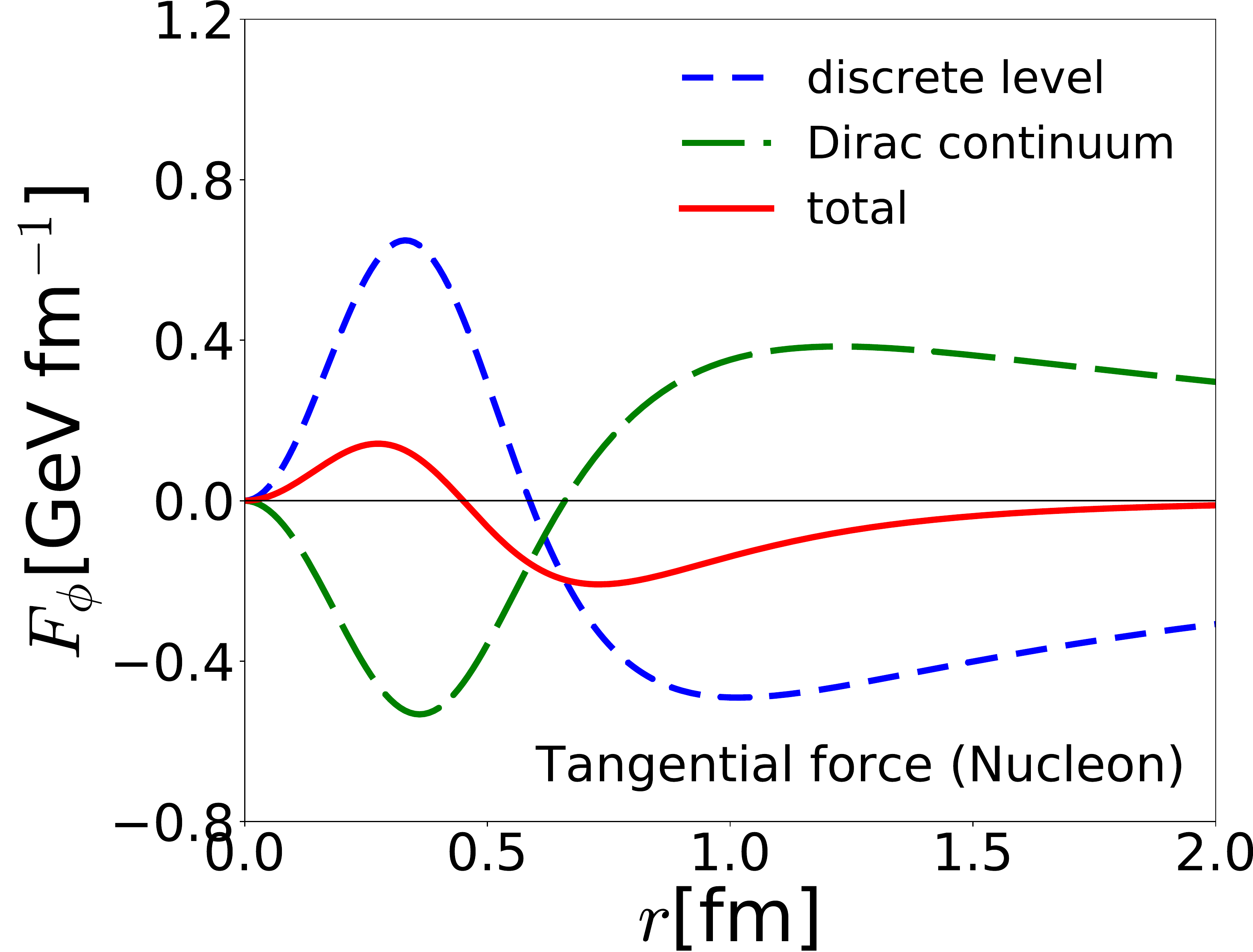}
\includegraphics[scale=0.28]{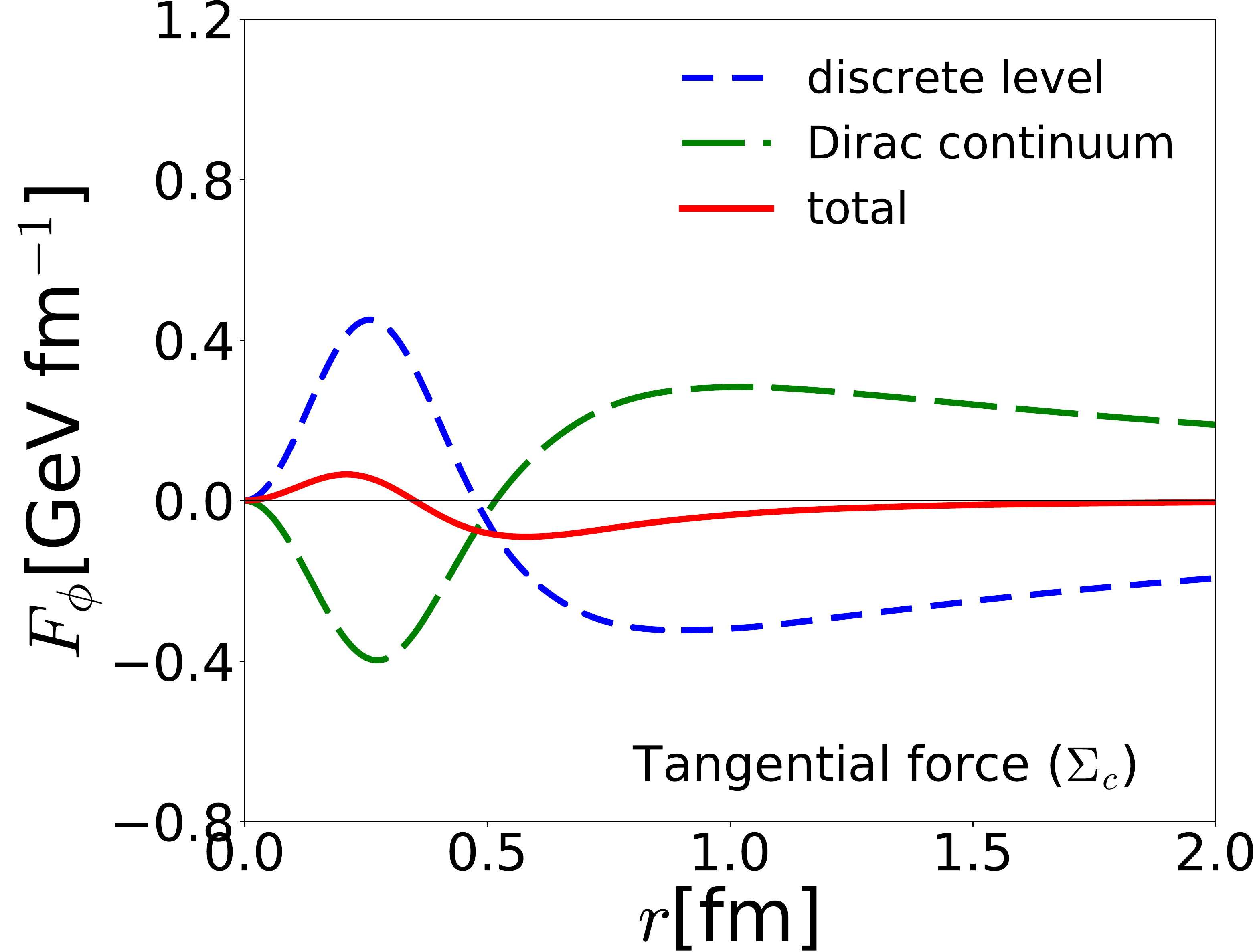}
\caption{Results for the normal and tangential force fields in the
  nucleon and $\Sigma_c$. In the upper left panel, the results for the
level and Dirac-continuum contributions to the normal force field in
  the nucleon is drawn in the short-dashed and long-dashed curves,
  respectively.  In the upper right panel, those in $\Sigma_c$ are
  drawn in the same notation. In the lower left (right) panel, we
  depict the results for the tangential force field in the nucleon
  ($\Sigma_c$) in the same notation.}
\label{fig:6}
\end{figure}
We now discuss the normal and tangential force fields defined in
Eq.~\eqref{eq:strong_force}, since they shed light on how a baryon
acquires its stability microscopically. In fact, the normal force
field is the same as the condition of the mechanical stability except
the spherical areal factor as shown in Eq.~\eqref{eq:force_comp}. We
first examine the numerical results for the normal and tangential
force fields as functions of $r$, which are illustrated in
Fig.~\ref{fig:6}. Concerning the normal force fields in the nucleon
and $\Sigma_c$, which are drawn in the left and right upper panels of
Fig.~\ref{fig:6} respectively, the level-quark contributions are
positive definite whereas the Dirac continuum parts are negative
definite. However, the magnitude of the level parts is stronger than
that of the Dirac continuum parts, which leads to the
fact that the normal force fields are positive definite. This implies
that $F_r$ are directed outwards. On the other hand, the tangential
force fields exhibit more complicated structures. First of all, the 
tangential force fields are symmetric in $\theta$ and $\phi$ as shown
in Eq.~\eqref{eq:force_comp}. Thus, we do not need to distinguish
$F_\phi$ from $F_\theta$. This interesting feature will be explicitly
shown in three dimensional figures soon. To guarantee the stability
of a baryon, the tangential force field should have at least one nodal
point. The reason comes from Eq.~\eqref{eq:stability2D} that is called
the 2D von Laue stability condition.  Indeed, the numerical results
for $F_\phi$ in the nucleon and $\Sigma_c$ reveal one node as shown in
the lower panel of Fig.~\ref{fig:6}. Interestingly, the behavior of
the level-quark contributions is opposite to the Dirac
  continuum-quark ones, 
which is similar to the case of the normal force fields. This means
that the direction of the tangential force field arising from the
level quarks is also opposite to that from the Dirac
continuum quarks. As a result, the inner part of the total tangential
force fields inside both the nucleon and $\Sigma_c$ rotates
counterclockwise from any viewpoint, whereas the outer part of
$F_\phi$ does clockwise. We will show this feature more explicitly later.
\begin{figure}[htp]
\includegraphics[scale=0.28]{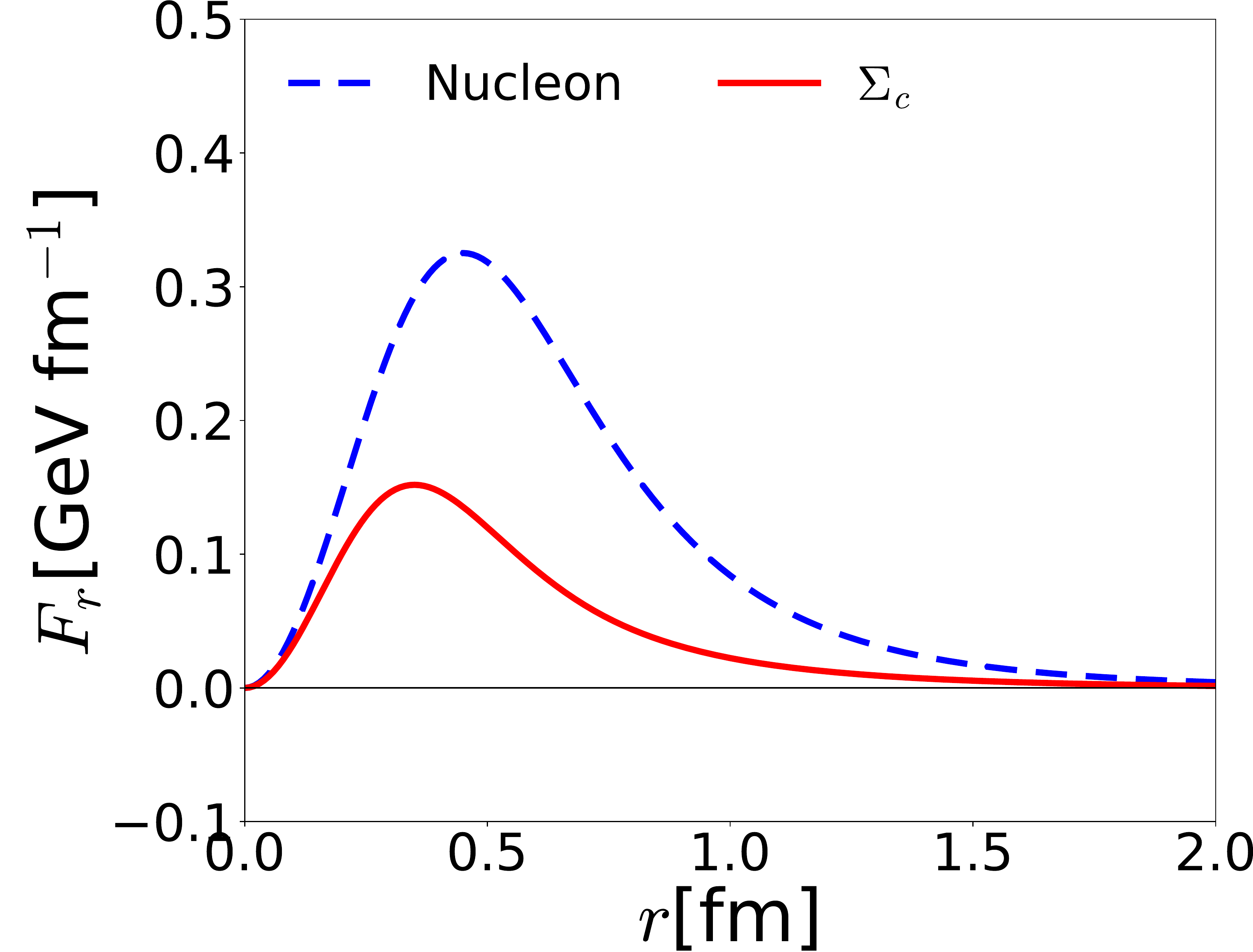}
\includegraphics[scale=0.28]{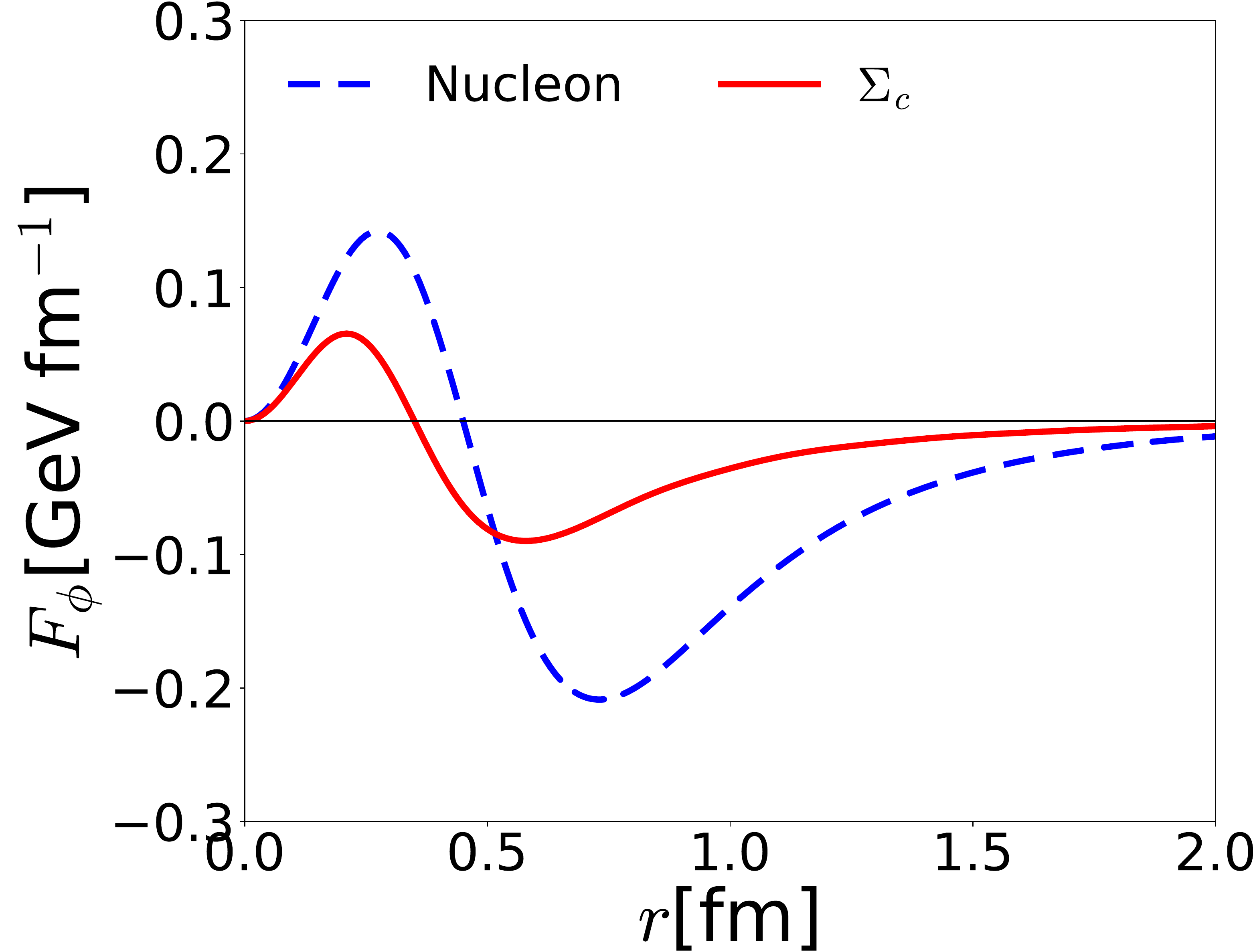}
\caption{Comparison of the normal and tangential force fields in the
  nucleon with those of $\Sigma_c$. The solid curves depict those of
  $\Sigma_c$ whereas the dashed ones draw those of the nucleon. }
\label{fig:7}
\end{figure}
In Fig.~\ref{fig:7}. we compare the numerical results for the normal
and tangential force fields in $\Sigma_c$ with those in the
nucleon. We find that both $F_r$ and $F_\phi$ in $\Sigma_c$ are weaker
and more compact than those in the nucleon.
\begin{figure}[htp]
\includegraphics[scale=0.59]{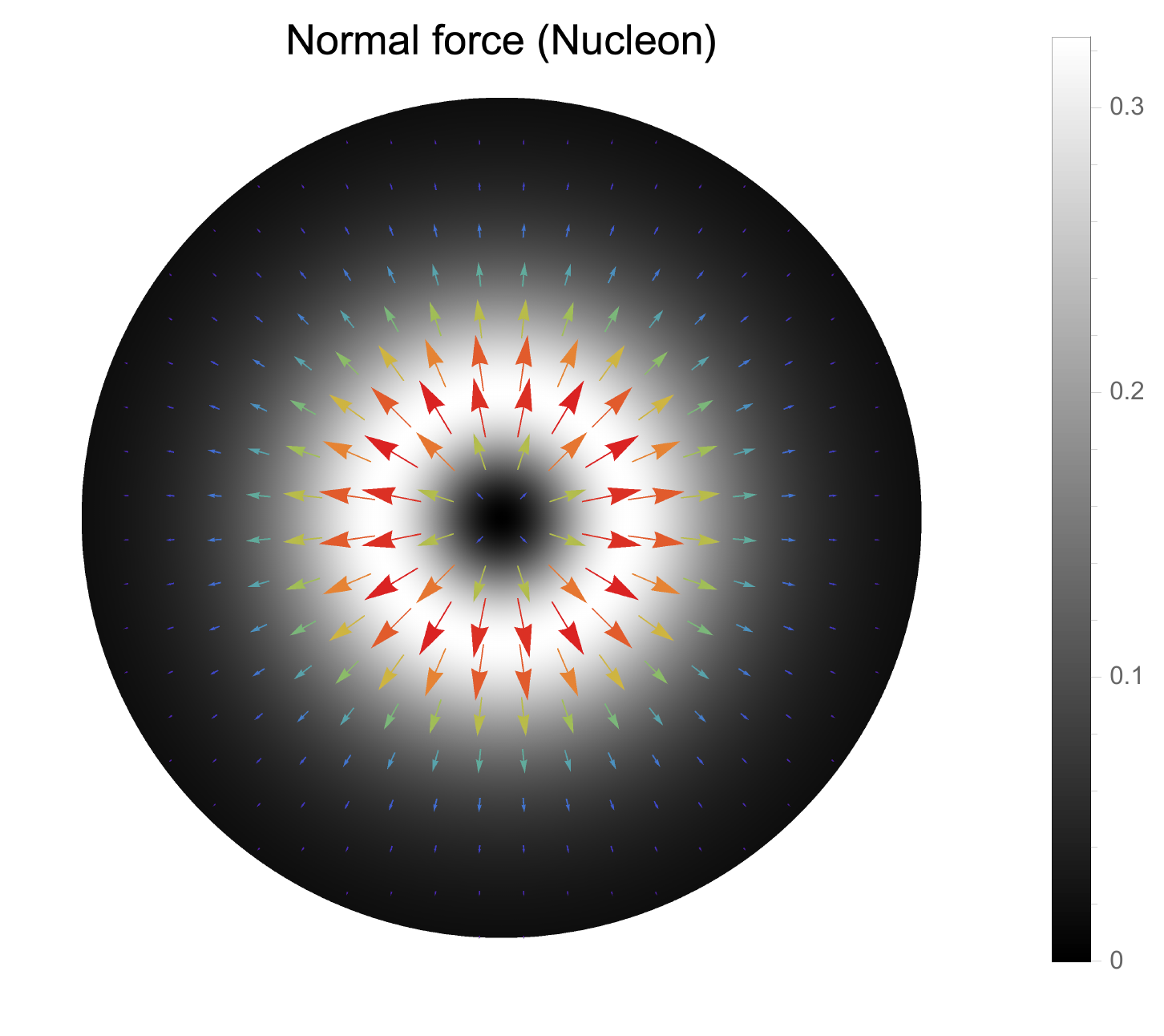}
\includegraphics[scale=0.59]{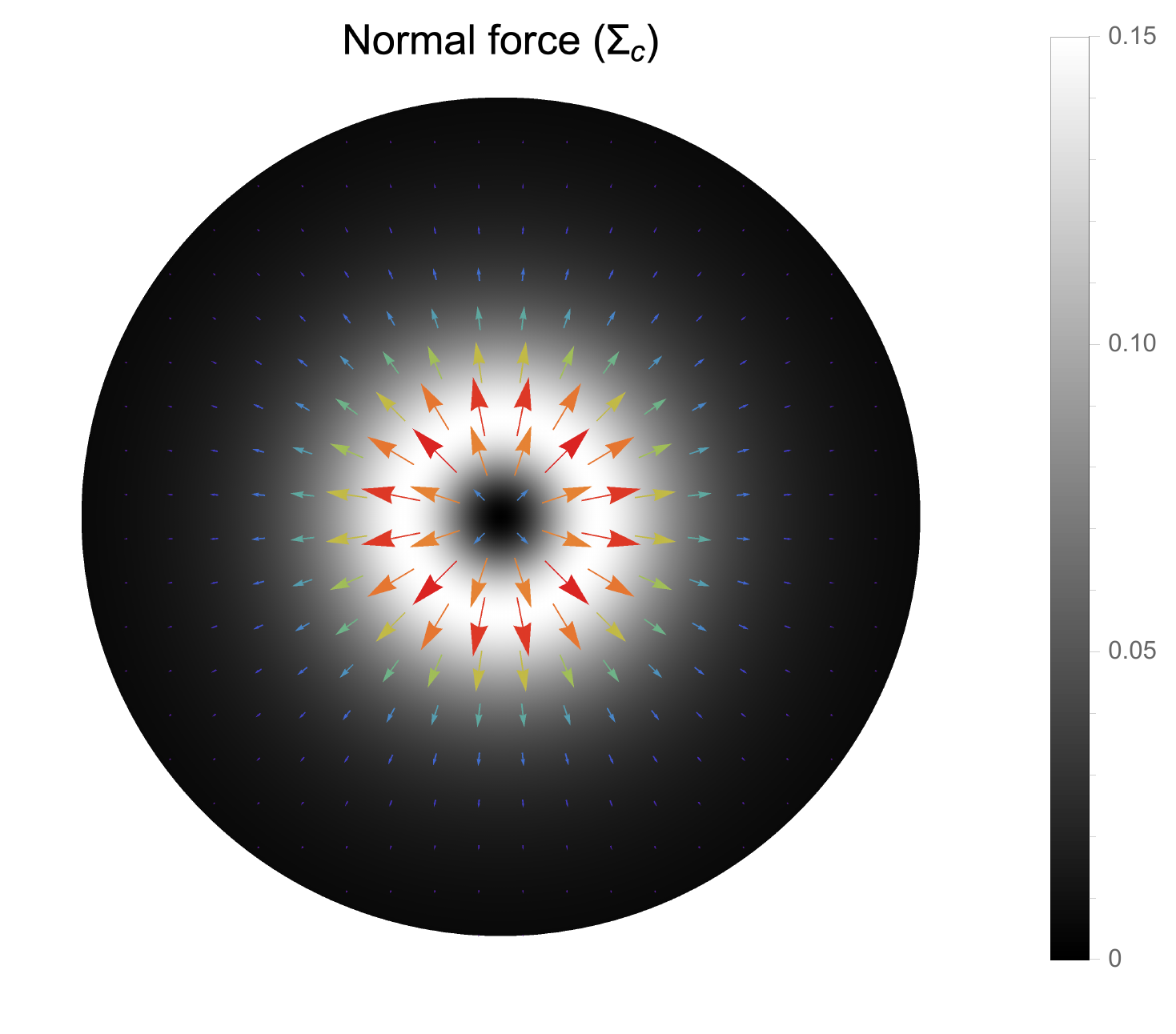}
\includegraphics[scale=0.59]{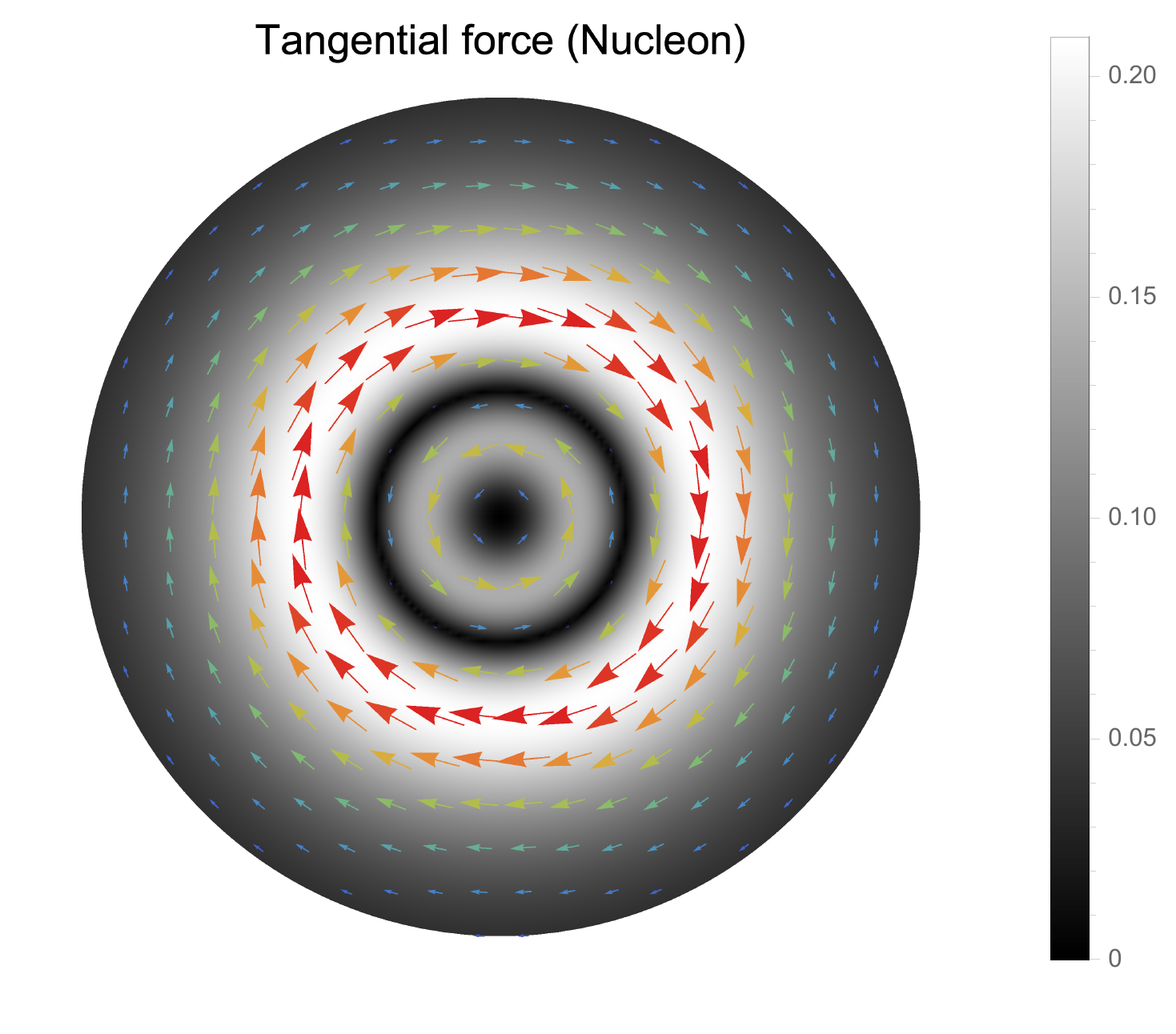}
\includegraphics[scale=0.59]{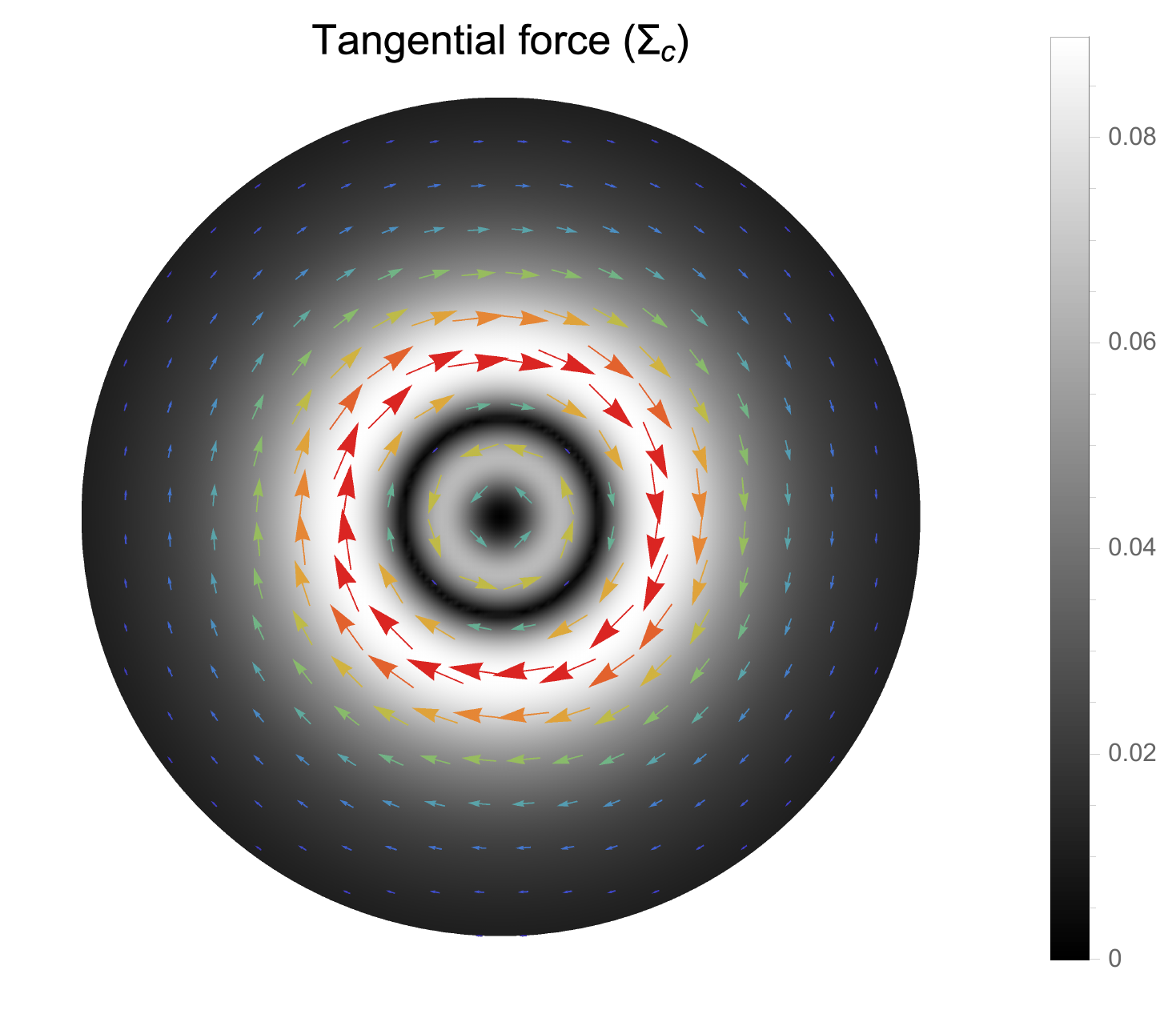}
\caption{Visualization of the normal force fields $4\pi
  r^{2} T^{ij}\bm{e}^{j}_{r}$ and tangential force fields
$4\pi r^{2} T^{ij}\bm{e}^{j}_{\phi}$. In the upper panel, we
visualize the normal force fields in the nucleon (left) and $\Sigma_c$
(right), respectively. In the lower panel, we show the tangential
force fields in the nucleon (left) and $\Sigma_c$ (right), respectively.
The radius of the disc is taken to be 1.5 fm, the color legend gives
the absolute value of the force fields in GeV/fm.}
\label{fig:8}
\end{figure}
As discussed previously, the upper panel of Fig.~\ref{fig:8} explicitly
visualizes the fact that the normal force fields are directed
outwards. On the other hand, the lower panel of Fig. ~\ref{fig:8}
demonstrates how the tangential force fields rotate inside both the
nucleon and $\Sigma_c$. The inner part of $4\pi r^{2}
T^{ij}\bm{e}^{j}_{\phi}$ rotates counterclockwise, whereas the outer
part of that does in the opposite direction.

\begin{figure}[htp]
\includegraphics[scale=0.44]{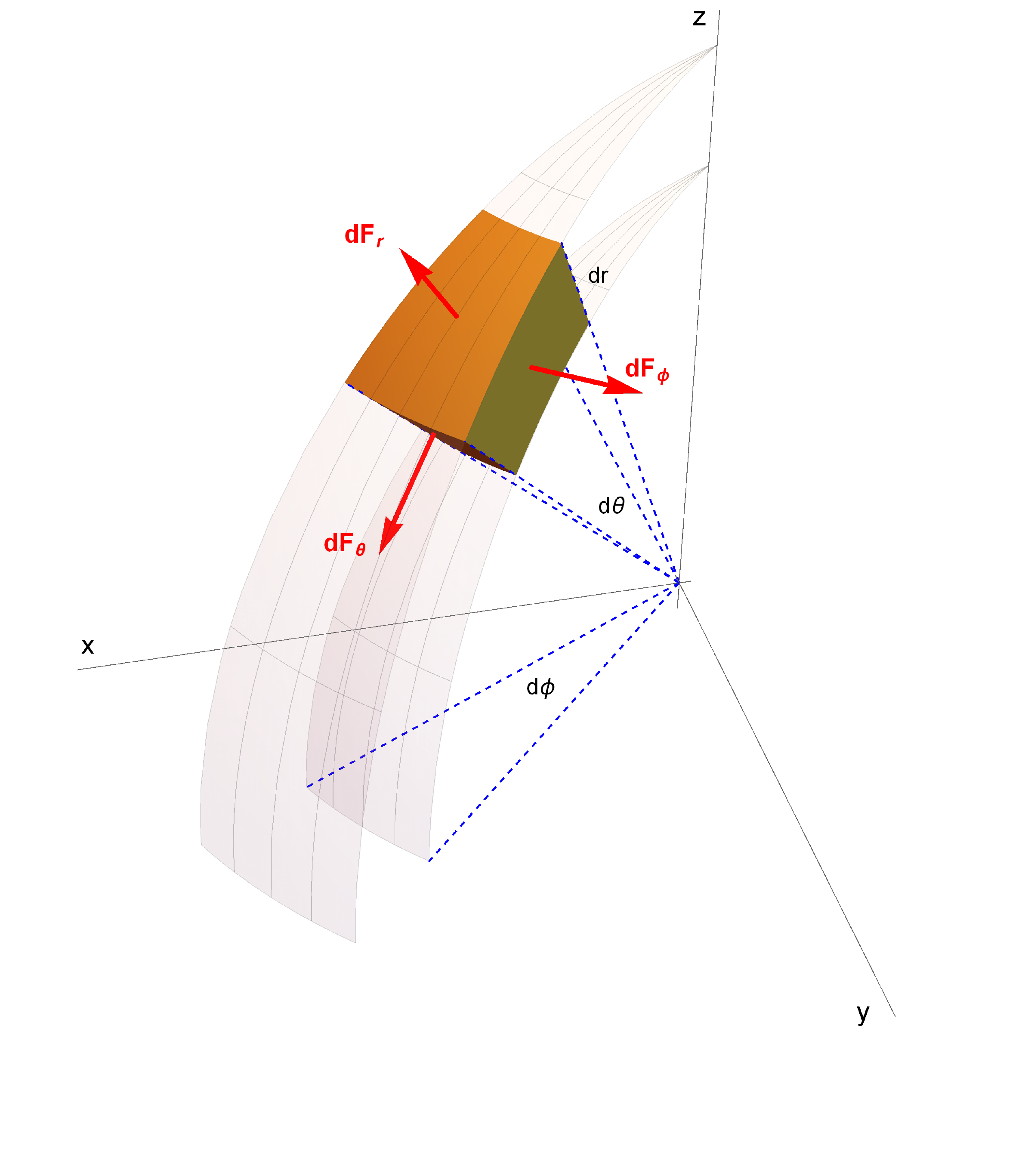}
\includegraphics[scale=0.70]{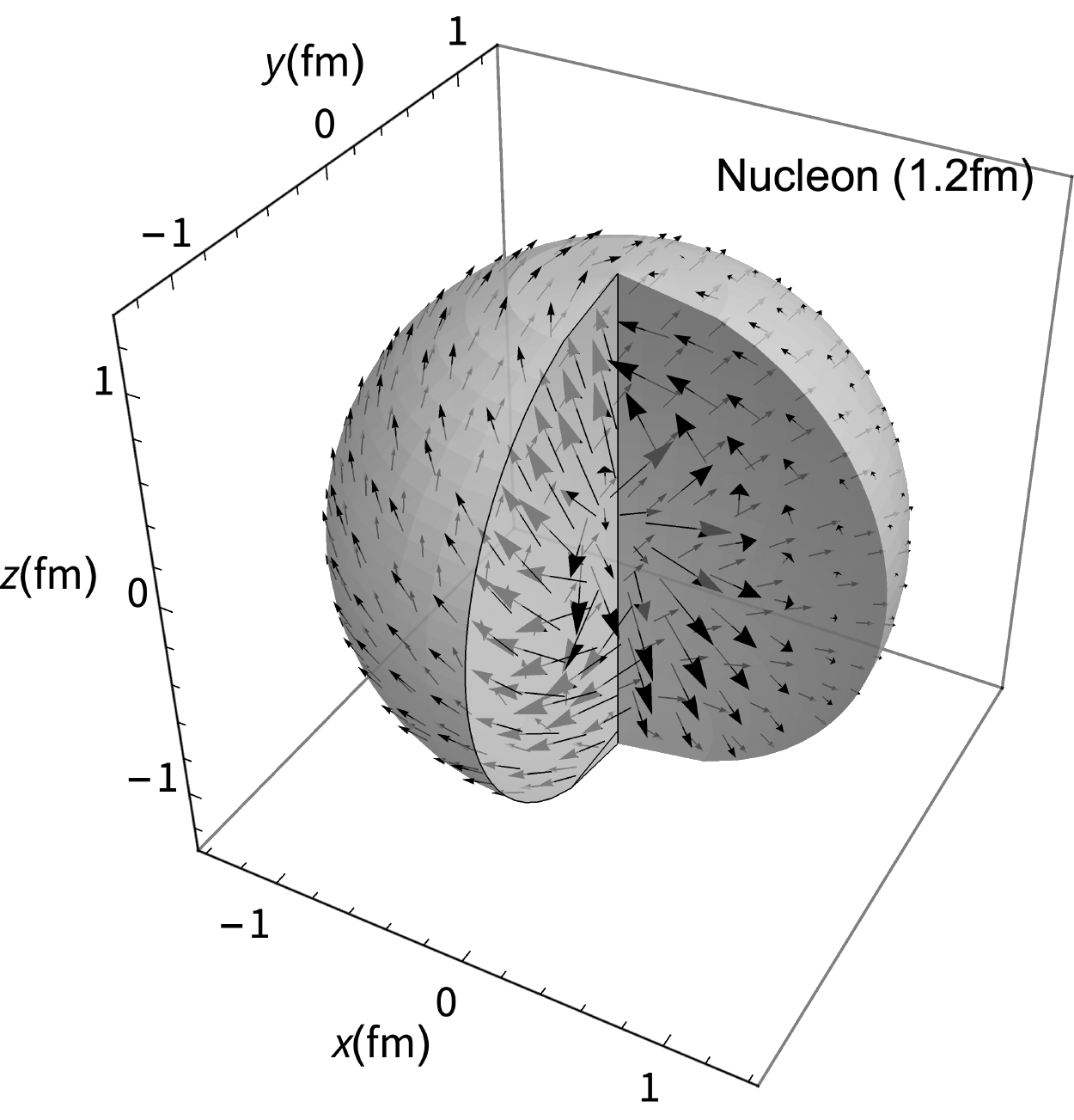}
\includegraphics[scale=0.70]{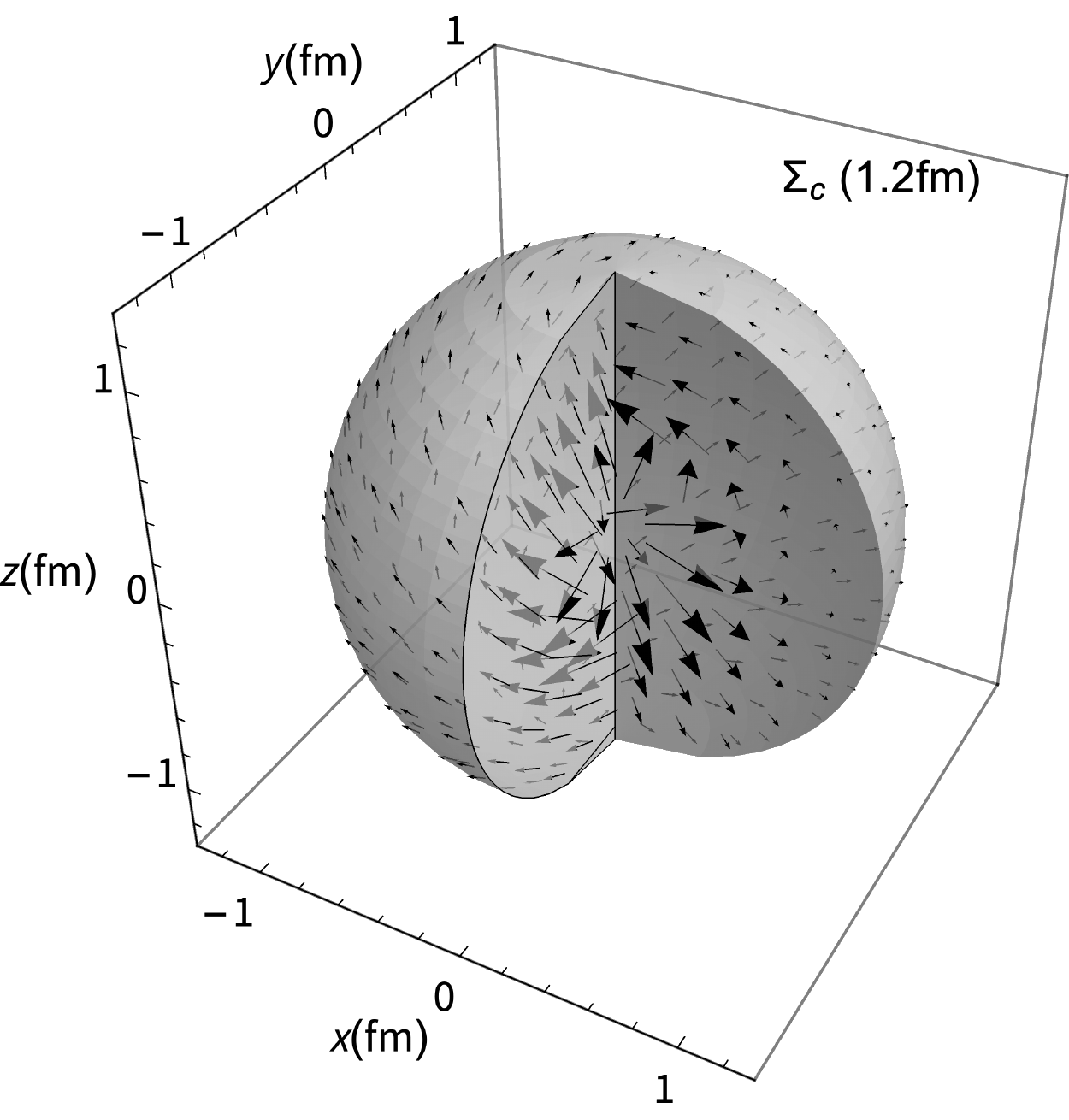}
\caption{In the upper panel, the infinitesimal force fields 
    $dF_{(r,\theta,\phi)}$ defined in Eq.~\eqref{eq:strong_force} are
    visualized as the arrows, which will be used in the 3D
    visualization of the strong force fields in Fig.~\ref{fig:10},
    Fig.~\ref{fig:11} and the lower panel of Fig.~\ref{fig:9}. In the
  lower left and right panels, the 3D visualization of the strong
  force fields ($\bm{F}$) for the nucleon and $\Sigma_c$ are
 respectively illustrated, respectively.} 
\label{fig:9}
\end{figure}
\begin{figure}[htp]
\includegraphics[scale=0.70]{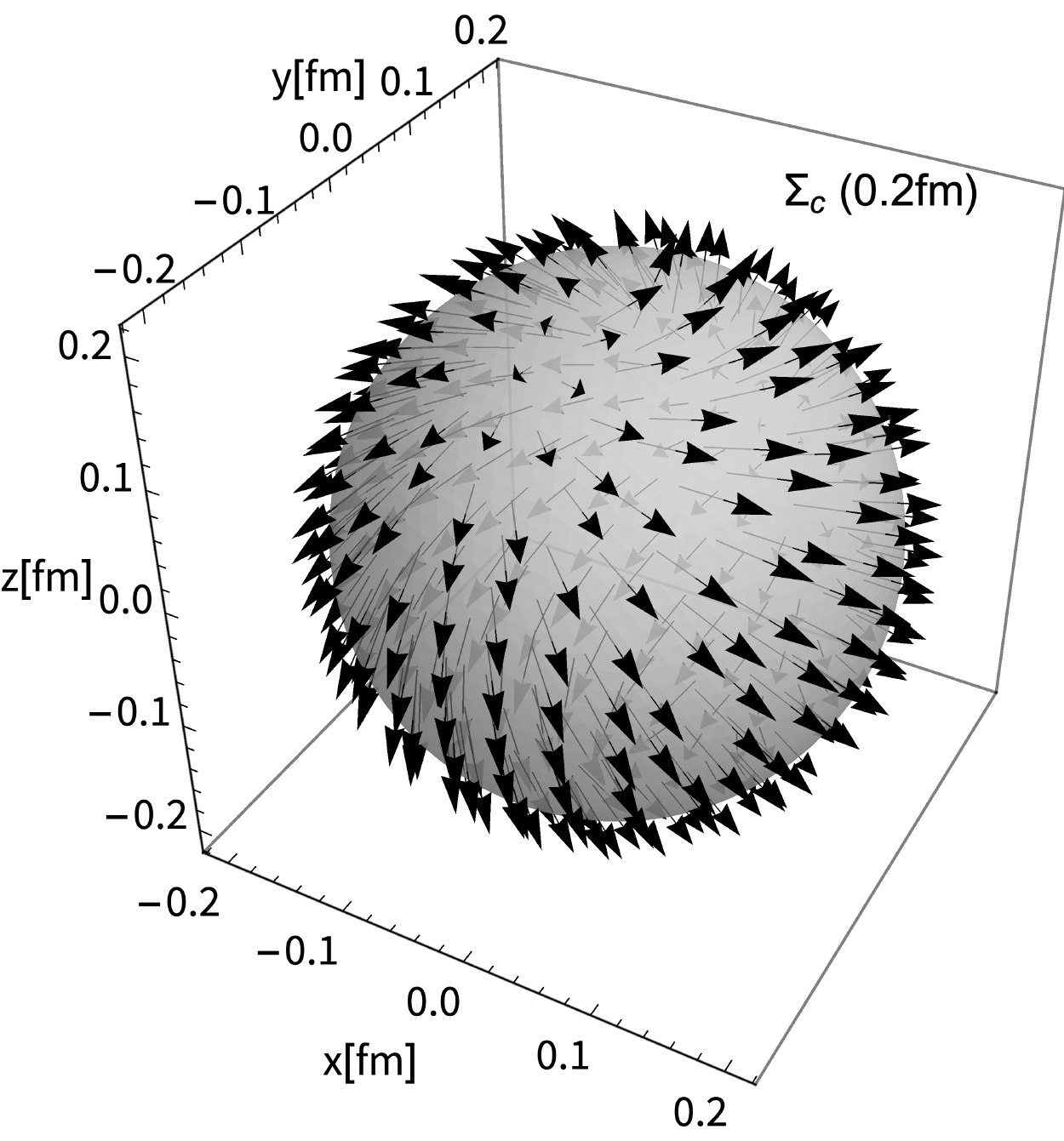}
\includegraphics[scale=0.70]{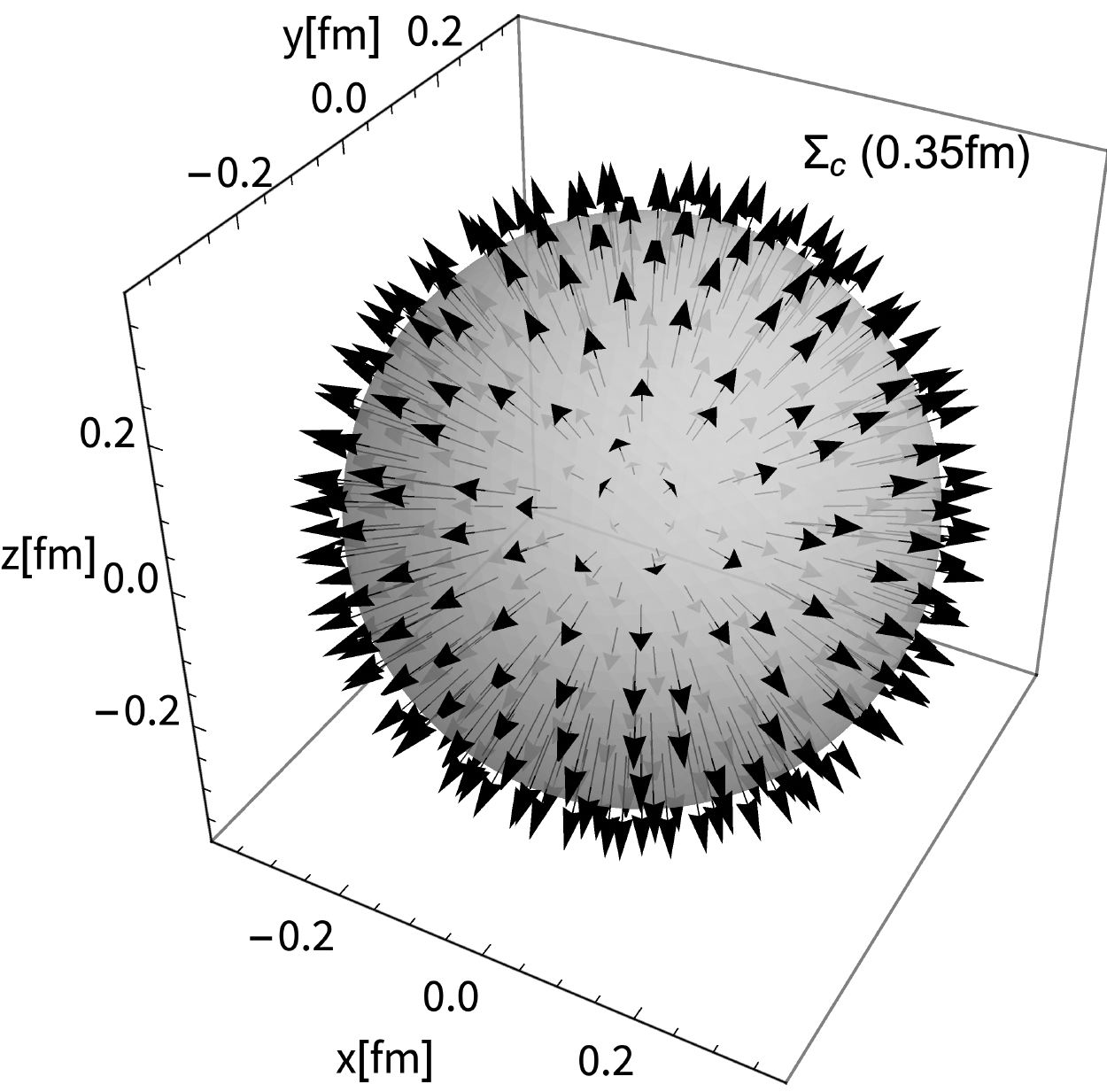}
\includegraphics[scale=0.70]{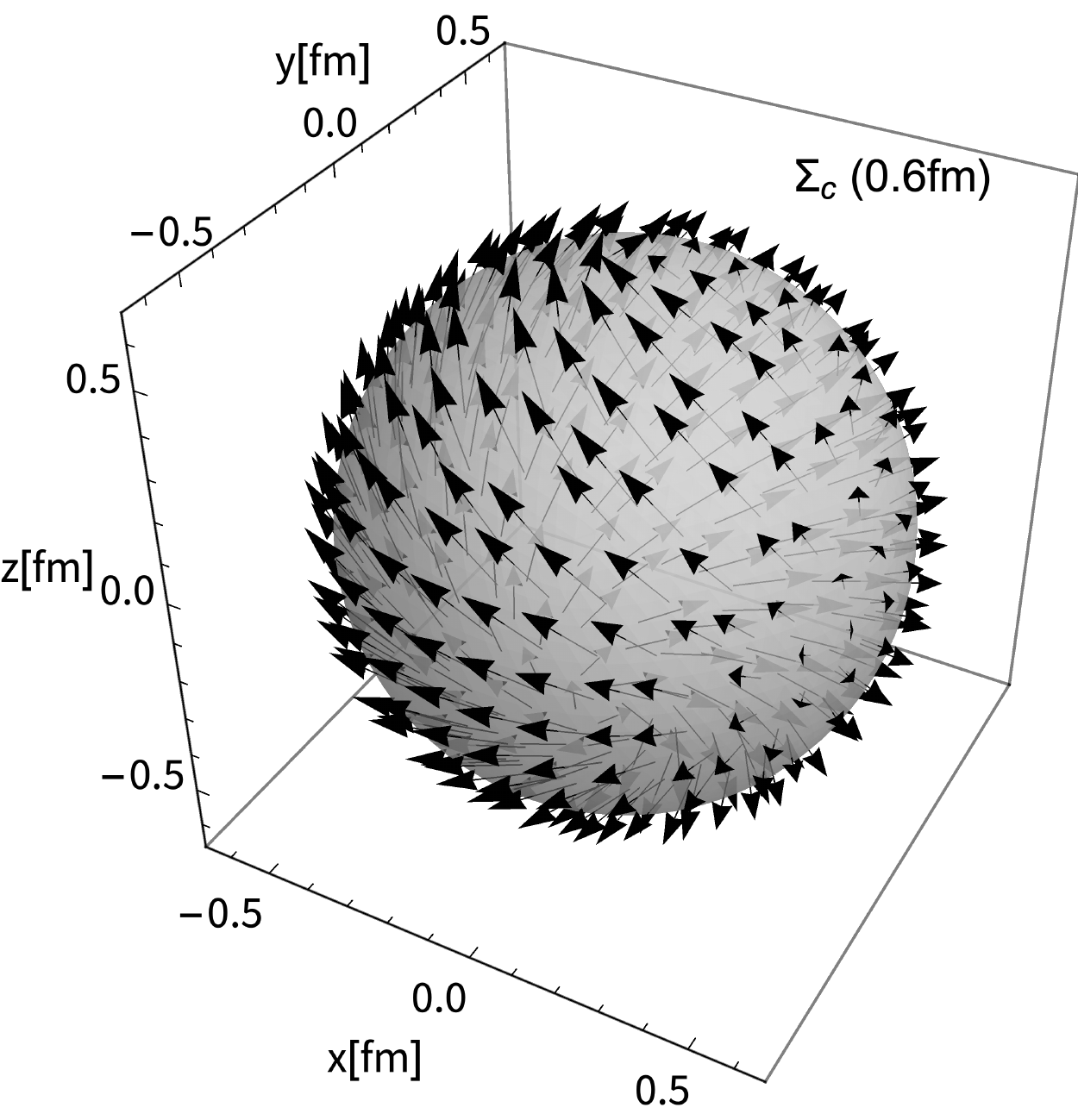}
\includegraphics[scale=0.70]{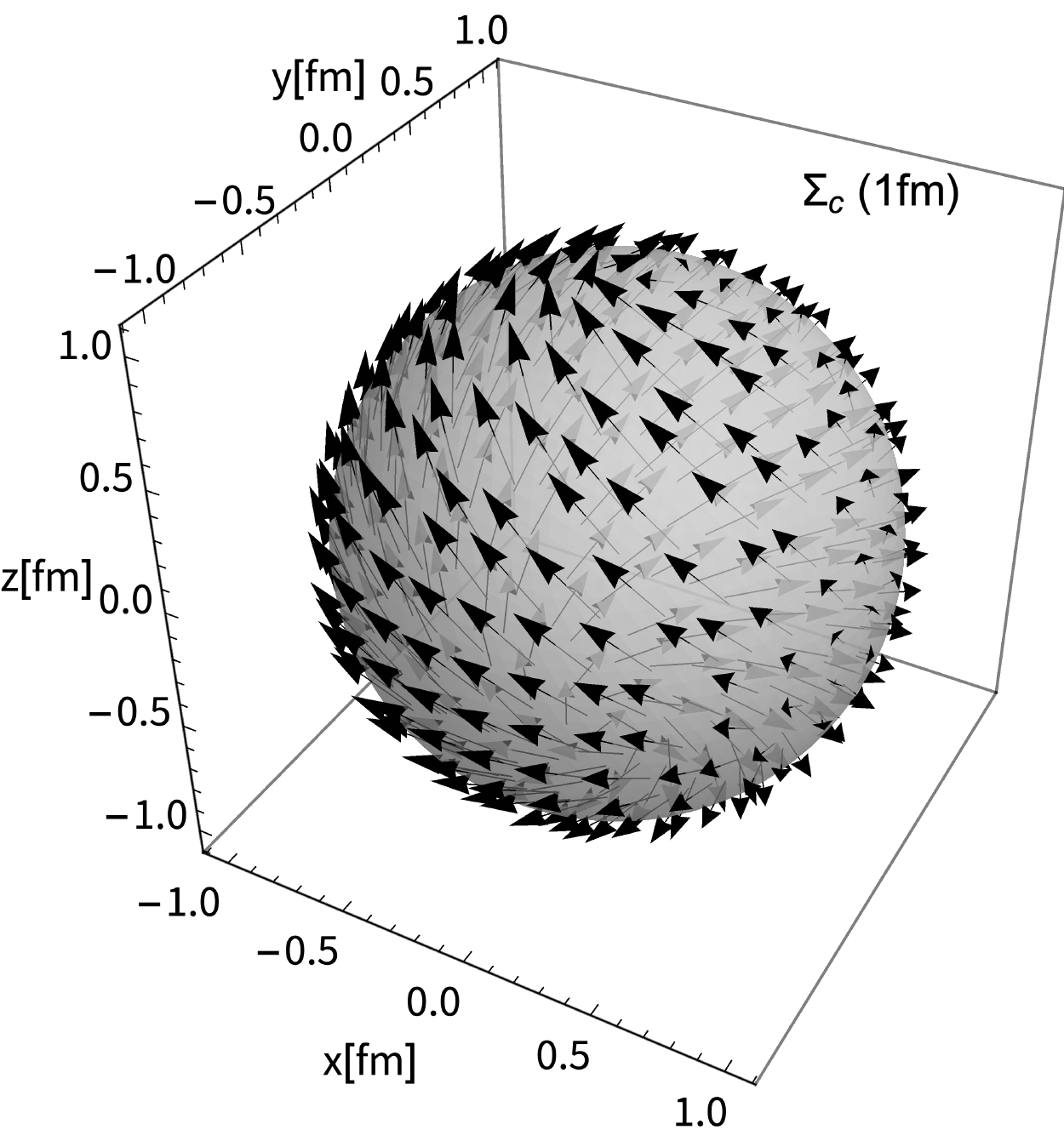}
\caption{3D visualization of the strong force field ($\bm{F}$) as a
  vector field inside $\Sigma_c$. We display the 3D force field
  exerting locally on each point in given shells with the distance
  from the center varied from  $0.2$ fm, $0.35$ fm, $0.6$ fm, and $1$
  fm. Note that at $0.35$ fm the tangential force field vanishes.}
\label{fig:11}
\end{figure}

Before illustrating the 3D visualization of the strong force fields,
we first define each strong force field acting on an infinitesimal
area. In the upper panel of Fig~\ref{fig:9}, the infinitesimal force
fields $dF_{(r,\theta,\phi)}$ defined in Eq.~\eqref{eq:strong_force} are
visualized as the arrows, which will be used in the 3D visualization
of the strong force fields. In the upper left and right panels of
Fig~\ref{fig:9}, we exhibit the 3D visualization of the strong force
fields ($\bm{F}$) for the nucleon and $\Sigma_c$,
respectively. 
\begin{figure}[htp]
\includegraphics[scale=0.70]{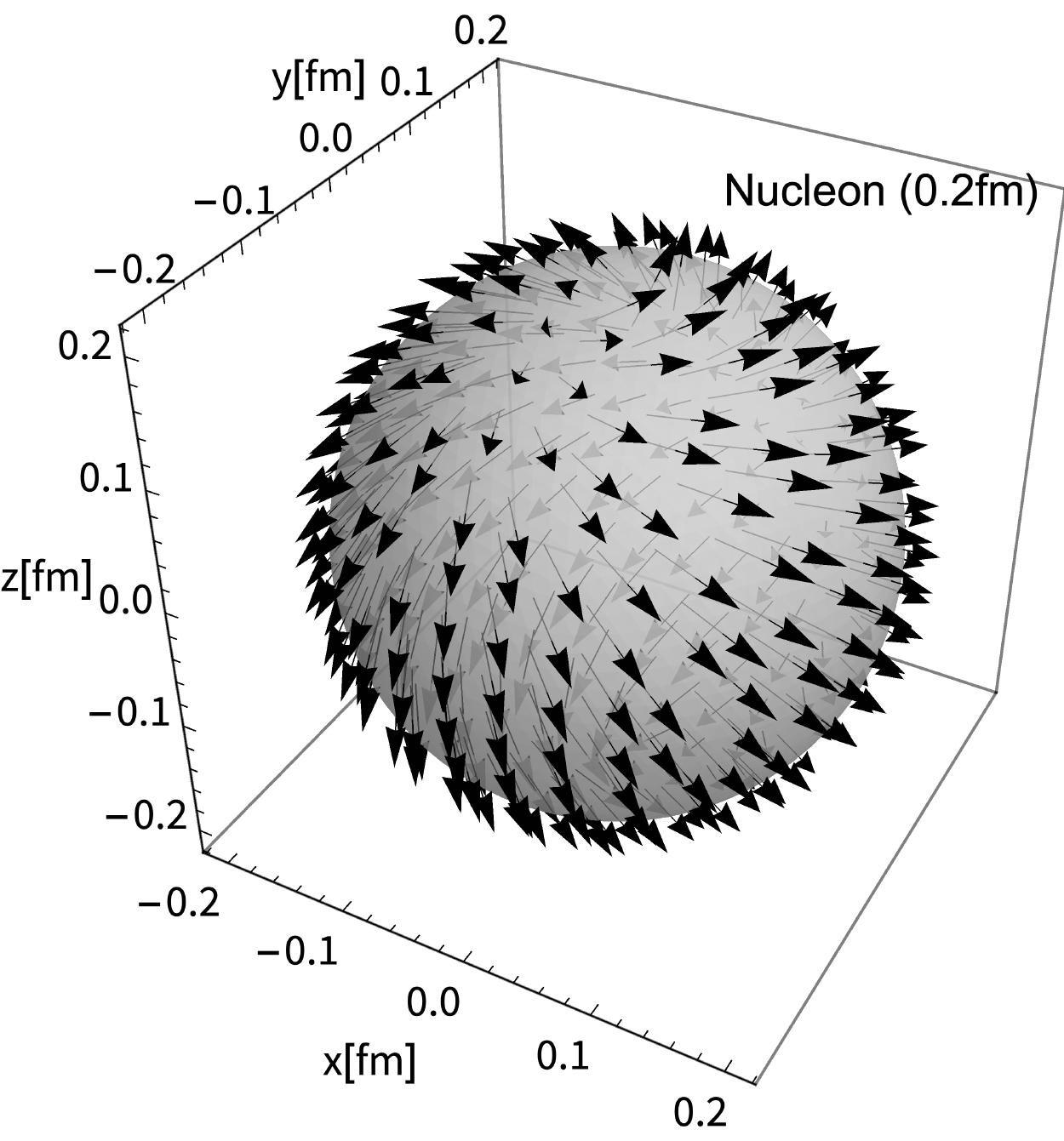}
\includegraphics[scale=0.70]{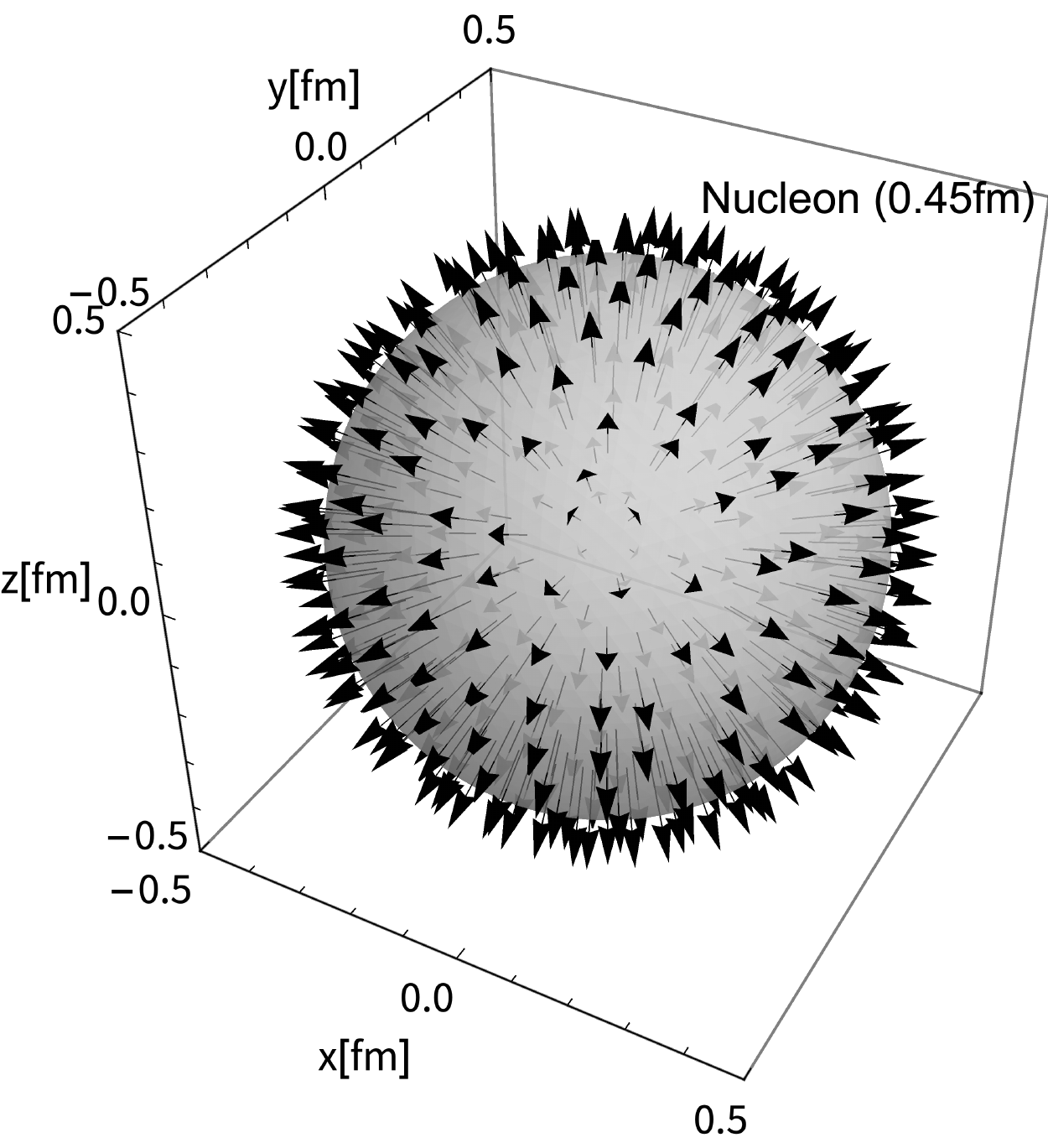}
\includegraphics[scale=0.70]{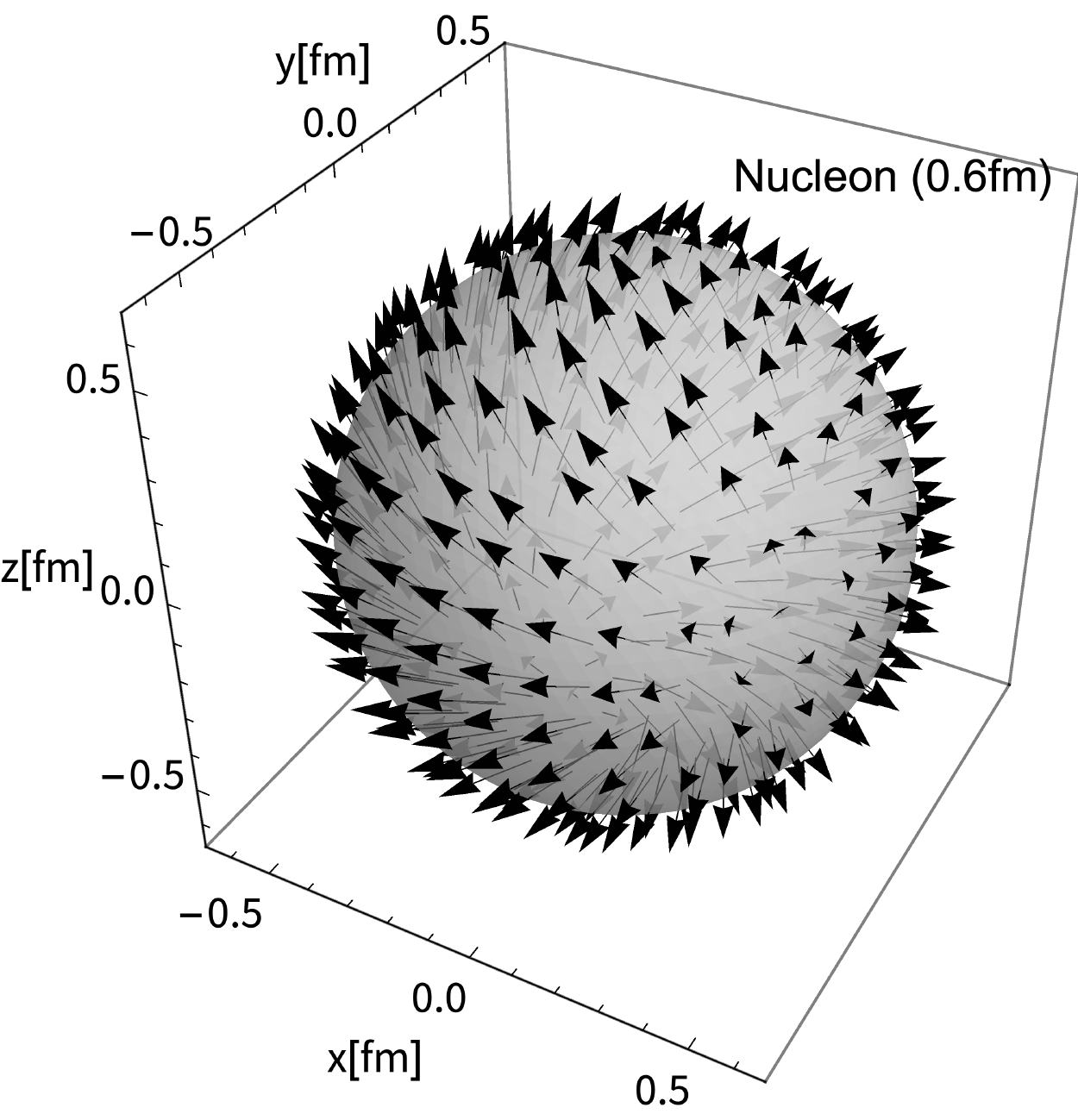}
\includegraphics[scale=0.70]{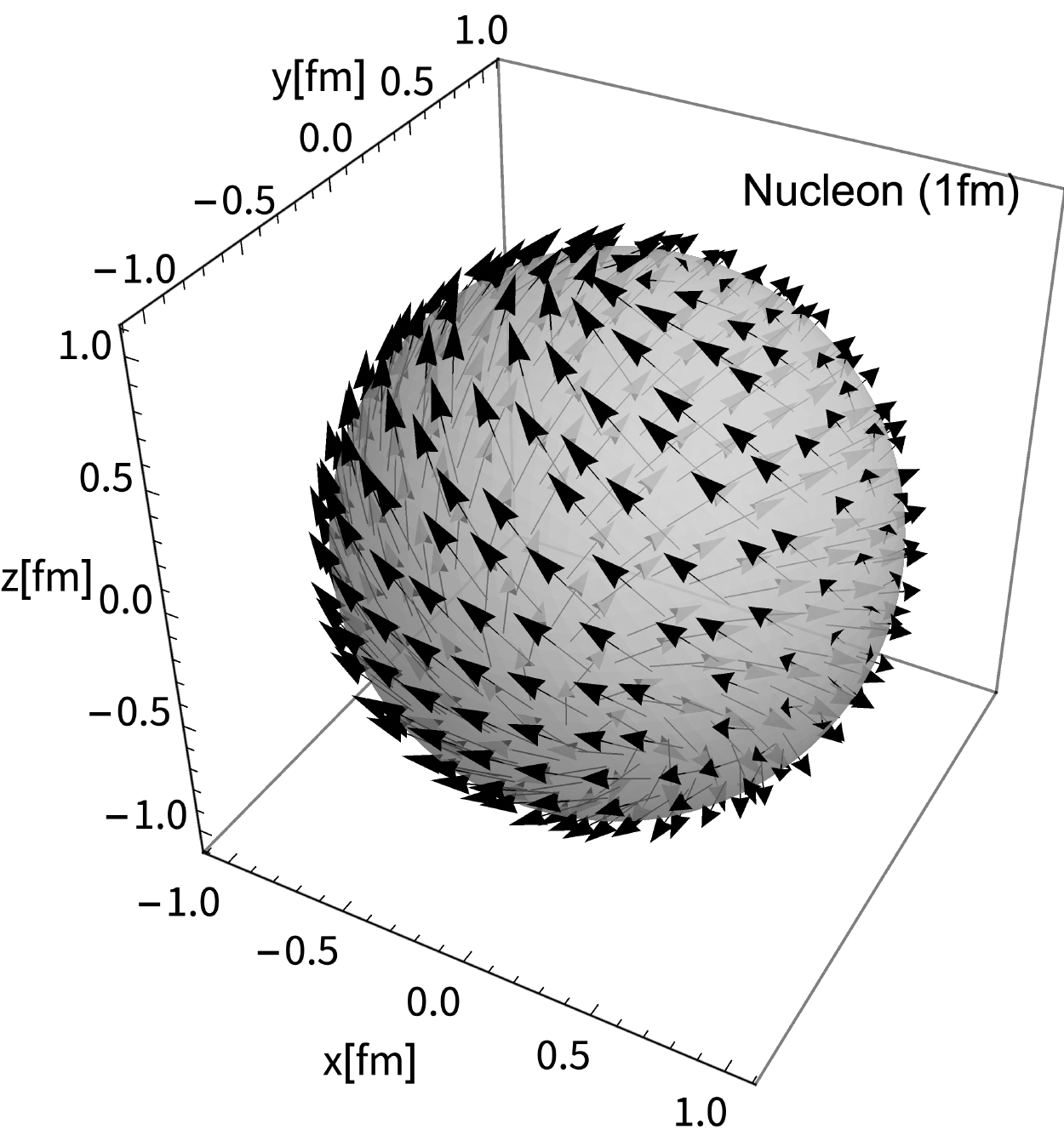}
\caption{3D visualization of the strong force field ($\bm{F}$) as a
  vector field inside a nucleon. We display the 3D force field
  exerting locally at each point in given shells with the distance
  from the center varied from  $0.2$ fm, $0.45$ fm, $0.6$ fm, and $1$
  fm. Note that at $0.45$ fm the tangential force field vanishes.}
\label{fig:10}
\end{figure}
In Fig.~\ref{fig:10} we illustrate how the strong force field inside a
nucleon undergoes the changes as the distance from its center
varies from $0.2$ fm to 1 fm. We want to emphasize that the force
field acts locally on each point of a 3D surface or a shell with a given
value of the distance. As shown in the upper left panel of
Fig.~\ref{fig:10}, the radial component of the strong force field
dominates over the tangential one. When the distance from the center
reaches $0.45$ fm, the tangential force field vanishes as already
shown in Fig.~\ref{fig:6}. As a result, the strong force field is
directed normally outwards at $0.45$ fm as displayed in the upper
right panel of Fig.~\ref{fig:6}. As $r$ further increases, however,
the signature of the tangential force field is changed, which
indicates that the direction of $F_\phi$ is reversed. This can be
easily understood by comparing the upper left panel of
Fig.~\ref{fig:10} with the lower left one. When $r$ becomes larger,
then $F_\phi$ dominates over $F_r$ as exhibited in the lower right
panel of Fig.~\ref{fig:10}.  In Fig.~\ref{fig:11}, we depict the 3D
visualization of the strong force field in the case of $\Sigma_c$. The
general behavior of the strong force field inside $\Sigma_c$ is very
similar to the nucleon case except that the magnitude of the strong
force field inside $\Sigma_c$ is weaker than that inside a nucleon.

\begin{figure}[htp]
\includegraphics[scale=0.28]{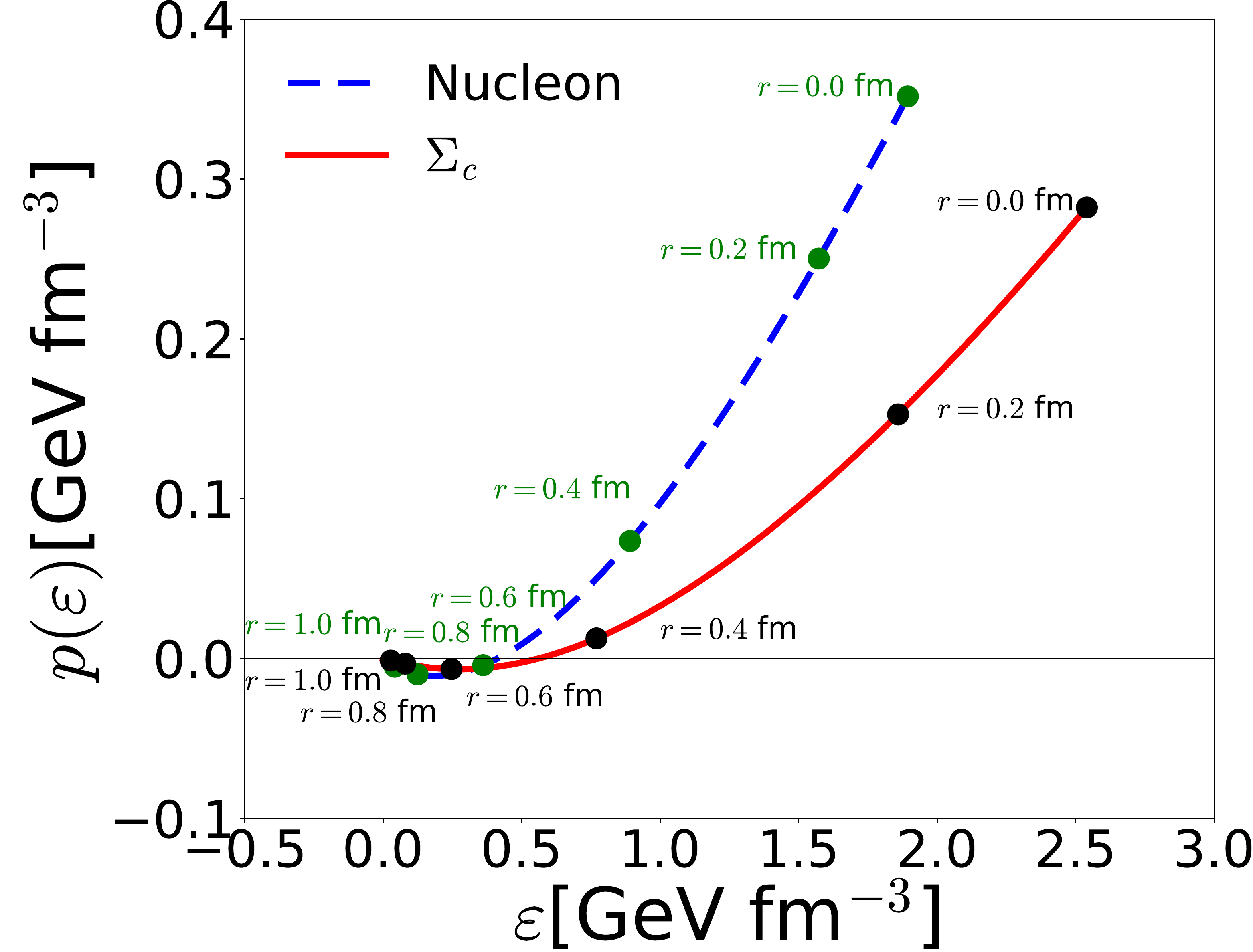}
\includegraphics[scale=0.28]{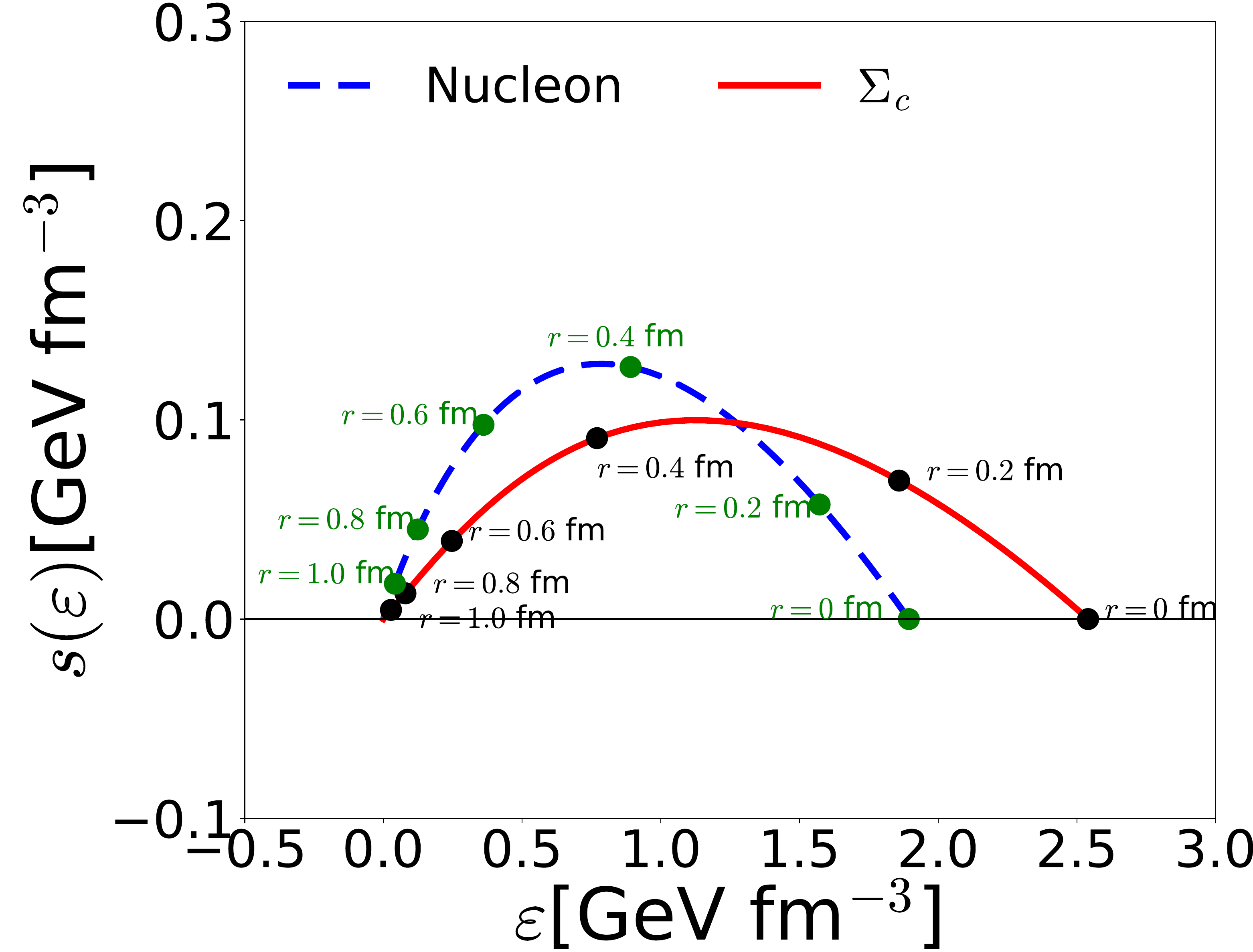}
\includegraphics[scale=0.28]{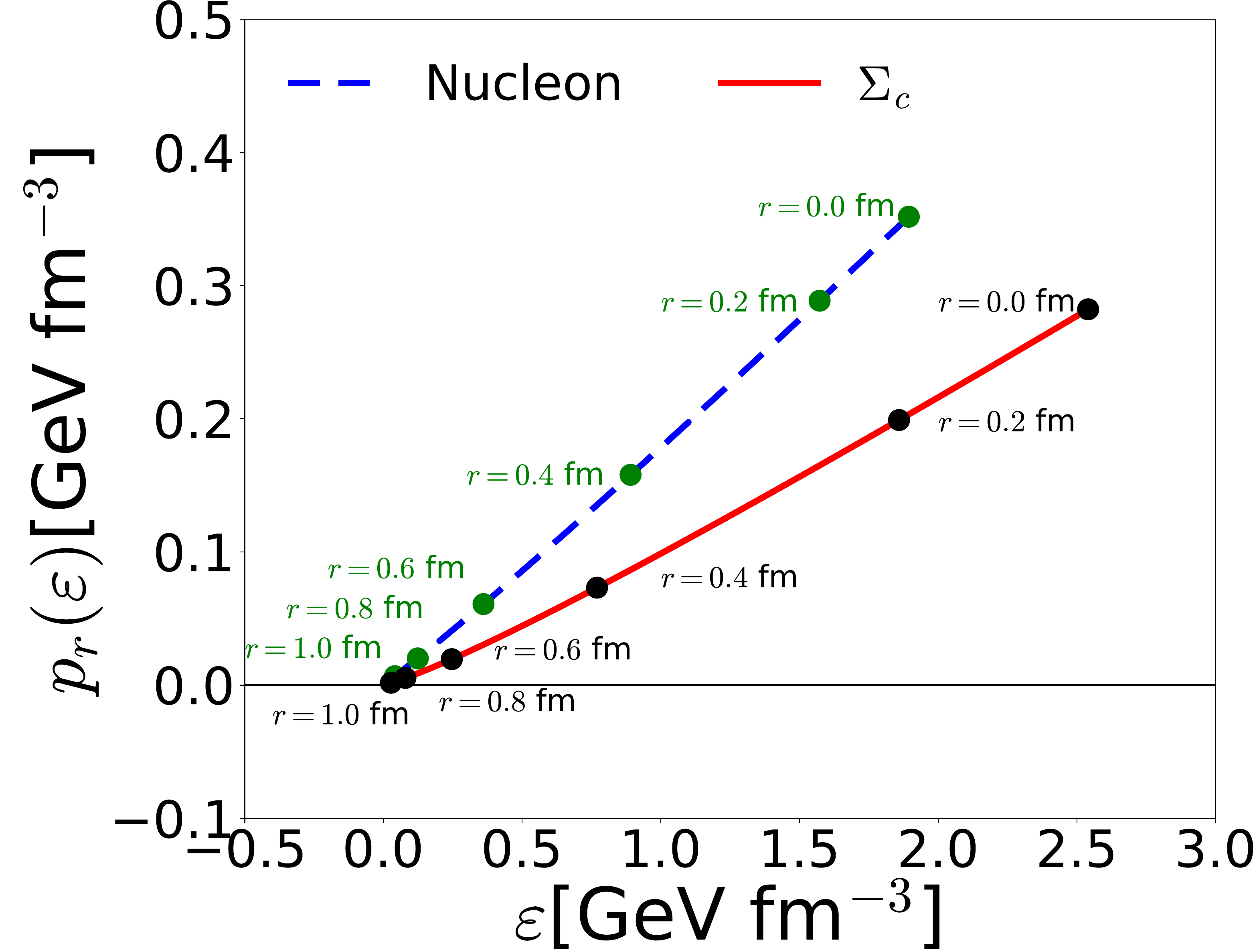}
\includegraphics[scale=0.28]{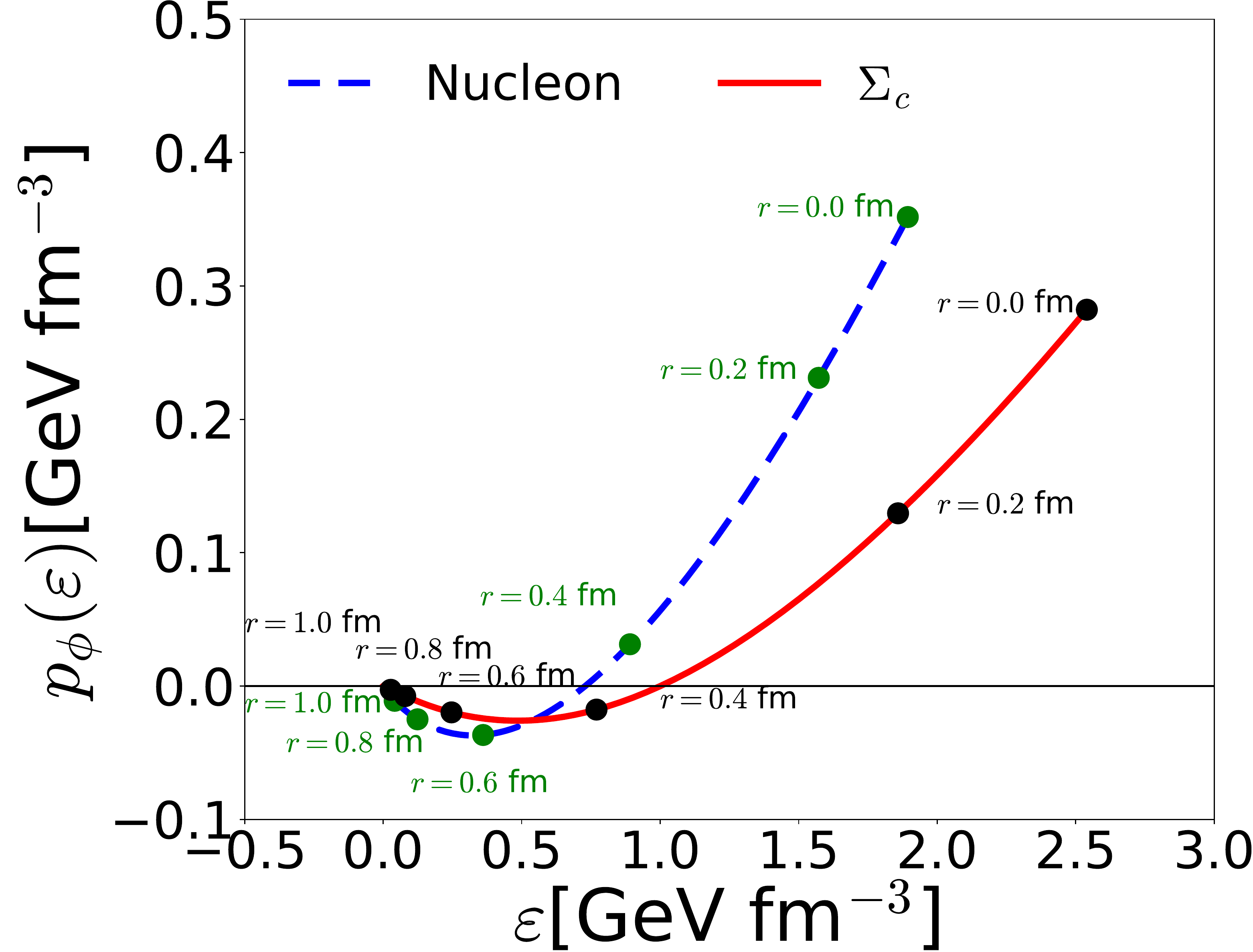}
\caption{The results for $p(\varepsilon)$, $s(\varepsilon)$,
$p_r(\varepsilon)$, and $p_\phi(\varepsilon)$. The solid curves
depict those of $\Sigma_c$ whereas the dashed ones draw those
  of the nucleon.}
\label{fig:12}
\end{figure}
Recently, the equation of state (EoS) inside a nucleon has been
conjectured in the hope that it may give a certain clue in
understanding the EoS inside compact stars~\cite{Lorce:2018egm,
  Polyakov:2018rew}.  Thus, we examine the EoS
inside a nucleon and $\Sigma_c$. In the upper left panel of
Fig.~\ref{fig:12}, we depict the pressure densities of the nucleon and
$\Sigma_c$ as functions of the energy density, $\varepsilon$. The
pressure density of the nucleon increases faster than that of
$\Sigma_c$. This result may arise from the fact that the pion mean
field for the nucleon is stronger than that for $\Sigma_c$. As we have
discussed the results for the energy densities in Fig.~\ref{fig:1} and
those for the pressure densities in Fig.~\ref{fig:3}, the energy
density of $\Sigma_c$ is stronger in the core part than that of the
nucleon, which originates from the different pion mean fields. On the
other hand, the pressure density of $\Sigma_c$ is overall weaker than
the nucleon one by the same reason. This leads to the fact that the
EoS inside the nucleon is stiffer than that inside $\Sigma_c$ again
due to the different pion mean fields. Interestingly, $p(\varepsilon)$
of both the nucleon and $\Sigma_c$ become negative and are saturated,
as $\varepsilon$ increases. Then $p(\varepsilon)$ starts to rise
monotonically as $\varepsilon$ further increases. A similar behavior
is also found in the results of $p_\phi(\varepsilon)$ drawn in the
lower right panel of Fig.~\ref{fig:12}.

Using the EoS for the nucleon and $\Sigma_c$, we can recapitulate the
stability conditions for the baryons. As already discussed in
Fig.~\ref{fig:1}, the energy density decreases, as $r$ increases,
being concentrated mainly on the inner part of the baryons. This means
that the region of smaller values of $\varepsilon$ corresponds to the
outer region of the baryons, in which the pressure densities are
negative. As discussed previously, the contribution of the Dirac
continuum is attractive whereas that of the discrete level is
repulsive. One can understand this feature from the results for the EoS
drawn in both the upper left panel and lower right panel of
Fig.~\ref{fig:12}. It is natural that in the region of the lower
energy density below approximately $\varepsilon \approx
0.5\,\mathrm{GeV\cdot fm^{-3}}$ the pressure densities should be
negative. As $\varepsilon$ increases, which means that we go into the
inner part of the baryons, the pressure density should become
positive. As a result, the pressure density has a saturation point
where $p(\varepsilon)$ starts to increase, as $\varepsilon$ increases.
This observation implies that the compliance with the von Laue
conditions for the stability of a baryon given in
Eqs.~\eqref{eq:stability} and~\eqref{eq:stability2D}
is related to the existence of the saturation point for $p(\varepsilon)$.
In the upper right panel of Fig.~\ref{fig:12}, we depict the results
for the shear-force densities as functions of $\varepsilon$. As
discussed already in Fig.~\ref{fig:5}, $s(r)$ turns out positive
for all the values of $r$. Note that the energy density is also
positive definite over all the regions. As a result, $s(\varepsilon)$
turns out positive as shown in the upper right panel of
Fig.~\ref{fig:12}. This indicates again that the stabilities of the
nucleon and $\Sigma_c$ are secured.
In the lower left panel of Fig.~\ref{fig:12}, we illustrate the
results for $p_r(\varepsilon)$. Since $\varepsilon$ is positive
definite over all the values of $r$, the results for $p_r(\varepsilon)$ imply
the local stability conditions $p_r(r)>0$. Interestingly, they exhibit
behaviors similar to the EoS for compact
stars~\cite{Paschalidis:2017qmb, Weber:2004kj, Ozel:2016oaf, Baym:1976yu, RikovskaStone:2006ta, Abhishek:2018xml}.

\subsection{Results for the gravitational form factors}
\begin{figure}[htp]
\includegraphics[scale=0.24]{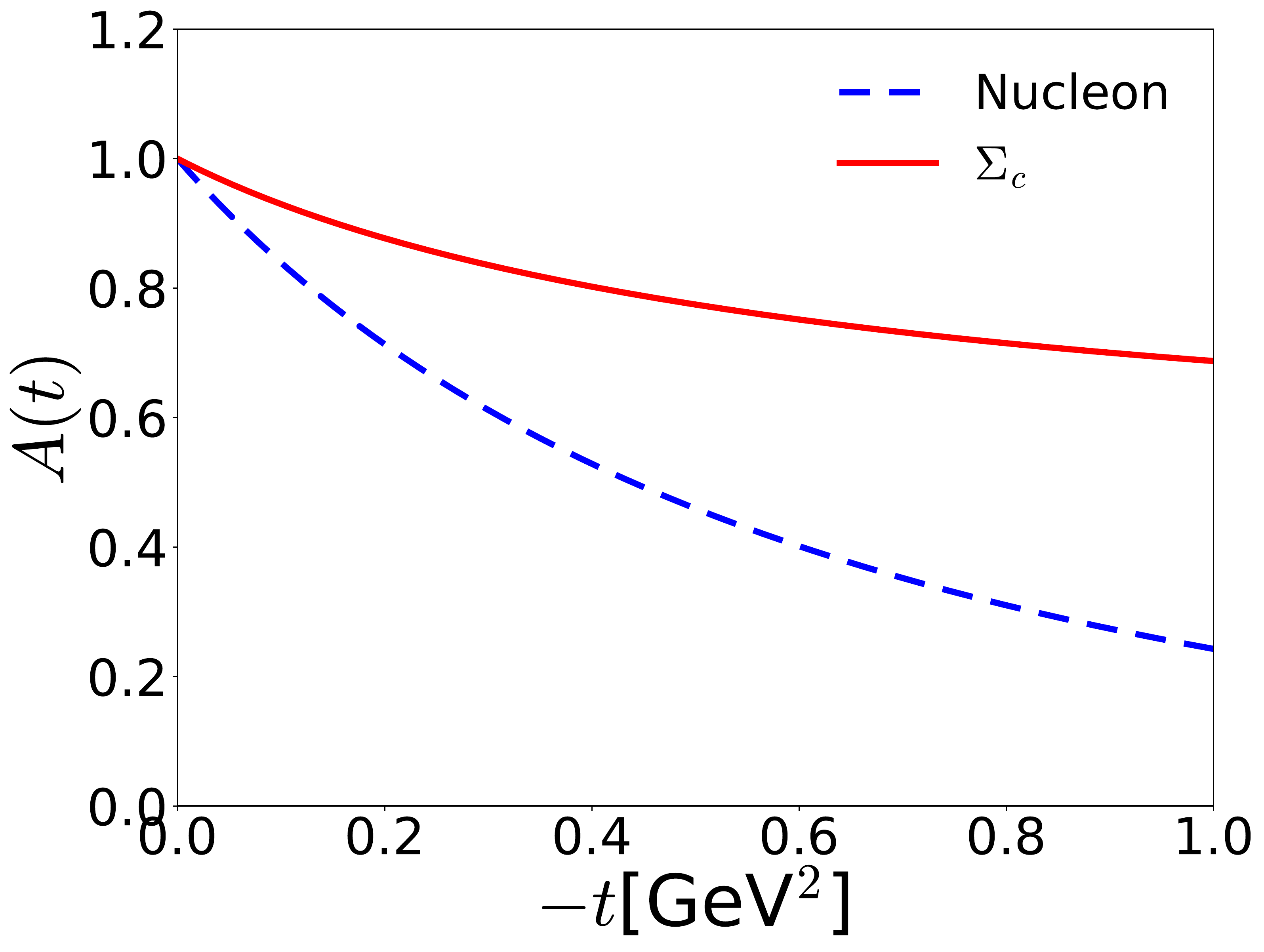}
\includegraphics[scale=0.23]{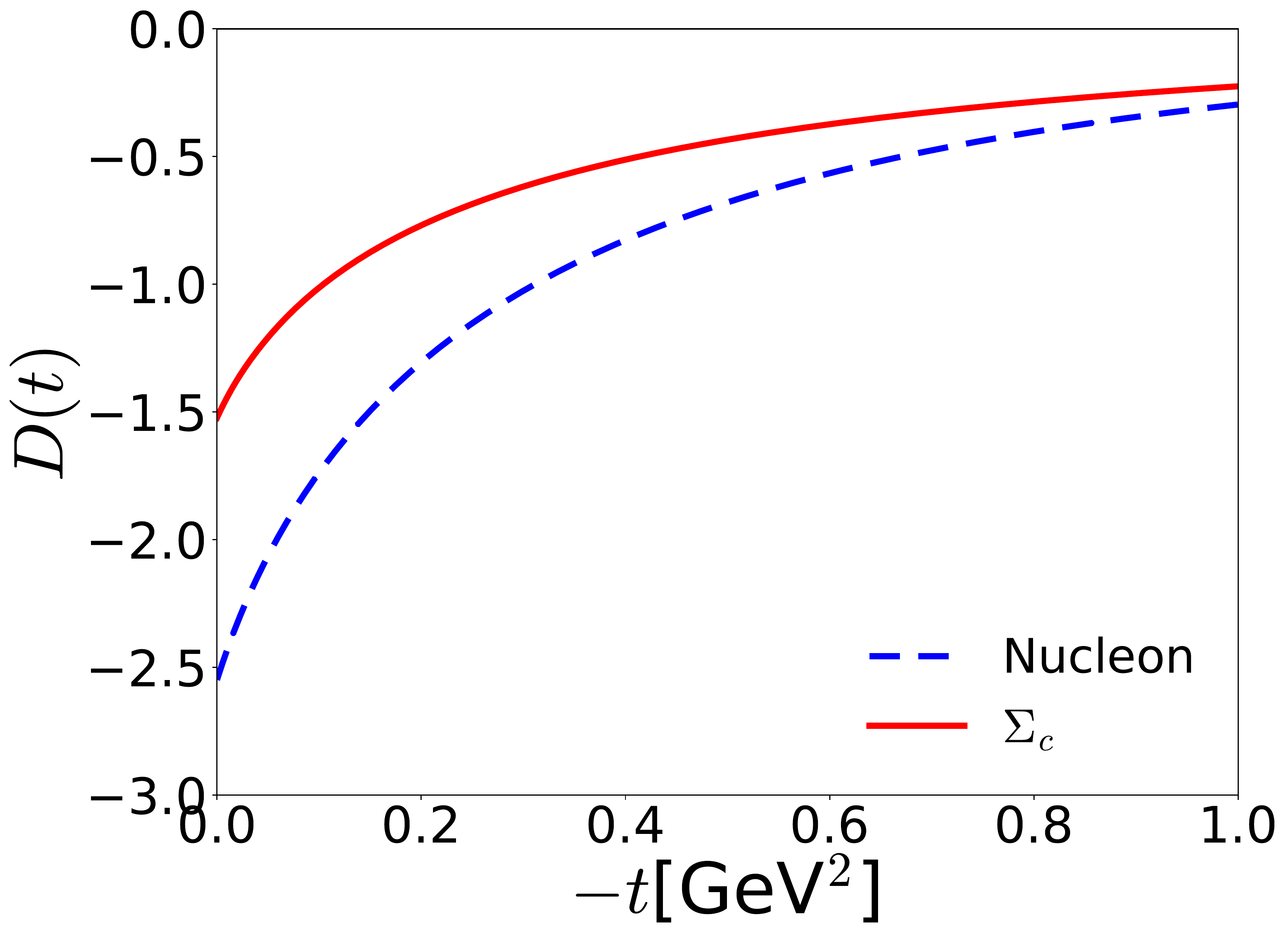}
\includegraphics[scale=0.23]{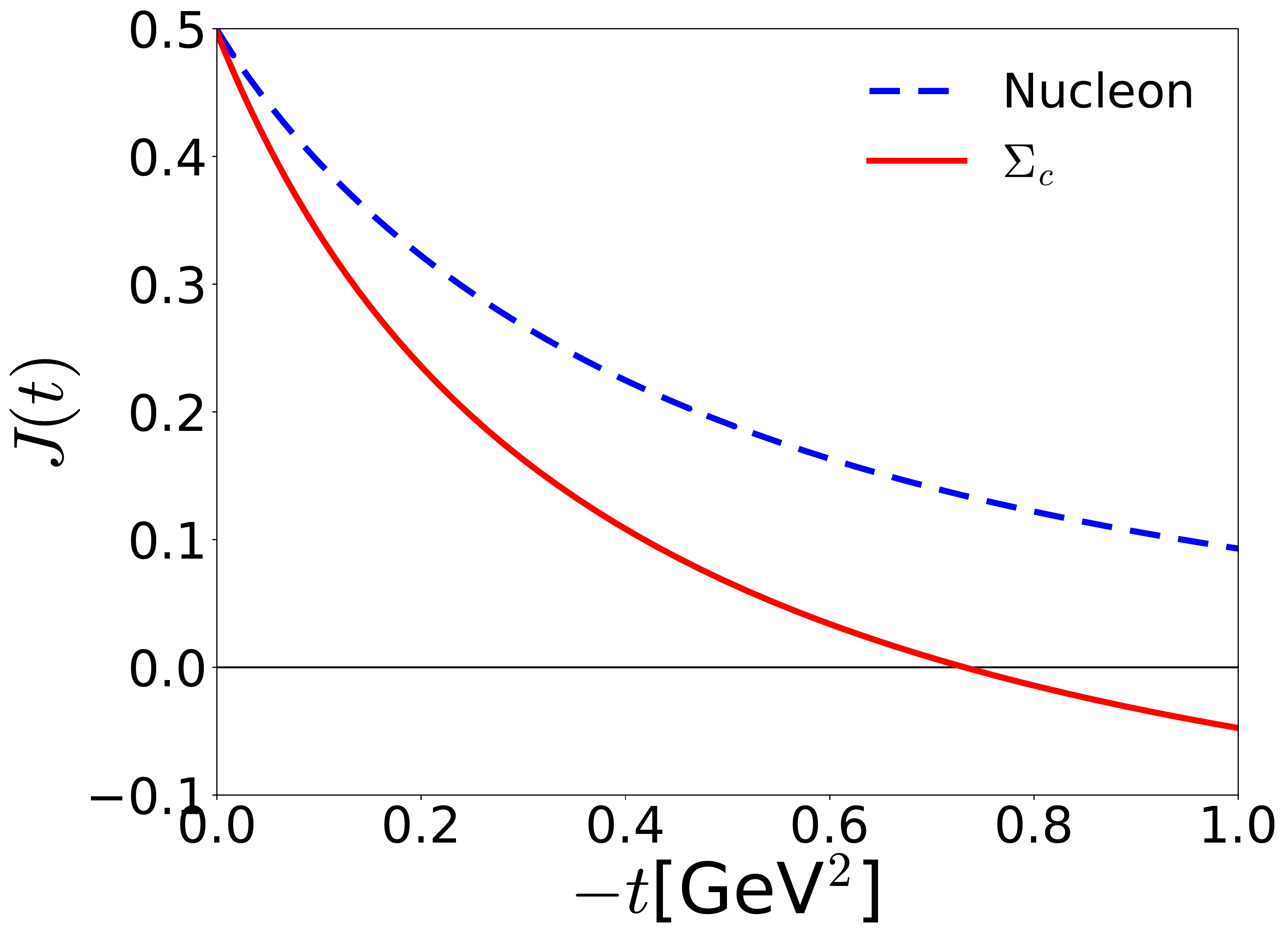}
\caption{Results for the gravitational form factors $A(t)$, $D(t)$,
  and $J(t)$. The solid curves depict those of $\Sigma_c$ whereas the
  dashed ones draw those of the nucleon.}
\label{fig:13}
\end{figure}
In Fig.~\ref{fig:13} we present the numerical results for the
gravitational form factors of $\Sigma_c$ in comparison with those of
the nucleon. In the upper left panel of Fig.~\ref{fig:13}, the results
for $A(t)$ show that the nucleon form factor falls off faster than
that of $\Sigma_c$. This means that $\Sigma_c$ is more
compact object than the nucleon. A similar feature was found in the
case of the electric form factors of $\Sigma_c$~\cite{Kim:2018nqf}.
The upper right panel depicts the results for the $D$-term form
factors of $\Sigma_c$ and the nucleon. Both the nucleon and $\Sigma_c$
form factors are negative, which ensures the stabilities of both the
baryons. In the lower panel, we illustrate the results for the $J(t)$
form factors. In contrast to $A(t)$, the result for $J(t)$ of
$\Sigma_c$ falls off faster than that for the nucleon. As we have
mentioned previously already, the main contribution to $J(t)$ of
$\Sigma_c$ comes from the solitonic part that has spin 1, so that the
total angular-momentum density of $\Sigma_c$ becomes larger than the
proton one, as shown in Fig.~\ref{fig:2}. This leads to the results
given in the lower panel of Fig.~\ref{fig:13}. In Table~\ref{tab:1},
we list the results for various observables for the nucleon and $\Sigma_c$.

\setlength{\tabcolsep}{5pt}
\renewcommand{\arraystretch}{1.5}
\begin{table}[htp]
\caption{Various observables for the nucleon and $\Sigma_{c}$:
  The energy densities at the center $\varepsilon(0)$, the mean square
  radii $\langle r^{2}_{E} \rangle$, $\langle r^{2}_{J} \rangle$,the
  pressure densities $p(0)$ at the center of the nucleon,
  the pressure densities $p(r_{0})=0$ at the point $r_0$ where they
  vanish, the $D$-term values $D$, the mean squared  radii of the
  trace of the EMT, and the mechanical radii $\langle
  r^{2}_{\mathrm{mech}} \rangle$. We compare the results using
  $m_\pi=140$ MeV with those in the chiral limit.
  The heavy quark mass is taken to be $m_{c}=1.27$~GeV. }
\centering
\label{tab:1}
\begin{tabular}{c c c c c c c c c c}
\hline
\hline
  &$m_{\pi}$ & $\varepsilon(0)$ & $\langle r^{2}_{E} \rangle$ &
  $\langle r^{2}_{J} \rangle$ & $p(0)$ & $r_{0}$ & $D(0)$ &
  $\langle r^{2}_{F} \rangle$ & $\langle r^{2}_{\mathrm{mech}} \rangle$  \\
  & $[$MeV$]$ & [GeV/fm$^{3}$] & [fm$^{2}$] & [fm$^{2}$] &
 [GeV/fm$^{3}$] & [fm] &  & [fm$^{2}$] & [fm$^{2}$]  \\
\hline
  \multirow{2}{*}{$N$}&$0$   & 1.66 & 0.65 & $\infty$ & 0.305 & 0.59
                                                 & -3.07 & 0.71 & 0.72 \\
  &$140$ & 1.89 & 0.54 & 1.02    & 0.352
                                       & 0.57 & -2.55 & 0.58 & 0.55 \\
\hline
  \multirow{2}{*}{$\Sigma_{c}$}&$0$   & 2.39 & 0.24 & $\infty$
                              & 0.242 & 0.45 & -1.79 & 0.28 & 0.64 \\
  &$140$ & 2.54 & 0.21 & 1.56 & 0.282
                                       & 0.46 & -1.52 & 0.25 & 0.45 \\
\hline
\hline
\end{tabular}
\end{table}

\section{Summary and conclusions}
In the present work, we aimed at investigating the stability
conditions and strong force fields for the nucleon and the singly
heavy baryon $\Sigma_c$, emphasizing the differences between them. 
In the chiral quark-soliton model, the pion mean field for $\Sigma_c$
is weaker than that for the nucleon, since the presence of the $N_c-1$
level quarks inside a singly heavy baryon create the pion mean field,
whereas the nucleon arises as a bound state of the $N_c$ level quarks
that are bound by the stronger $N_c$ pion mean field. This difference
will be inherited into the results for $\Sigma_c$. We found that the
$N_c-1$ pion mean fields should be evaluated
self-consistently. Otherwise, the stability conditions for $\Sigma_c$
will be broken.  Starting from the matrix elements of the
energy-momentum tensor current for the nucleon and $\Sigma_c$, we   
were able to derive the four different densities:
the energy densities, the total angular-momentum densities, the
pressure densities, and the shear-force densities. The energy density
of $\Sigma_c$ is narrower than that of the nucleon, which indicates
that the singly heavy baryon $\Sigma_c$ is a more compact object than
the nucleon. The total angular-momentum density of $\Sigma_c$ is
wider and stronger than the nucleon one. This can be understood by the
fact that the spin distribution of $\Sigma_c$ is dominated by the 
$N_c-1$ soliton that is formed as a spin 1 state whereas the spin
distribution of the nucleon arises from the $N_c$ soliton with spin
1/2. We found that the present results for the
pressure and shear-force densities satisfy both the global and local
stability conditions. As already shown in the previous work in the
chiral quark-soliton model~\cite{Goeke:2007fp}, the global stability
condition for the nucleon is also secured by the balance between the
leavel-quark and Dirac-continuum contributions. The results for
$\Sigma_c$ also satisfy the stability conditions in the same manner. 
The shear-force densities turned out positive definite over all the
values of the distance from the center 
of the baryons.

The strong force fields are defined in terms of the
pressure and shear-force densities, which are exerted on the shell of
$\Sigma_c$ and the nucleon, given a distance from the center of the
baryons. The strong force fields can be decomposed into the normal and
tangential components, which are expressed by the normal and
tangential densities. These two densities are written in terms of the
pressure and shear-force ones. In particular, the positivity of the normal 
pressure density is identified as the local stability condition. The 
results for the normal pressure densities of the nucleon and
$\Sigma_c$ fulfill the local stability condition for them. The
tangential components of the strong force fields for the nucleon and
$\Sigma_c$ have at least one nodal point as in the case of the
pressure densities. This indicates that the tangential force fields
should change the direction. Thus, they swirl counterclockwise in the
inner parts of both the nucleon and $\Sigma_c$, whereas they circulate
around oppositely in the outer regions of the nucleon and
$\Sigma_c$. We also examined the equations of state with the
conjecture that the results for them may shed light on the inner
structure of compact stars.

Finally, we presented the results for the
gravitational form factors of the nucleon and $\Sigma_c$. The mass
form factor of the $\Sigma_c$ falls off slower than that of the
nucleon, which implies that $\Sigma_c$ is a more compact object than
the nucleon. As expected from the discussion of the stability
conditions, the $D$-term form factors of both the nucleon and
$\Sigma_c$ turn out to be negative. The total angular-momentum form
factor of $\Sigma_c$ falls off faster than that for the nucleon in
contrast to the case of the energy form factors. The reason comes from
the fact that the main contribution to the total angular-momentum form
factor of $\Sigma_c$ is governed by the solitonic part with spin
1.

The present work can be extended to the grativational form factors of
the baryon sextet and decuplet. To do that, we
need to consider explicitly the strange-quark contributions and to see
how the strange quarks come into play in understanding the stabilities
of the baryons with spin 3/2. The corresponding investigations are
under way.  

%-------------------------------------------------
\begin{acknowledgments}
%-------------------------------------------------
The present work was supported by Basic Science Research Program
through the National Research Foundation of Korea funded by the
Ministry of Education, Science and Technology
(Grant-No. 2018R1A2B2001752 and 2018R1A5A1025563). J.-Y.K is supported
by the Deutscher Akademischer Austauschdienst(DAAD) doctoral scholarship.
Work of MVP is supported in part by BMBF (Grant No. 05P18PCFP1).
\end{acknowledgments}

\appendix

\section{Regularization functions\label{app:a}}
The proper-time regularization functions used in Eq.~\eqref{eq:EMT_light}, Eq.~\eqref{eq:EMT_total} and Eq.~\eqref{eq:EMT_dens}
are defined by 
\begin{align}
R_{1}(E_{n},\Lambda)&=\frac{1}{4\sqrt{\pi}}
                             \int^{\infty}_{\Lambda^{-2}}
                             \frac{du}{u^{3/2}}e^{-uE^{2}_{n}}, \cr 
R_{2}(E_{n},\Lambda)&=\frac{1}{4\sqrt{\pi}}
                             \int^{\infty}_{\Lambda^{-2}}
                             \frac{du}{u^{1/2}}E_{n}e^{-uE^{2}_{n}},
                             \cr 
R_{3}(E_{n},E_{m},\Lambda)&=\frac{1}{4\sqrt{\pi}}
                                   \int^{\infty}_{\Lambda^{-2}} 
 \frac{du}{u^{1/2}}\left(\frac{e^{-uE^{2}_{n}}-e^{-uE^{2}_{m}}}{u^{3/2}(E^{2}_{m}-E^{2}_{n})} 
 - \frac{E_{n}e^{-uE^{2}_{n}}-E_{m}e^{-uE^{2}_{m}}}{u^{1/2}(E_{m}+E_{n})}\right). \cr
\end{align}

%=========================================================

\end{document}